\documentclass{aa}
\pdfoutput=1
\usepackage{graphicx}
\usepackage[varg]{txfonts}
\usepackage[breaklinks]{hyperref}

\usepackage{multirow}  
\usepackage{enumitem}  
\usepackage{subfigure}
\usepackage{epstopdf} 
\usepackage{color}
\usepackage{longtable}
\usepackage[normalem]{ulem}

\newcommand{\teff}{\ensuremath{{T_{\rm eff}}}}           
\newcommand{\logg}{\ensuremath{\log g}}                  
\newcommand{\loggf}{\ensuremath{\log gf}}                
\def\kms{$\mathrm {km~s}^{-1}$}

\def\vsini {$v\sin i$}
\def\met {$\mathrm{[Fe/H]}$}                                                       
\def\vmic {$V_{\rm micro}$}
\def\vmac {$V_{\rm macro}$}
\def\chisq{$\mathrm{\chi^2_{red}}$}
\def\liseven {$\mathrm{^{7}Li}$}
\def\lisix {$\mathrm{^{6}Li}$}
\def\ali{$A(\mathrm{Li})$}
\def\alionednlte{$A(\mathrm{Li})_{\rm \,1DNLTE}$}
\def\aliTd{$A(\mathrm{Li})_{\rm \,3DNLTE}$}
\def\qli{$q(\mathrm{Li})$}
\def\qliTd{$q(\mathrm{Li})_{\rm \,3DNLTE}$}

\def\iso{$\mathrm{^{6}Li/^{7}Li}$}
\def\isoTd{$\mathrm{^{6}Li/^{7}Li}_{\rm \,3DNLTE}$}
\def\isoonednlte{$\mathrm{^{6}Li/^{7}Li}_{\rm \,1DNLTE}$}
\def\delA{$\Delta A_{\rm \,3DNLTE-1DLTE}$}
\def\delq{$\Delta q_{\rm \,3DNLTE-1DLTE}$}
\def\delAstar{$\Delta^{\ast}A_{\rm \,3DNLTE-1DLTE}$}
\def\delqstar{$\Delta^{\ast}q_{\rm \,3DNLTE-1DLTE}$}
\newcommand{\linfor}{{\sf Linfor3D}}
\newcommand{\nlte}{{\sf NLTE3D}}
\newcommand{\cobold}{{\sf CO$^5$BOLD}}
\newcommand{\mlhd}{{\sf LHD}}
\begin{document}

   \title{3D non-LTE corrections for Li abundance and \iso\ isotopic ratio in solar-type stars \thanks{The table with the 3D NLTE corrections is available only in electronic form at the CDS ... }}

   \subtitle{I. Application to HD\,207129 and HD\,95456  }

   \author{
                G. \,Harutyunyan\inst{1}, 
                M.\,Steffen\inst{1},
                A.\,Mott\inst{1},
                E.\,Caffau\inst{2},
                G. \,Israelian\inst{3},
                J.I. Gonz\'{a}lez Hern\'{a}ndez\inst{3}, 
                \and K.G.\,Strassmeier\inst{1}
        }

        \institute{
        Leibniz-Institut f\"ur Astrophysik Potsdam, An der Sternwarte 16, 14482
        Potsdam, Germany, \email{gharutyunyan@aip.de}
        \and
        GEPI, Observatoire de Paris, PSL Research University, CNRS, Univ. Paris           Diderot, Sorbonne Paris Cit\'e, Place Jules Janssen,
        92190 Meudon, France
        \and Instituto de Astrofísica de Canarias, 38200 La Laguna, Tenerife,                 Spain \\
}

\authorrunning{G. Harutyunyan et al.}
\titlerunning{3D non-LTE corrections for the Li abundance and \iso\ ratio}

\date{Received DD MM YY/\,Accepted DD MM YY}

\abstract
{Convective motions in solar-type stellar atmospheres induce Doppler shifts
  that affect the strengths and shapes of spectral absorption 
  lines and create slightly
  asymmetric line profiles. One-dimensional (1D) local thermodynamic equilibrium (LTE) studies of elemental abundances are not
  able to reproduce this phenomenon, which becomes particularly important when
  modeling the impact of isotopic fine structure, like the subtle depression 
  created by the \lisix\ isotope on the red wing of the \ion{Li}{i} resonance 
  doublet line.
}
{ The purpose of this work is to provide corrections for the lithium
    abundance, \ali, and the \iso\ ratio that can easily be applied to correct
    1D LTE lithium abundances in G and F dwarf stars of approximately solar
    mass and metallicity for three-dimensional (3D) and non-LTE (NLTE) effects.
}   
{The corrections for \ali\ and \iso\ are computed using
grids of 3D NLTE and 1D LTE synthetic lithium line profiles, generated
from 3D hydro-dynamical \cobold\ and 1D
hydrostatic model
atmospheres, respectively. For comparative purposes, all calculations are
performed for three different line lists representing the  \ion{Li}{i}
$\lambda$670.8\,nm spectral region. The 3D NLTE corrections are then
approximated by analytical expressions as a function of the stellar 
parameters (\teff, \logg, \met, \vsini, \ali, \iso).
These are applied to adjust the 1D LTE isotopic lithium abundances in two solar-type stars, HD\,207129 and HD\,95456, for which high-quality HARPS observations are available.
}
{The derived 3D NLTE corrections range
between $-0.01$ and $+0.11$\,dex for \ali, and between $-4.9$ and $-0.4$\% for
the \iso\ ratio, depending on the adopted stellar parameters. We confirm that
the inferred \lisix\ abundance depends critically on the strength
of the \ion{Si}{i} 670.8025~nm line. Our findings show a
general consistency with recent works on lithium abundance corrections. After
the application of such corrections, we do not find a significant amount of
\lisix\ in any of the two target stars.}
{ In the case of the \iso\ ratio, our corrections are always
  negative, showing that 1D LTE analysis can significantly overestimate the
  presence of \lisix\ (up to 4.9 percentage points) in the atmospheres of
  solar-like dwarf stars. These results emphasize the importance of
  reliable 3D model atmospheres combined with NLTE line
  formation for deriving precise isotopic lithium abundances. Although
  3D NLTE spectral synthesis implies an extensive computational effort,
  the results can be made accessible with parametric tools like the ones presented in this work.   
}


\maketitle


\section{Introduction}
The lithium abundance, \ali\footnote{$A(\mathrm{X})$ =
$\log(N(\mathrm{X})/N(\mathrm{H})) + 12$, where X is the chemical element.}, 
and the \iso\ isotopic ratio measured in stellar
atmospheres can provide valuable information contributing to understanding
different problems in astrophysics such as stellar structure and
evolution, stellar activity, exoplanetary system evolution and even
cosmology. A reliable determination of the lithium content in stellar
atmospheres is important, and apart from requiring high-resolution stellar spectra with high
signal-to-noise ratios (S/Ns), it also requires realistic stellar atmosphere
models allowing a sound theoretical interpretation of both the strength and the
shape of the Li lines.

The stable isotope 
\liseven\ is among the few light elements that have been produced in
  the Big Bang. In contrast, the other stable isotope, \lisix, is not produced
  in significant amounts in standard Big Bang nucleosynthesis
  \citep[e.g.,][]{thomas1993}, but mainly through spallation and $\alpha +
  \alpha$ fusion reactions triggered by high-energy collisions between
  Galactic cosmic rays and He, C, N, O nuclei in the interstellar medium
  \citep[e.g.,][]{meneguzzi1971,prantzos1993,prantzos2012}. 
  Additional \liseven\ is produced
  by the same mechanism, and partly by stellar nucleosynthesis as well. The
  current \iso\ ratio in the local interstellar medium (ISM) is measured to be $\sim13\%$
  (\citealt{k2009}, see also \citealt{howk2012}), while the solar system
  (meteoritic) \iso\ ratio is $8.2\%$ \citep{lodders2003}. 

Both $^6$Li and
$^7$Li are very fragile with respect to nuclear reactions with protons and are
destroyed in stellar interiors at temperatures above $\sim2.0\times10^6$\,K
and $\sim2.5\times10^6$\,K, respectively
\citep[e.g.,][]{pinsonneault1997}. According to standard stellar evolution
models, \lisix, being the more fragile of the two isotopes, undergoes an
intense destruction during pre-main sequence evolution
\citep[e.g.,][]{forestini1994,proffitt1989, montalban2002}.  The standard
models predict that by the time a solar-type solar-metallicity star reaches
the main sequence its $^6$Li has already been destroyed by nuclear reactions
at the base of the extended, completely mixed convective envelope.
In non-standard stellar evolution, where additional processes such 
as rotational mixing, atomic diffusion, internal gravity waves, and the 
effect of magnetic fields are included in order to explain the observed 
lithium abundances in stellar atmospheres (see \cite{talon2005} and references 
therein), lithium destruction is generally even enhanced compared to the 
standard models.

This suggests that any significant detection of the fragile \lisix\ isotope in
the atmosphere of a solar-type star would most probably indicate an external
pollution process, for example by planetary matter accretion
\citep{israelian2001} or alternative sources like stellar flares
\citep{montes1998}. \lisix\ can also be produced by superflares around stars
with hot Jupiters \citep{cuntz2000}.  Therefore, it is of great interest to
measure the presence of the $^6$Li isotope in solar-type stars with effective
temperatures between 5900 and 6400 K (see \cite{montalban2002}) and, in case
of a positive detection, to investigate its origin.

\cite{israelian2001} measured the \iso\ ratio in the giant-planet
hosting star HD~82943 and found a significant amount of \lisix\ in its
atmosphere (\iso\ = 12\%, reduced to 5\% after being remeasured by
\citealt{israelian2003}). This finding was interpreted as evidence for a
planetary material accretion on the surface of the star. \iso\ has been
measured afterwards by several other groups in small samples of main-sequence
solar-metallicity stars with and without known planets \citep{reddy2002,
  mandell2004, ghezzi2009, pavlenko2018}, but no significant detections of
\lisix\ have been reported.

The use of one-dimensional (1D) model atmospheres can lead to erroneous
results in terms of lithium isotopic ratios, due to the fact that the missing
convective line asymmetry must be compensated by a spurious \lisix\ abundance
to fit the observed line profile. The result is an overestimation of the
\iso\ ratio using 1D model atmospheres (e.g., \citealt{cayrel2007},
\citealt{steffen2012}). Furthermore, not only Li but also all other
spectral lines have asymmetric profiles. The asymmetry depends on several
factors such as the mean depth of formation and atomic mass. The use of
detailed three-dimensional (3D) hydrodynamical model atmospheres allows us to
obtain more reliable abundances and isotopic ratios, in particular since they
are able to treat the convective motions responsible for the line asymmetries
in a much more realistic way compared to standard 1D model atmospheres. In
addition, in the case of lithium, the line formation must be treated
considering departures from local thermodynamic equilibrium (LTE) that are
known to be significant, especially in metal-poor stars \citep{cayrel2007}, 
affecting not only the strength of the spectral line but also its Doppler 
shift.

On the other hand, 3D non-LTE (NLTE) calculations are computationally demanding and not
easily accessible to the scientific community. For this reason, it is
important to provide some simplified approach to correct the results of any 1D
LTE spectral analysis of the lithium line region for both 3D and NLTE effects.
Such an approach was suggested by \cite{steffen2010a, steffen2010b,
  steffen2012}. In \cite{steffen2012}, the authors used a grid of 3D model
atmospheres to derive a polynomial approximation for deriving the 3D NLTE
correction of the \iso\ ratio as a function of \teff, \logg\ and metallicity
(\met\ from $-3$ to 0).

A similar approach was followed by \cite{sbordone2010} for the lithium
abundance. They derived analytical approximations to convert the equivalent
widths (EW) of the lithium 670.8\,nm line into \ali\ (and vice versa) in 1D
LTE, 1D NLTE and 3D NLTE, which, in principle, can be used to infer the 3D
NLTE abundance correction by comparing the 1D LTE and 3D NLTE \ali\ for a
given EW, \teff, \logg\ and \met.

Several authors have previously performed 1D NLTE analyses of \ali,
  covering a large range of stellar parameters (e.g., \citealt{carlsson1994},
  \citealt{pavlenko1996}, \citealt{takeda2005}, \citealt{lind2009}). We
  compare our 3D NLTE and 1D NLTE results with some of the previous works in
  Sect.~\ref{sec_1dnle}. In the context of very metal-poor stars, where no
  blend lines interfere, the most advanced full 3D NLTE analyses of \iso\ have
  been performed by \cite{lind2013}, who showed that previous detections of
  \lisix\ based on 1D and 3D LTE modeling could not be confirmed when
  reanalyzed with refined 3D NLTE methods. On the other hand,
  \cite{mott2017} found clear evidence for the presence of \lisix\ in the
  active subgiant HD\,123351, both in 1D LTE and in a full 3D non-LTE analysis
  of very-high-quality spectra. Since the latter investigations use dedicated 
  3D model atmospheres to analyze individual stars, they do not provide 
  a systematic grid of isotopic abundance corrections.

In this work, we use a grid of synthetic spectra computed from 3D
hydrodynamical \cobold\ models and 1D hydrostatic \mlhd\ model atmospheres for
a typical range of stellar parameters for solar-type stars. The intention is
to provide a set of 3D NLTE corrections for the lithium abundance, \ali, and
in particular for the \iso\ ratio that can be directly used to correct the
results of 1D LTE analyses.

The ultimate purpose of this work is to facilitate the analysis of
high-resolution and high-S/N spectra of a large sample of
solar-type stars with and without known giant planets. In case of any positive
detection of the fragile $^6$Li isotope in the atmosphere of our target stars,
we aim to investigate its possible source (e.g., planetary material accretion,
flare production, etc.) by looking for correlations between the derived
lithium content and the presence of a giant planet or planetary system.

This manuscript consists of two major parts. In
 Sect.~\ref{sec_methods}, the methods and the results of the computations of
 3D NLTE corrections for \ali\ and \iso\ are presented, and the analytical
 expressions for their quick evaluation are provided. Section~\ref{sec_stars}
 explains the analysis and gives the results for \ali\ and \iso\ measurements
 in two solar-type stars, HD~95456 and HD~207129, using their HARPS spectra. A
 summary and our conclusions are presented in Sect.~\ref{Sec:conclusion}.


\begin{table}
\caption{\cobold\ 3D hydrodynamical model atmospheres used in this work.}
\label{models}
\begin{center}
\begin{tabular}{lrrrcc}
\hline\hline
\\
Model $N$ & $T_{\rm eff}$ & $\log g$ & [Fe/H] & $N_{\rm snap}$ & $N_{\rm bins}$  \\
&  [K]\\
\hline\\

1   &$ 5850 $ & $ 4.0 $& $ -1.0 $ &$ 20 $ & $ 6$ \\
2   &$ 5920 $ & $ 4.5 $& $ -1.0 $ &$  8 $ & $ 6$ \\
3   &$ 6260 $ & $ 4.0 $& $ -1.0 $ &$ 20 $ & $ 6$ \\
4   &$ 6240 $ & $ 4.5 $& $ -1.0 $ &$ 20 $ & $ 6$ \\
5   &$ 6500 $ & $ 4.0 $& $ -1.0 $ &$ 20 $ & $ 6$ \\
6   &$ 6460 $ & $ 4.5 $& $ -1.0 $ &$ 19 $ & $ 6$ \\
\\\hline\\
7   &$ 5920 $ & $ 4.0 $& $ -0.5 $ &$ 20 $ & $ 6$ \\
8   &$ 5900 $ & $ 4.5 $& $ -0.5 $ &$ 20 $ & $ 6$ \\
9   &$ 6250 $ & $ 4.0 $& $ -0.5 $ &$ 20 $ & $ 6$ \\
10  &$ 6230 $ & $ 4.5 $& $ -0.5 $ &$ 20 $ & $ 6$ \\
11  &$ 6520 $ & $ 4.0 $& $ -0.5 $ &$ 20 $ & $ 6$ \\
12  &$ 6490 $ & $ 4.5 $& $ -0.5 $ &$ 20 $ & $ 6$ \\
\\\hline\\
13  &$ 5930 $ & $ 4.0 $& $  0.0 $ &$ 18 $ & $ 5$ \\
14  &$ 5870 $ & $ 4.5 $& $  0.0 $ &$ 19 $ & $ 5$ \\
15  &$ 6230 $ & $ 4.0 $& $  0.0 $ &$ 20 $ & $ 5$ \\
16  &$ 6230 $ & $ 4.5 $& $  0.0 $ &$ 20 $ & $ 5$ \\
17  &$ 6490 $ & $ 4.0 $& $  0.0 $ &$ 20 $ & $ 5$ \\
18  &$ 6460 $ & $ 4.5 $& $  0.0 $ &$ 20 $ & $ 5$ \\
\\\hline\\
19  &$ 5870 $ & $ 4.0 $& $  0.5 $ &$ 20 $ & $ 5$ \\
20  &$ 5900 $ & $ 4.5 $& $  0.5 $ &$ 20 $ & $ 5$ \\
21  &$ 6190 $ & $ 4.0 $& $  0.5 $ &$ 20 $ & $ 5$ \\
22  &$ 6350 $ & $ 4.5 $& $  0.5 $ &$ 20 $ & $ 5$ \\
23  &$ 6410 $ & $ 4.0 $& $  0.5 $ &$ 20 $ & $ 5$ \\
24  &$ 6390 $ & $ 4.5 $& $  0.5 $ &$ 20 $ & $ 5$ \\
\\
\hline
\end{tabular}
\tablefoot{$N_{\rm snap}$ is the number of selected representative snapshots,
  and $N_{\rm bins}$ is the number of opacity bins.}
\end{center}
\end{table}

\section{3D NLTE corrections}
\label{sec_methods}


\begin{figure}[htb!]
\centering
\includegraphics[clip=true,width=250pt]{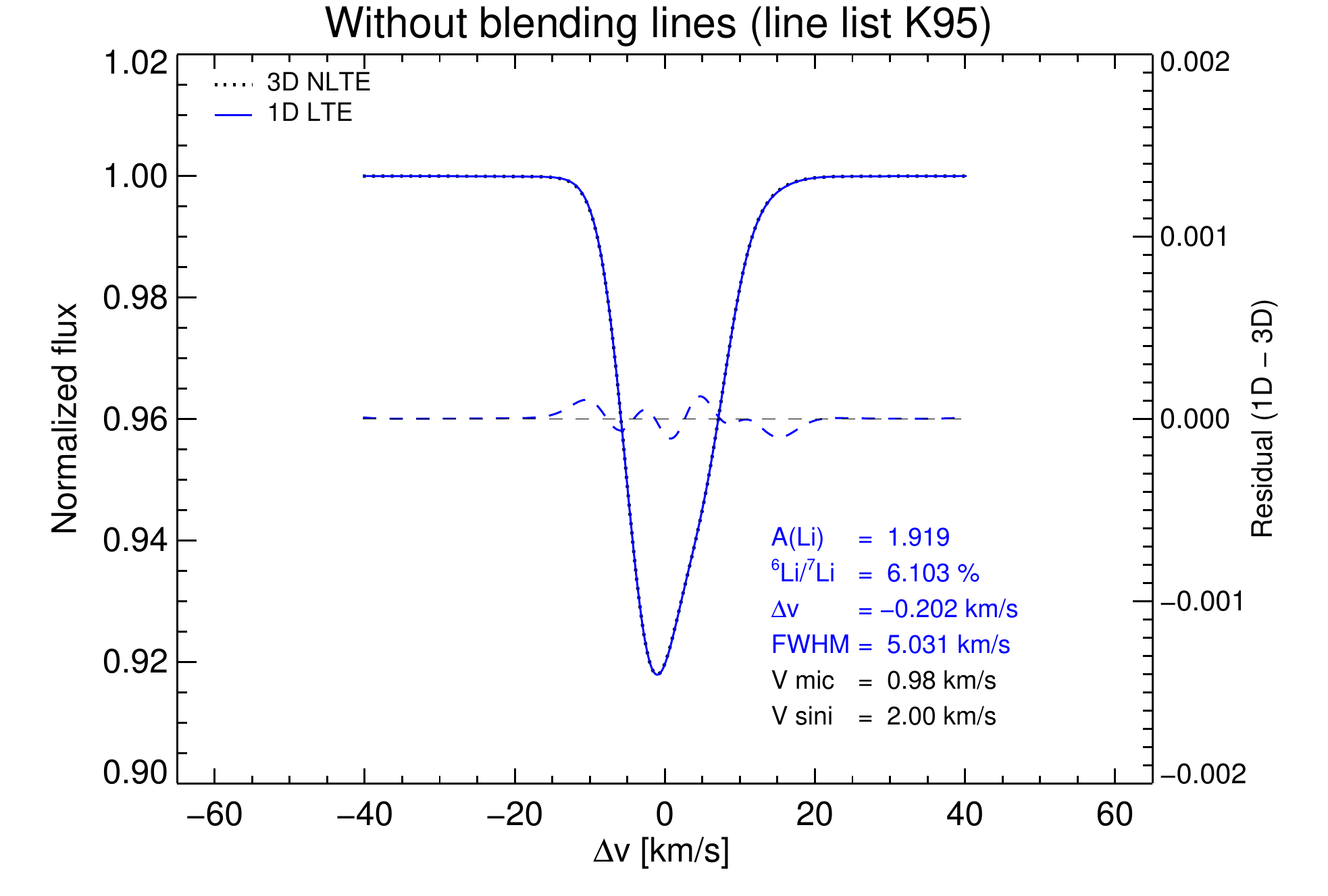}\\
\includegraphics[clip=true,width=250pt]{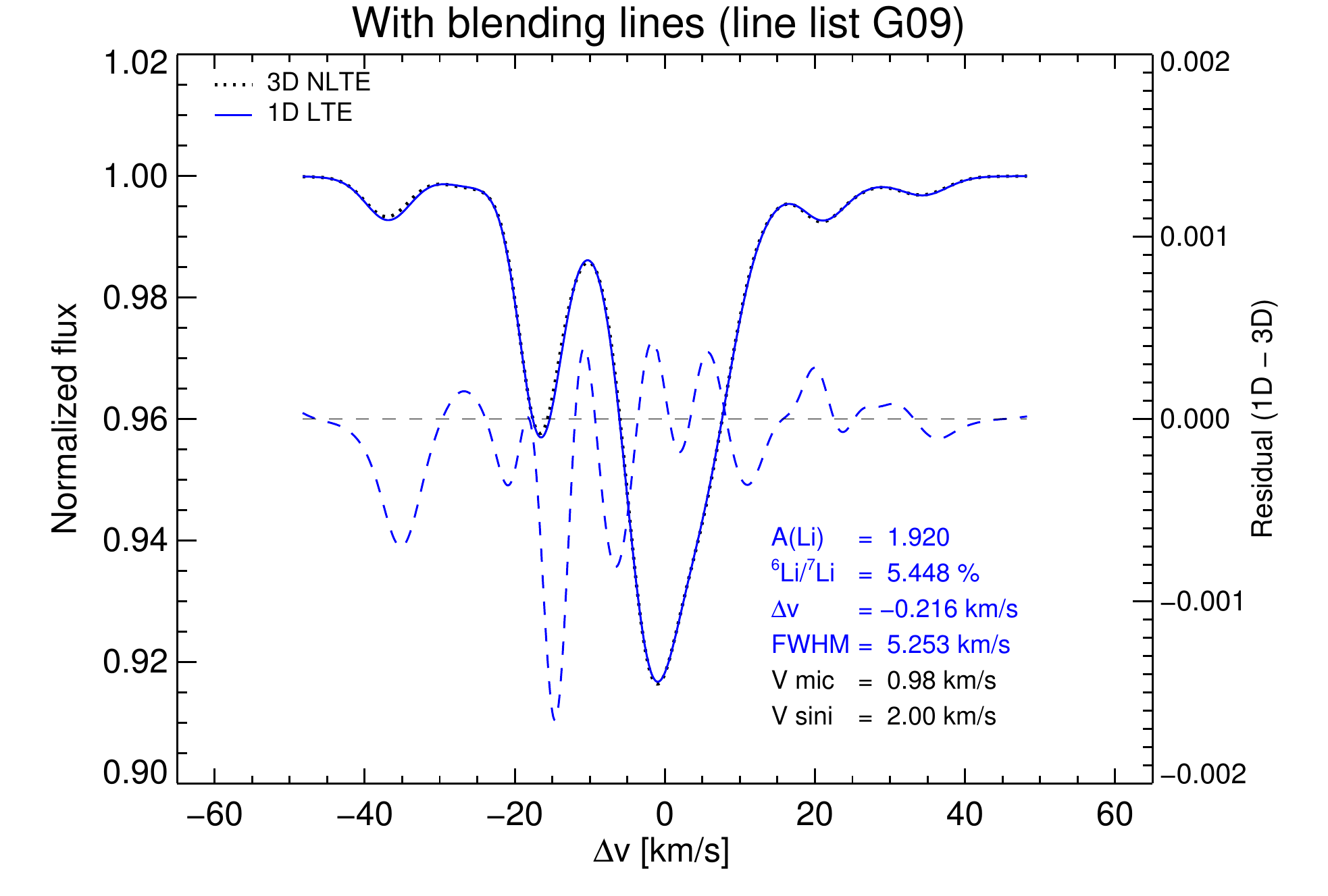}
\caption{The impact of the blending lines in the \ion{Li}{I} $\lambda$670.8\,nm region on the resulting 3D NLTE corrections. The best fitting 1D LTE spectrum (blue continuous line) is superimposed on the 3D
NLTE spectrum (black dotted line) computed for \teff=5870 K, \logg=4.5,
\met=0, \textbf{\aliTd}=2.0, \textbf{\isoTd}=5\% and representing the `observation'. The synthetic
spectra are computed adopting only the \ion{Li}{i} doublet including hyperfine 
structure (line list K95, \textit{upper panel}), and the line list G09 (\textit{lower panel}),
which includes also other atomic blends. The right y-axis gives the scale of
the residuals (blue dashed line). We measure \ali=1.92 and \iso=6.1\% for line list K95, and \ali=1.92 and \iso=5.4\% in case of line list G09.}
\label{fit}
\end{figure}

\subsection{Model atmospheres}
For deriving our 3D NLTE corrections, we adopted a subset of the CIFIST 3D
hydrodynamical model atmospheres grid \citep{cifist}, computed with the 
\cobold\ code \citep{cobold1,cobold2}. In total, 24 3D model atmospheres were 
used,
covering three different effective temperatures (\teff), two surface gravities
(\logg) and four metallicities (\met). Table~\ref{models} lists the assigned
number of each model (Model $N$) and its effective temperature, surface gravity,
metallicity, number of the representative snapshots ($N_{\rm snap}$), and the
number of opacity bins ($N_{\rm bins}$) for treating the radiative energy 
transport in the 3D simulations. The final \teff\ of each
\cobold\ hydrodynamical model atmosphere is determined after the
selection of a number of snapshots from their averaged radiative surface flux
(this \teff\ is given in Table~\ref{models}) and usually differs slightly from the intended \teff\ value. Therefore, the model atmospheres in our grid do not have exactly the same effective temperatures for each nominal temperature point of the grid.

To each \cobold\ model in our grid a 1D \mlhd\
model atmosphere \citep{lhd} is associated, having the same effective
temperature, metallicity, and surface gravity. Using these particular 1D model
atmospheres for comparison with the 3D models is advantageous since they employ the same opacity tables and
equation of state as the \cobold\ models. Such a differential analysis
minimizes unphysical (numerical) intrinsic differences between 1D and 3D models and isolates the true 3D effects. The 1D \mlhd\ model atmospheres employ
the mixing-length theory to describe convection, and a typical mixing length
parameter $\alpha_{\rm MLT}$ = 1.0 is used in this work. Several authors
\citep[e.g.,][]{klevas2016, mott2017} have shown (as verified in 
Sect.\,\ref{sec_MLT} below) that the choice of this parameter is not critical 
for our \ali\ and \iso\ studies. This is because, in the framework of the 
mixing-length theory, the line-forming layers of the considered
stellar atmospheres are not strongly affected by convection.

\subsection{Spectral synthesis}
\label{sec_synthesis}
For each 3D \cobold\ and 1D \mlhd\ model atmosphere, a corresponding grid of
synthetic spectra for the  \ion{Li}{i}  $\lambda$670.8\,nm region has been
computed using the spectral synthesis code Linfor3D \citep{steffen2015}. The
NLTE lithium line profiles for the 3D case were computed for combinations of 
three different Li abundances (\textbf{\aliTd}=1.5, 2.0, 2.5) and three different
\iso\ isotopic ratios (\textbf{\qliTd}=0\%, 5\%, 10\%); hereafter \qli$=n(^6$Li$)/n(^7$Li).

A 17-level lithium model atom
including 34 bound-bound transitions was adopted for the computation of 
the NLTE departure coefficients. This model atom was initially developed by
\cite{cayrel2007} and further updated and used by several authors
\cite[e.g.,][]{sbordone2010, steffen2012, klevas2016, mott2017}. Individually for each of the three assumed lithium 
abundances, NLTE departure coefficients were computed with the code 
NLTE3D \citep{steffen2015}. 

In the 1D case, a series of LTE synthetic line profiles were computed for
combinations of nine Li abundance values (from 1.00 to 3.00 with a step
of 0.25), nine different \iso\ isotopic ratios (from 0\% to 16\% with a step
of 2\%), and three different microturbulence velocities(\vmic), centered on
the value obtained from the analytic expression derived from a set of 3D model 
atmospheres by \cite{dutraferreira2016}, with an offset of 
$\pm$ 0.5\,$\mathrm{km s^{-1}}$.  

To examine the dependence of our results on the list of blend lines used in
the spectral synthesis, we replicated the computation of the full grids of 3D
NLTE and 1D LTE line profiles using three different line lists. The first line
list includes only the  \ion{Li}{i}  atomic lines adapted from
\cite{kurucz1995}. The original hyperfine structure was simplified to include
six \lisix\ and six $^{7}$Li components (hereafter, K95; as given in
\cite{mott2017}). The second line list was constructed by \cite{ghezzi2009},
including $31$ blend lines in addition to Li (hereafter, G09),
whereas the third line list with $36$ additional blend lines has 
been taken from \cite{melendez2012} (hereafter, M12). In the case of G09 and 
M12, we have replaced the  \ion{Li}{i} hyper-fine structure by the same data 
as in K95.

Moreover, to be able to distinguish between the contributions of 3D and NLTE
effects in the combined 3D NLTE corrections, we computed 1D NLTE spectra with
the same methods and for the same grid as the 3D NLTE spectra but only for
\alionednlte\ = 2.0 and \isoonednlte\ = 5\%, adopting the line list K95. The 
1D NLTE spectra were computed using the 1D \mlhd\ model atmospheres together 
with NLTE departure coefficients computed with NLTE3D.

All the synthetic spectra are computed in the wavelength range between 670.69
and 670.87\,nm  for the K95 line list, and between 670.672 and 670.888\,nm in
case of line lists G09 and M12, to be able to include in the synthesized spectral region the blends other than lithium that are present in the line lists.

\subsection{3D-1D fitting procedure}
\label{sec_fitting3d}
At first we derived the 1D LTE lithium abundance and \iso\ isotopic ratio by fitting each 3D NLTE line
profile with the grid of pre-computed 1D LTE spectra. This is done for
each value of \aliTd\ and \qliTd\ of the 3D NLTE grid. All 3D spectra were
broadened with a Gaussian instrumental profile of \textit{FWHM}=3.0\,\kms, 
a value close to the resolution of the HARPS spectrograph. The fitting was
performed through interpolation within the 1D LTE grid, driven by the 
least-squares fitting algorithm MPFIT \citep{markwardt}. Four free parameters were varied to
achieve the best fit ($\chi^{2}$ minimization): \ali , \qli, the FWHM of the
Gaussian line broadening, and a global wavelength shift ($\Delta v$). The
Gaussian broadening accounts for instrumental broadening and  macroturbulence
velocity ($V_{\rm macro}$) in the 1D case. 

In addition, to investigate the possible dependence of our results on the
rotational broadening, we applied identical rotational broadening to the 3D
and 1D  spectra, ranging from \vsini\ = 0 to 6\,\kms\ with a step of
2\,\kms. The rotational broadening is applied using the flux convolution
approximation, assuming a typical limb darkening coefficient of $\epsilon=0.6$
\citep[e.g.,][]{gray2005}.

\subsection{3D NLTE corrections for \ali\ and \iso}
\label{sec_corr}

We define the 3D NLTE correction of the lithium abundance, \delAstar, as the
difference between the \aliTd\ value assumed in the 3D NLTE synthesis and the
\ali\ value obtained from the best 1D LTE fit to the 3D NLTE
spectrum. Similarly, the 3D NLTE correction for the lithium isotopic ratio, 
\delqstar, is defined as the difference between the assumed \isoTd\ ratio of
the 3D NLTE spectrum and the best-fitting 1D LTE \iso\ value. 
The asterisk indicates that the corrections are computed on a 
grid of given 3D NLTE Li abundances.


\begin{figure*}[htb!]
\centering
\includegraphics[clip=true,width=250pt]{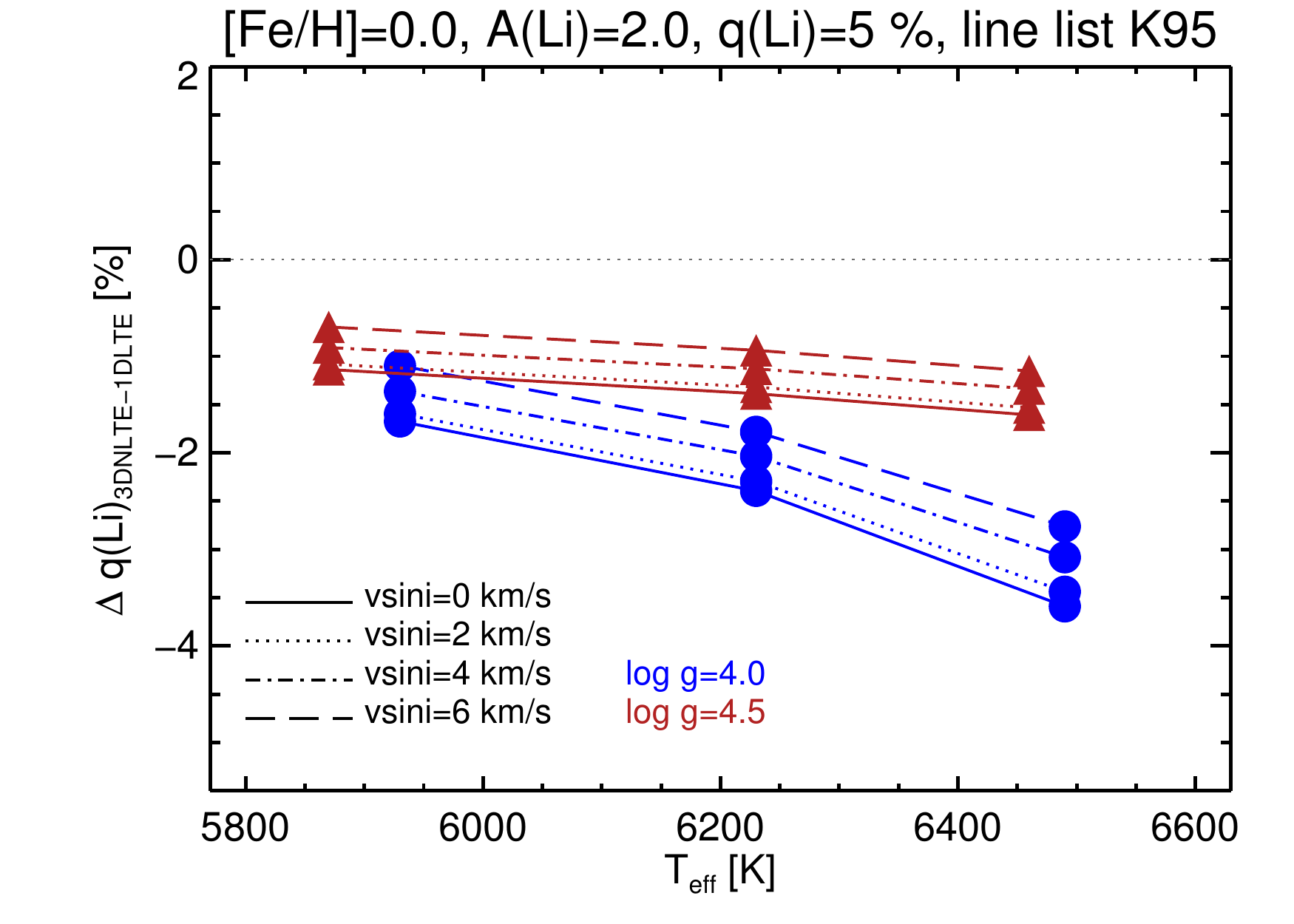}
\includegraphics[clip=true,width=250pt]{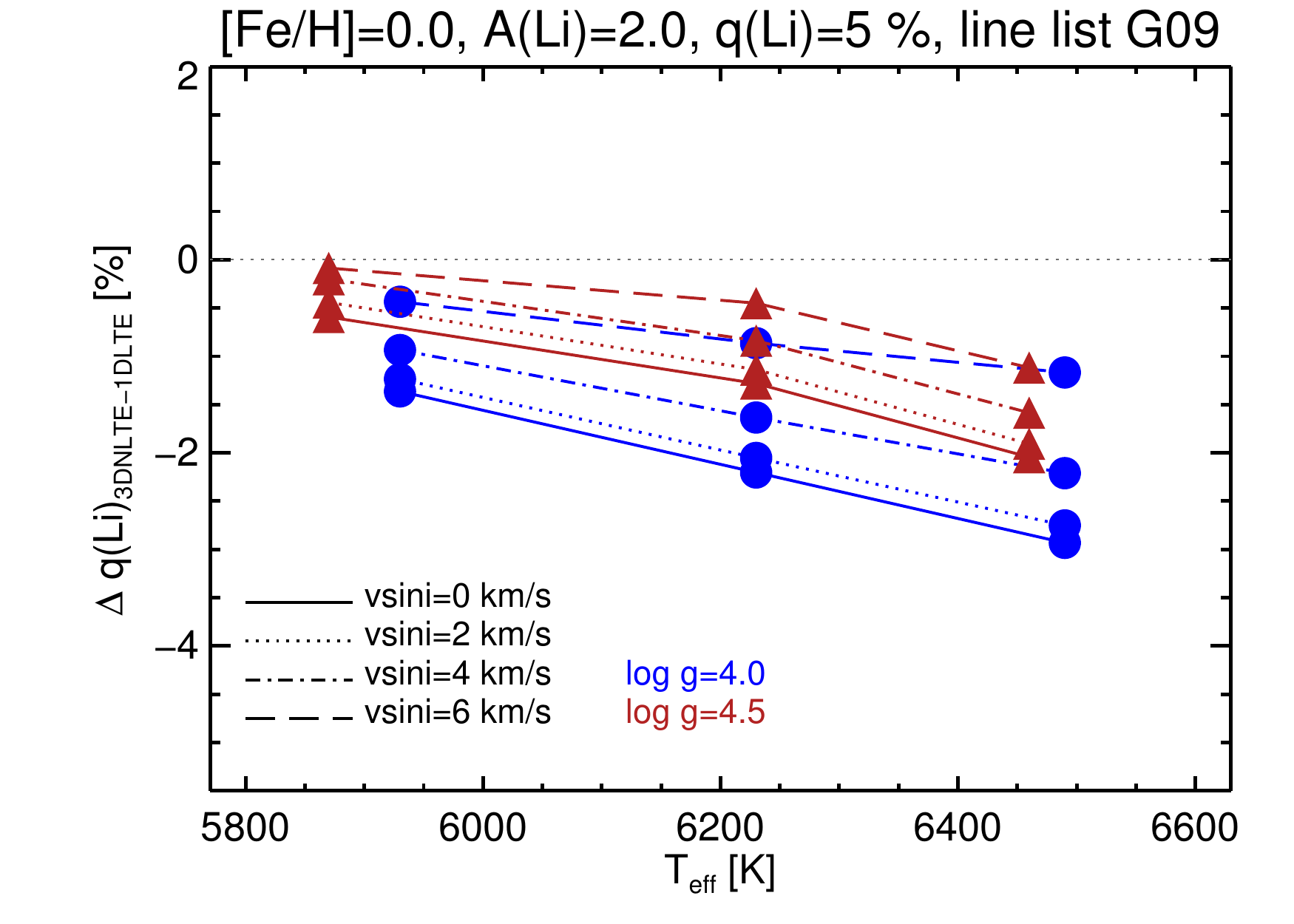}\\
\includegraphics[clip=true,width=250pt]{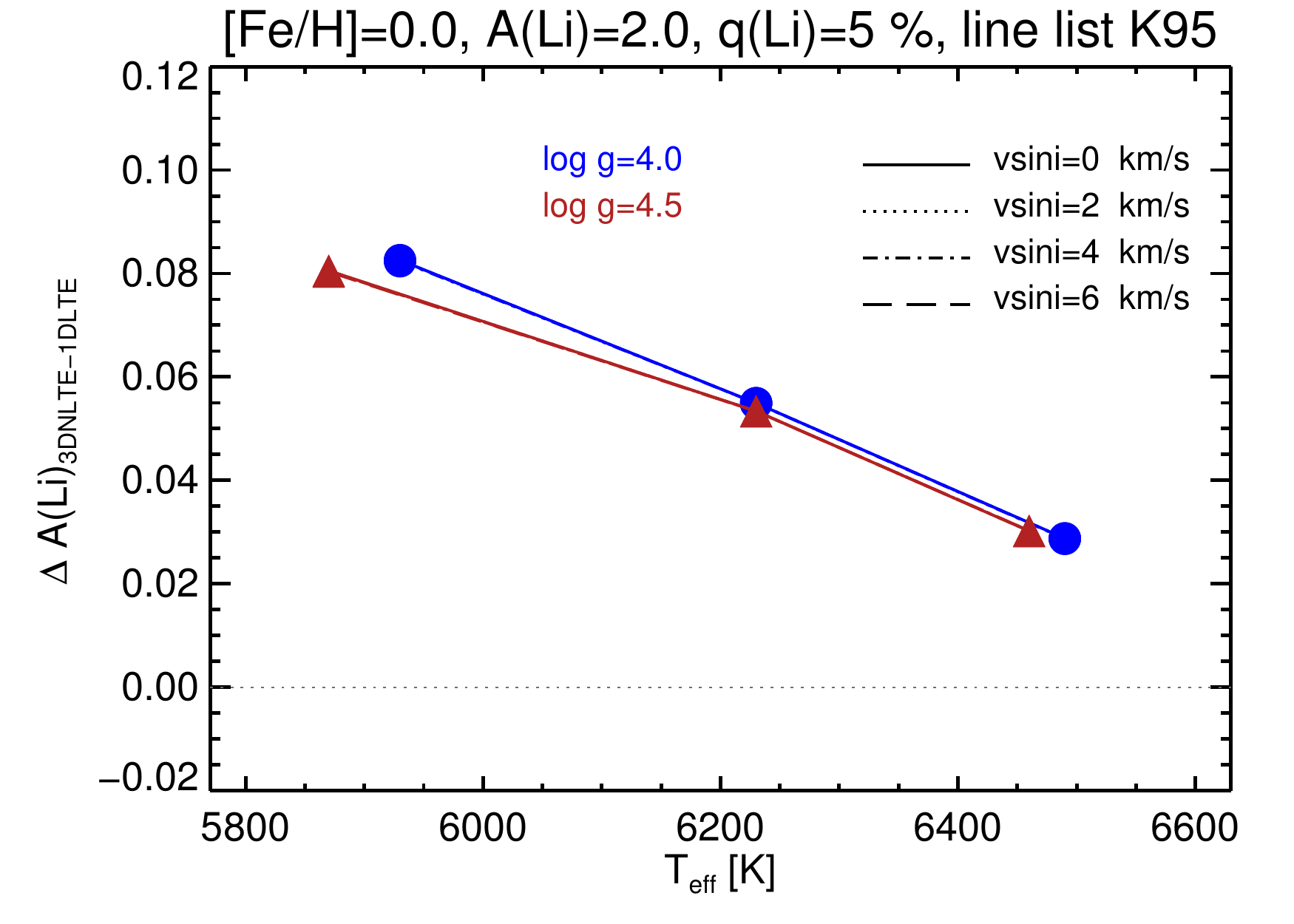}
\includegraphics[clip=true,width=250pt]{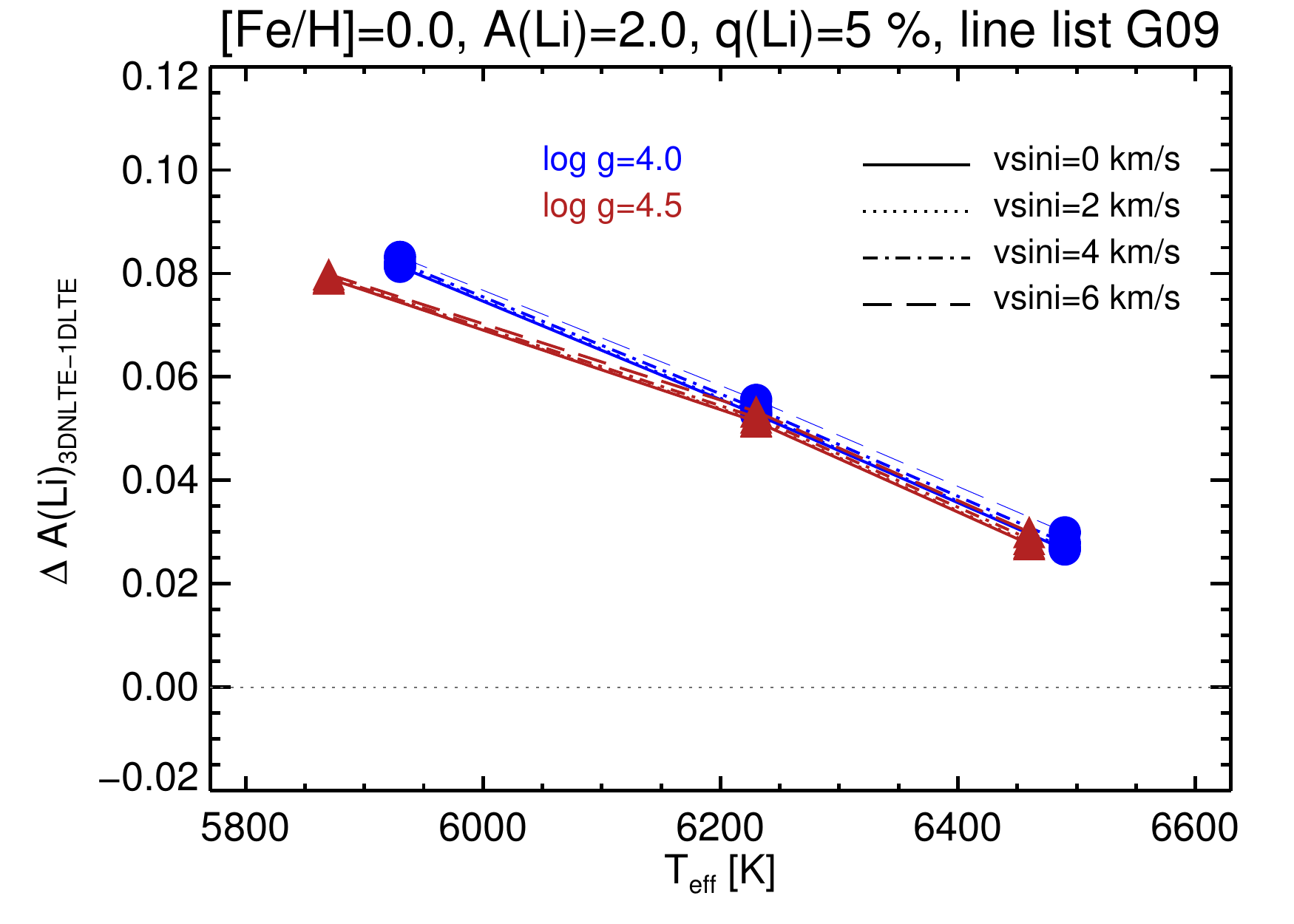}
\caption{3D NLTE corrections \delq\ (upper panels) and  \delA\ (lower panels) 
vs. $T_{\rm eff}$ for \met=0, \ali=2.0, \qli=5\%, obtained with line lists 
K95 (left panels) and G09 (right panels). The blue circles and the red 
triangles correspond to $\log g$=4.0 and $\log g$=4.5, respectively. The 
computed 3D NLTE corrections for different $v\sin i$ values are connected 
with lines of different styles (see legend). }
\label{corr_plot}
\end{figure*}

\subsubsection{Fitting the pure lithium feature}
\label{sec_corr1}
As an example, we show in Fig.\,\ref{fit} (top panel) the best fit to the pure
lithium 3D NLTE spectrum achieved for \teff =5870\,K, \logg =4.5, \met =0,
\aliTd =2.0 and \qliTd =5\,$\%$, computed adopting the Li hyperfine components
from the line list K95. The \iso\ isotopic ratio obtained by the best 1D LTE
fit is \qli=6.1$\%$, thus overestimating the true 3D NLTE \iso\ ratio of 5$\%$
by 1.1 percentage points, and underestimating the lithium abundance by 
$\sim$0.08\,dex. 

The 3D NLTE corrections (\delAstar\ and \delqstar) are computed for a grid of
three \aliTd\ and three \qliTd\ values for each 3D model atmosphere given in
Table 1. These corrections are meant for correcting the 1D LTE results without
knowing the 3D NLTE \aliTd\ and \isoTd\ of the observed spectrum, and
therefore, they should not depend on the 3D NLTE values of \aliTd\ and
\qliTd\ but on the measured 1D LTE values. Therefore, we converted our 3D NLTE
corrections such that they depend on the 1D LTE lithium abundance, \ali, and
1D LTE isotopic ratio, \qli, instead of depending on the true (3D NLTE) values
\aliTd\ and \qliTd. To do this, we used the best-fitting 1D LTE values of the
lithium abundance and the \iso\ isotopic ratio for each \aliTd\ and
\qliTd\ point of our 3D NLTE grid and interpolated the 3D NLTE corrections to
the 1D LTE values \ali\ = 1.5, 2.0, 2.5 and \qli\ = 0, 5 and 10\%. All the
subsequent plots and tables presented in this work will be using these 3D NLTE
corrections as a function of \ali\ and \qli, denoted as \delA\ and \delq,
respectively.

Figure\,\ref{corr_plot} shows the 3D NLTE corrections plotted versus \teff\ for
\iso\ (upper left panel) and \ali\ (lower left panel) for \met\ = 0, \ali\ =
2.0 and \qli\ = 5\%. The corrections for surface gravities 4.0 and 4.5 are
plotted as blue circles and red triangles, respectively, and are connected
with lines of different style for different \vsini\ values.  For a given
\logg, the 3D NLTE corrections of the \iso\ ratio (\delq) become larger (more
negative) for higher temperatures, and they are larger for the lower \logg. We
note that \delq\ depends also on \vsini, \ali, and \qli: for a given
\logg\ and \teff, they become smaller (more positive) for higher \vsini, and
for a given stellar parameters they will be higher for stars with higher
lithium abundance and isotopic ratio. 
The dependence of \delq\ on the \vmic\ used for 1D
LTE spectral synthesis is negligible, and we take the central value of
\vmic\ as representative for this work.

On the other hand, the 3D NLTE corrections for the lithium abundance,
\delA, become larger for lower effective temperatures. There is only a
slight dependence on the surface gravity, and their dependence on the
\vsini\ can be neglected (see Fig.\,\ref{corr_plot}, lower left panel).
The \delA\ corrections do not show variations for different \vmic\ and
\qli\ values, whereas they are decreasing slightly with larger \ali\ values.

Similar plots for the other three metallicities of our grid are given in
Appendix A. For full details, we provide electronically a table of
the corrections for \teff\ = 5900, 6300, 6500\,K, \logg\ = 4.0 and 4.5,
\met\ = $-$1.0, $-$0.5, 0.0, 0.5, \ali\ = 1.5, 2.0, 2.5, \qli\ = 0, 5, 10\%, and
\vsini\ = 0.0, 2.0, 4.0, 6.0\,\kms, based on the pure lithium line
  list K95.

\subsubsection{The impact of the mixing-length parameter}
\label{sec_MLT}
To investigate the dependence of our results on the choice of the mixing
length parameter for 1D \mlhd\ models, we computed \delA\ and
\delq\ using the models N 1--6 and N 13--18 with $\alpha_{\rm MLT}$ = 0.5
and 1.5 (N 3, 15 and 17 only with $\alpha_{\rm MLT}$ = 0.5). As expected, the
choice of this parameter did not affect the \delq\ results, while
\delA\ varied slightly by $\pm$ 0.01\,dex over the considered range of 
 $\alpha_{\rm MLT}$.

\subsubsection{The impact of blend lines}
\label{sec_corr2}

The bottom panel of Fig.\,\ref{fit} shows the best fit to the 3D NLTE
spectrum computed for the same parameters as in Sect.\,\ref{sec_corr1}, but
now adopting line list G09 that includes atomic and molecular blends
partly overlapping with the Li feature. In this case, the 3D NLTE correction
of \iso\ isotopic ratio is somewhat smaller ($-0.45$ percentage points), while
the 3D NLTE correction for \ali\ is similar to the K95 case ($\sim+$0.08\,dex).

Initially, we were expecting that the fitting results would depend only 
weakly on the adopted line list. Comparison of the lower panels of 
Fig.\,\ref{corr_plot} shows that this is basically the case for the
\ali\ correction. However, while the \iso\ corrections derived with line list 
G09 are qualitatively similar to those obtained from fitting the pure lithium 
spectrum, there are quantitative differences that depend on a variety of 
factors.

As shown in Fig.\,\ref{fit} (bottom), the blends in this spectral region may
be quite strong (e.g., the \ion{Fe}{i} line at $\sim$670.74\,nm) and can play
an even greater role in the fitting procedure than the \ion{Li}{i} line
itself. We recall that the main fitting parameters are \ali\ and \qli, both
only influencing the Li line, while the two remaining fitting parameters,
global Gaussian line broadening and line shift, act on the all spectral
lines. However, to keep the dimensionality of the problem manageable, 
the strength and wavelength shift of the individual blend lines 
is not adjusted in the fitting procedure. This deficiency of our method can 
lead to meaningless fitting results, because poorly reproduced stronger blend 
lines determine the global line broadening and shift of the best 
fit, and therefore indirectly dictate an ill-defined solution for the \iso\ ratio. 
This is especially true for the case of high metallicity (\met\,=\,+0.5) and 
low lithium abundance (\ali\,=\,1.5), where the \iso\ corrections obtained 
with line list G09 are essentially useless.

It is therefore hardly surprising that line lists G09 and M12 produce somewhat
different fitting results (compare Figs.\,\ref{corr_plot} and \ref{a4}), which
moreover depend strongly on the wavelength range selected for fitting the
\ion{Li}{i} 670.8\,nm spectral region, as well as on whether the continuum
placement is a free or a fixed parameter.

The above-mentioned difficulties do not exist if we derive the 3D NLTE
corrections using the line list K95 which includes only the \ion{Li}{i}
components. We also notice that the results obtained with this line list are
much less sensitive to the value of \vsini\ than in the case of line lists G09
and M12. We argue that the fitting results obtained by considering only
\ion{Li}{i} lines must be very similar to those one would derive if all blend
lines were fitted individually in the line lists G09, M12, or any other 
line list representative of the Li region. This argument is 
illustrated and supported by the investigations presented in Appendices
 \ref{ap_test_Li_Si} and \ref{ap_bestfit_sun}, where we show that a line
list adjusted to fit a given spectrum in 1D generally produces a less
satisfactory fit when used for 3D modeling. An equally good fit in 3D can only
be achieved by readjusting the line list to account for 3D abundance
corrections and differential wavelength shifts of the blends. When using
the 1D and 3D fine-tuned line lists, respectively, the differential 1D-3D
lithium isotopic abundances become largely independent of the blend lines
and are well approximated by the corrections obtained with line list K95 
(only \ion{Li}{i}).

\subsection{Analytical expressions}
\label{sec_analytic}
We derived analytical expressions for the corrections derived from the 
``only Li'' line list K95 to be able to
numerically evaluate \delA\ and \delq\ as a function of 
the stellar parameters for any target star within the range of our grid.

As mentioned above, the \ali\ abundance corrections for all the models in our
grid show only a weak dependency on the surface gravity. Assuming a \logg\ of
4.0 or 4.5, the resulting \ali\ corrections are very similar, as shown in the
lower panels of Fig.\,\ref{corr_plot}. Within the modeling uncertainties
(see below), the \logg\ dependence may be neglected. For this reason, we 
assumed \delA\ to be independent of \logg\ while deriving the analytic
approximation.

Somewhat unexpectedly, model atmosphere $N2$, with \teff=5920\,K, \logg=4.5, 
\met=$-1$, results in a smaller $\Delta$\ali\ than expected from the
trends of the corrections derived using other models in our grid (see
Figs.\,\ref{a1}--\ref{a3} in Appendix \ref{appendix:A}). A possible explanation
for this discrepancy is that this model was computed for a smaller number of
snapshots (8) in comparison to the other models (20). We decided not to use
this result and instead consider it to be equal to the correction computed for
the model $N1$ (with the same \teff\ but \logg = 4.0) in deriving the numerical
fitting function for \delA.


\begin{table}
\caption{Numerical coefficients in Eqs.\,(\ref{eq_li}) and (\ref{eq_iso}) derived for the line list K95 with representative 1D LTE values of \ali\,=\,2.0, \qli\,=\,5\%, and \vsini\,=\,2.0\,\kms.}
\label{table_coeff}
\begin{center}
\begin{tabular}{c r | c r }
\hline
\hline
\noalign{\smallskip}
Coeff.  &   Value &  Coeff.  &  Value           \\ 
\noalign{\smallskip}
 \hline \noalign{\smallskip}
    $C_{0}$  &   0.074205      &    $c_{000}$ & $   -1.5994$            \\
    $C_{1}$  &  $-0.607931$    &    $c_{010}$ & $    1.0167$       \\
    $C_{2}$  &   1.518850      &    $c_{001}$ & $   -0.4877$         \\
    $C_{3}$  &   0.053601      &    $c_{011}$ & $    0.2553$            \\
    $C_{4}$  &   0.023510      &    $c_{002}$ & $   -0.1548$ \\
    $C_{5}$  &  $-0.159625$    &    $c_{012}$ & $    0.3447$    \\
             &                 &    $c_{100}$ & $   -7.7112$\\
             &                 &    $c_{110}$ & $    6.0203$     \\
             &                 &    $c_{101}$ & $    4.6808$    \\
             &                 &    $c_{111}$ & $    1.0406$     \\
             &                 &    $c_{102}$ & $    6.0582$    \\
             &                 &    $c_{112}$ & $  -12.9981$            \\
             &                 &    $c_{200}$ & $  -96.3931$      \\
             &                 &    $c_{210}$ & $  191.2326$        \\
             &                 &    $c_{201}$ & $ -142.0070$           \\
             &                 &    $c_{211}$ & $  223.2101$          \\ 
             &                 &    $c_{202}$ & $  -68.1011$               \\
             &                 &    $c_{212}$ & $  223.1616$       \\
   \noalign{\smallskip}
\hline
\end{tabular}
\end{center}
\end{table}


\begin{figure}[htb!]
\centering
\includegraphics[clip=true,width=250pt]{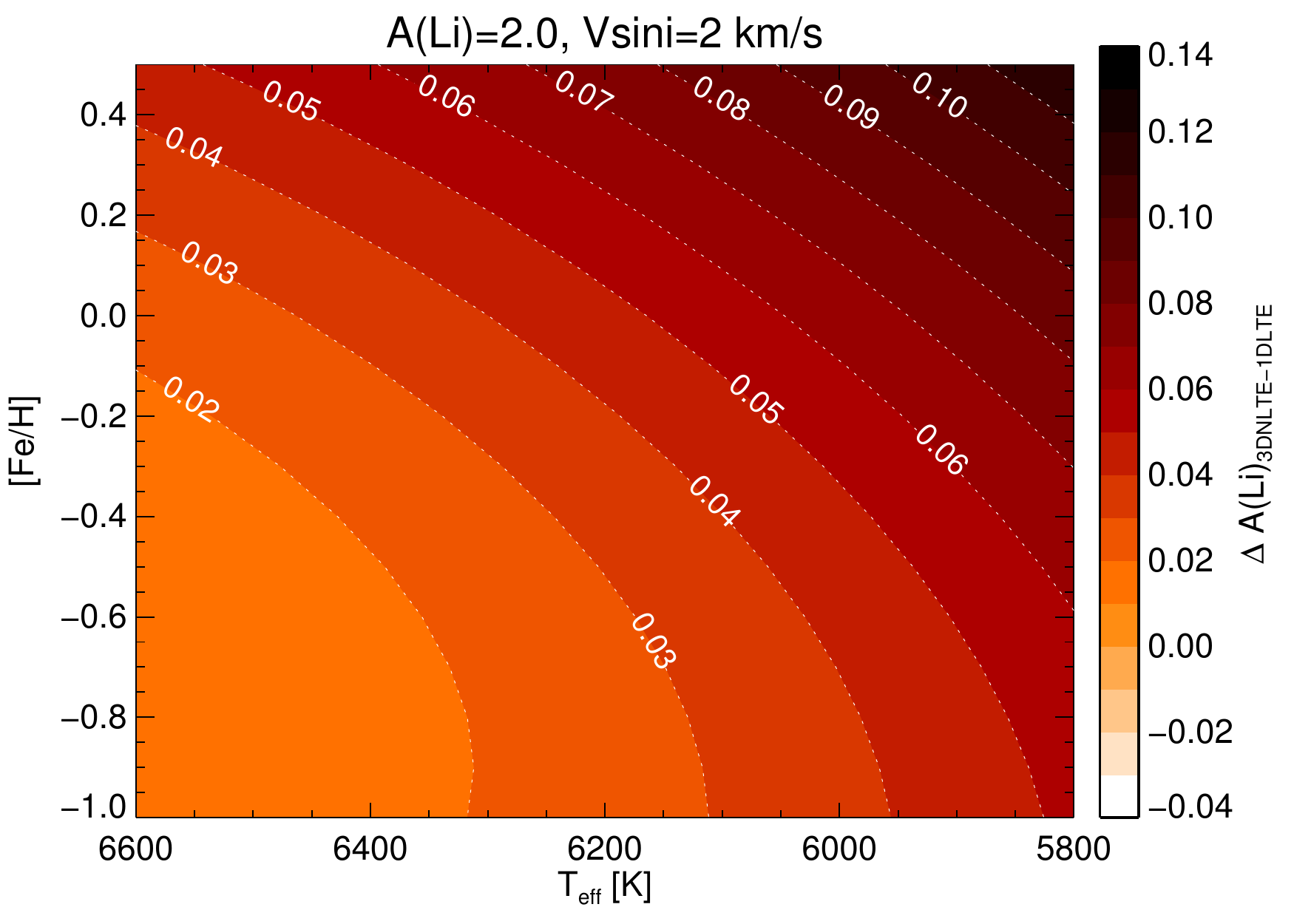}
\includegraphics[clip=true,width=250pt]{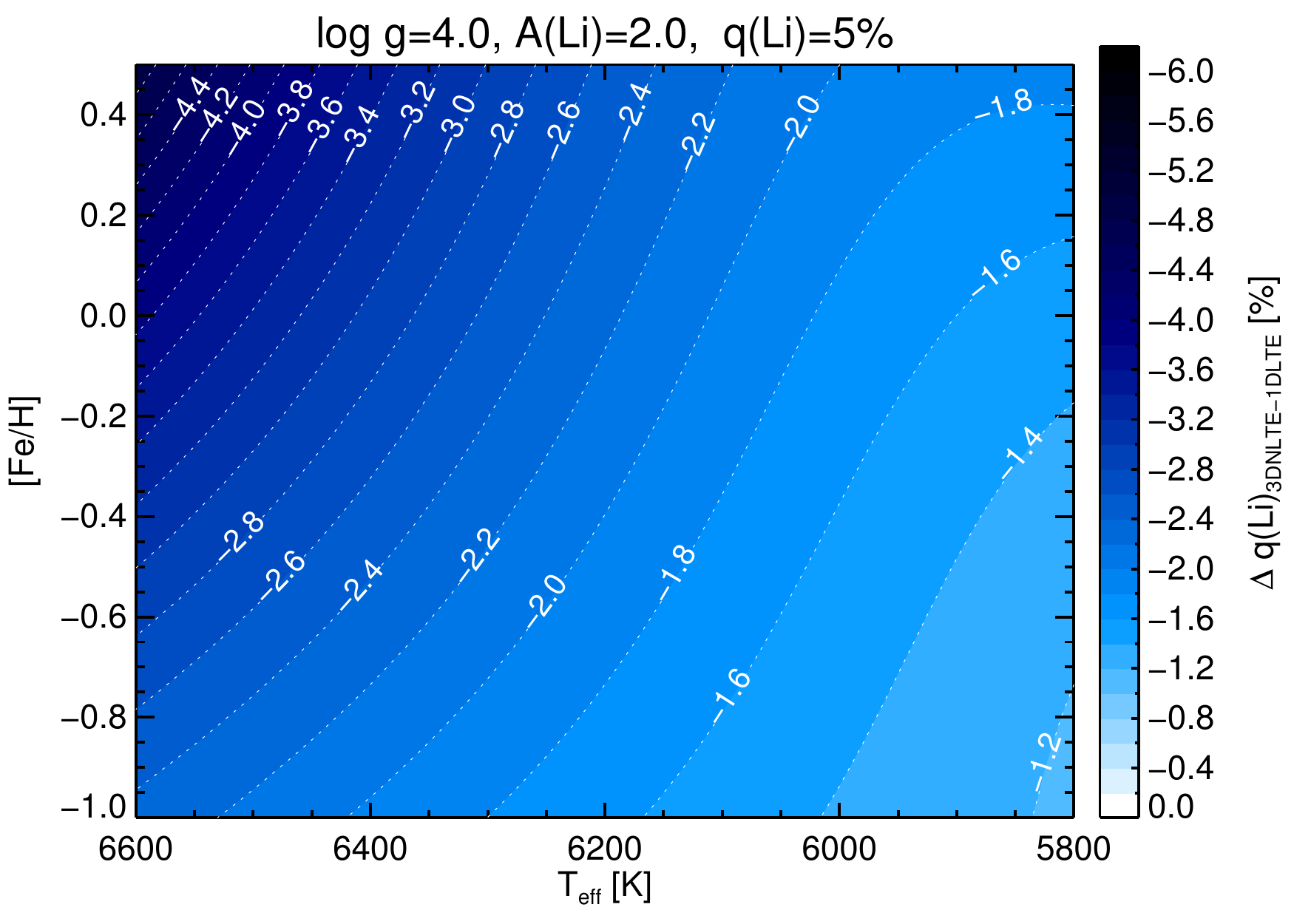}
\includegraphics[clip=true,width=250pt]{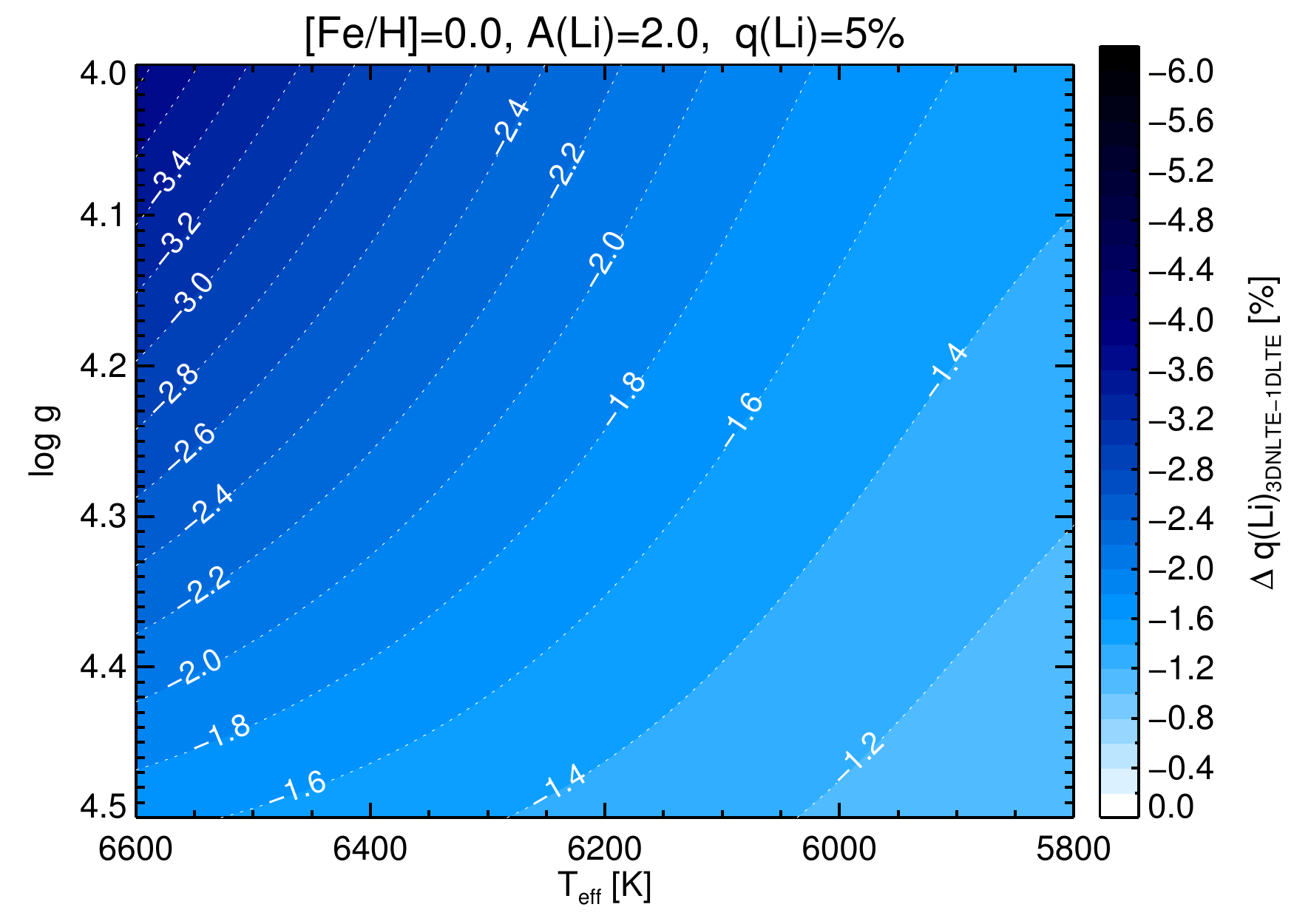}
\caption{
Contour plots of 3D NLTE \ali\ corrections  \delA\ (upper
  panel) and \iso\ ratio  corrections \delq\ (middle panel) in the
\teff\ -- \met\ plane. The same 3D NLTE \iso\ corrections in the 
\teff\ -- \logg\ plane are shown in the bottom panel. The contours are
computed for [Fe/H]=0, \vsini\,=\,2\,\kms, \ali\,=\,2.0, and
\qli\;=5\%, employing the analytical expressions in
 Eqs.\.(\ref{eq_li}) and (\ref{eq_iso}) with the coefficients given in 
Table\,\ref{table_coeff}.}
\label{contours}
\end{figure}

We provide a link to a web page\footnote{https://pages.aip.de/li67nlte3d} that uses the
analytical expressions developed in this work and allows the user to compute
the \delq\ corrections as a function of \teff, \met, \logg, \vsini,
\ali\ and \iso, and the \delA\ corrections as a function of \teff, \met\ and \ali. However, below we also provide simplified analytical expressions for a quick evaluation of the 3D NLTE corrections.

The simplified expressions for \delA\ and \delq\ provided in this
section are based on all available data points for \vsini\ = 2.0 \kms,
\ali\,=\,2.0, \qli\,=\,5\%, and are functions of \teff, (\logg), and
\met.  The two fitting functions are described by the following equations:


\begin{equation} \label{eq_li}
\Delta A_{\rm \,3DNLTE-1DLTE} = C_{0}+C_{1} x+C_{2} x^2 + C_{3} z + C_{4} z^2 + C_{5}xz\, ,
\end{equation}

\begin{equation} \label{eq_iso}
\Delta q_{\rm \,3DNLTE-1DLTE} = \sum\limits_{i=0}^{2}
\sum\limits_{j=0}^{1} \sum\limits_{k=0}^{2} c_{ijk} x^i y^j z^k\, , 
\end{equation}
where $x\equiv(T_{\rm eff}-T_{\rm eff}^\ast$)/$T_{\rm eff}^\ast$, $y\equiv \log g-\log g^\ast$,
and $z\equiv$\met, with $T_{\rm eff}^\ast$\,=\,$5900$\,K and 
$\log g^\ast$\,=\,$4.0$. The valid
parameter ranges are $0 \le x \le 0.1$; $0 \le y \le 0.5$; $-1.0 \le z \le +0.5$. The formulae consist of  six ($C_{0-5}$) and 18 ($c_{ijk}$) numerical 
coefficients for $\Delta A$ and $\Delta q$, respectively
(explicitly given in Table\,\ref{table_coeff}).

To visualize the resulting 3D NLTE corrections generated
with this tool, we present in Fig.\,\ref{contours} the contours of \delA\ 
in the \teff\ -- \met\ plane (upper panel), and contour plots of \delq\ 
in the  \teff\ $-$ \met\ and \teff\ $-$ \logg\ plane (middle and lower panel, 
respectively). 

The functional forms provided in this work allow a quick estimate of 3D
NLTE corrections for a limited range of stellar parameters. For a complete
overview and more precise 3D NLTE corrections, we refer to the
online table and to the above-mentioned web page. To give an 
example, our full analytical expressions give the following corrections 
for HD~82943 (\teff=6025\,K, \logg=4.53, \met=+0.30; \citealt{israelian2003}): 
\delA\ = $+0.08$\,dex and \delq\ = $-1.7$ percentage points.

The root mean square (rms) difference between the global fitting function 
used by the above-mentioned web page and the input data points of the regular grid is
$0.007$\,dex for \delA\ and $0.07$ percentage points for \delq. These numbers
represent the mean fitting function errors of the two quantities for 
stellar parameters that lie close to our grid points.

In order to evaluate the interpolation errors of our
analytical expressions outside of the grid points, we computed \delA\ and
\delq\ for intermediate \teff, \logg\ and \ali\ values by using two additional
3D model atmospheres with \met\ = 0.0, \logg\ = 4.3 and \teff\ = 6110 and
6430 K. The 3D NLTE spectra have been computed for \aliTd\ = 1.5, 1.75, 2.0,
2.25, 2.5 and for the same \iso\ and \vsini\ values as in case of the original
grid. The rms difference between the global fitting function and these new 
intermediate data points is again $0.007$\,dex for \delA\ and $0.10$ 
percentage points for \delq. Even if we do not have the possibility to compute 
3D NLTE corrections for intermediate points in \met, due to the lack of 3D
hydrodynamical model atmospheres, we assume that the mean interpolation
errors of $0.01$\,dex for $\Delta A$ and $0.10$ percentage points for $\Delta
q$ are valid for the whole parameter space covered by our grid. The largest 
errors in $\Delta A$ of $\pm 0.015$\,dex are encountered at low temperatures
(\teff $\,\approx 5900$\,K), while the largest errors in $\Delta q$ 
of $\pm 0.2$ percentage points are mainly incurred at high temperatures 
(\teff $\,\approx 6500$\,K).

\subsection{1D NLTE corrections and comparison with other works}
\label{sec_1dnle}
\begin{figure*}[htb!]
\centering
\includegraphics[clip=true,width=250pt]{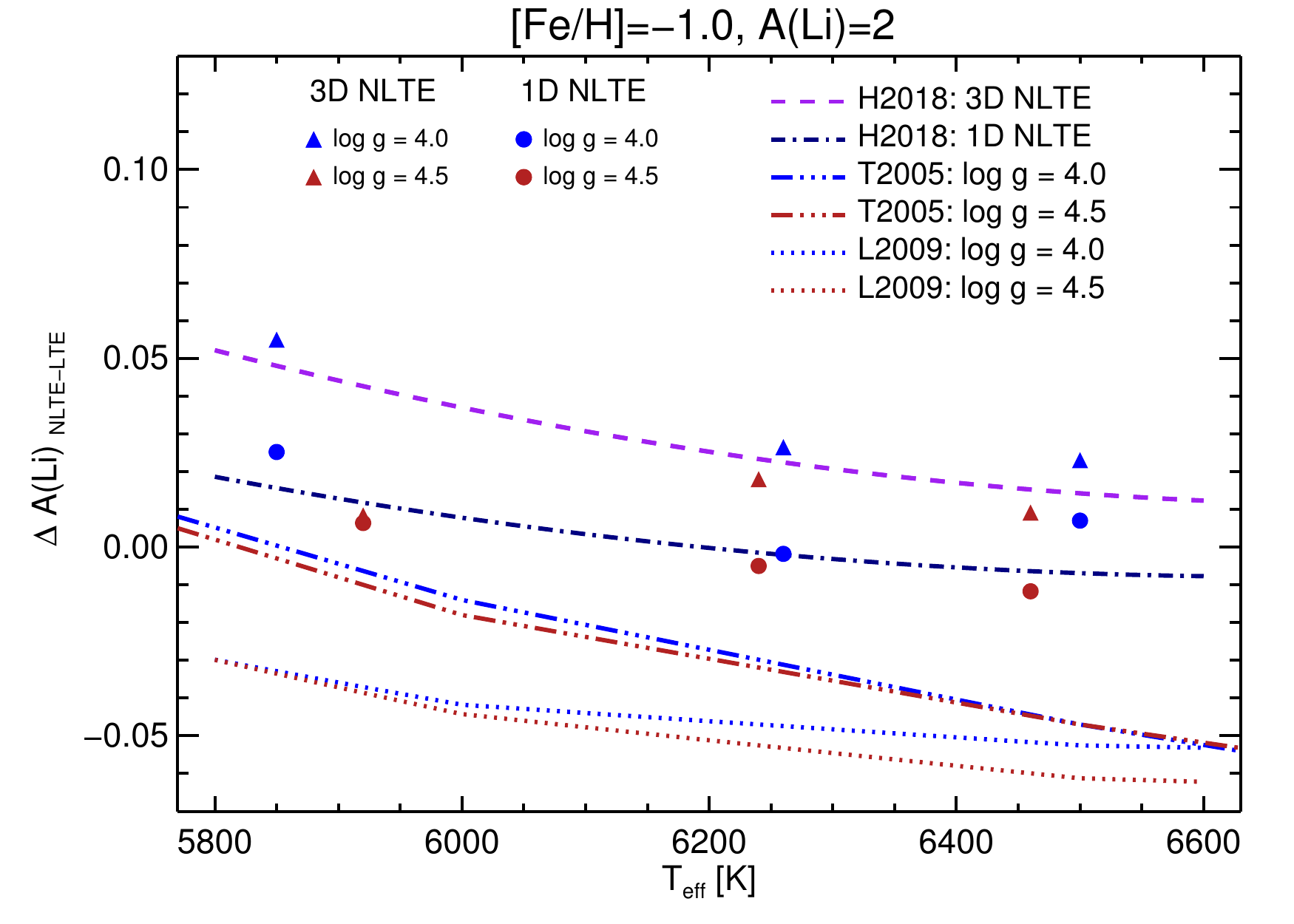}
\includegraphics[clip=true,width=250pt]{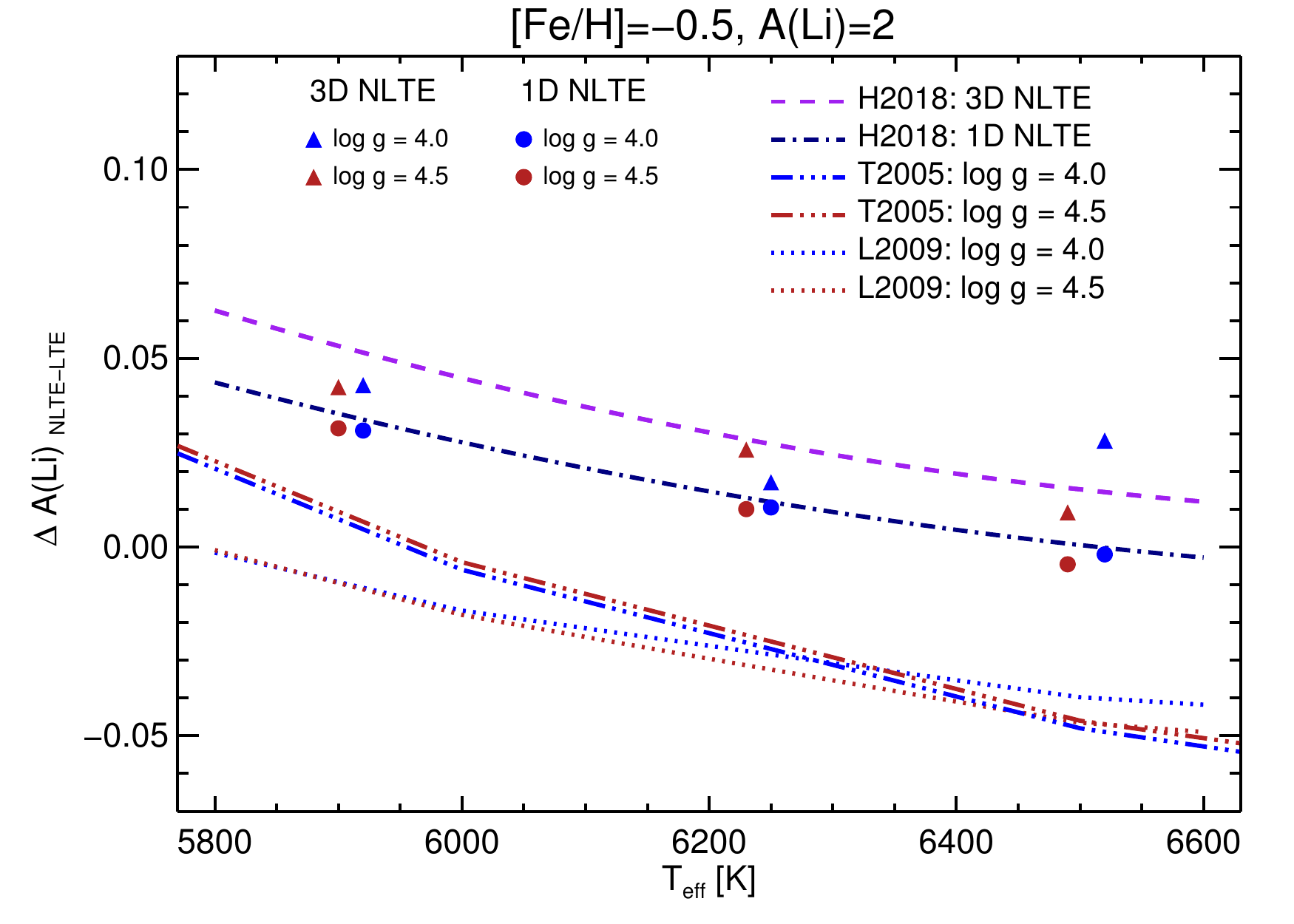}\\
\includegraphics[clip=true,width=250pt]{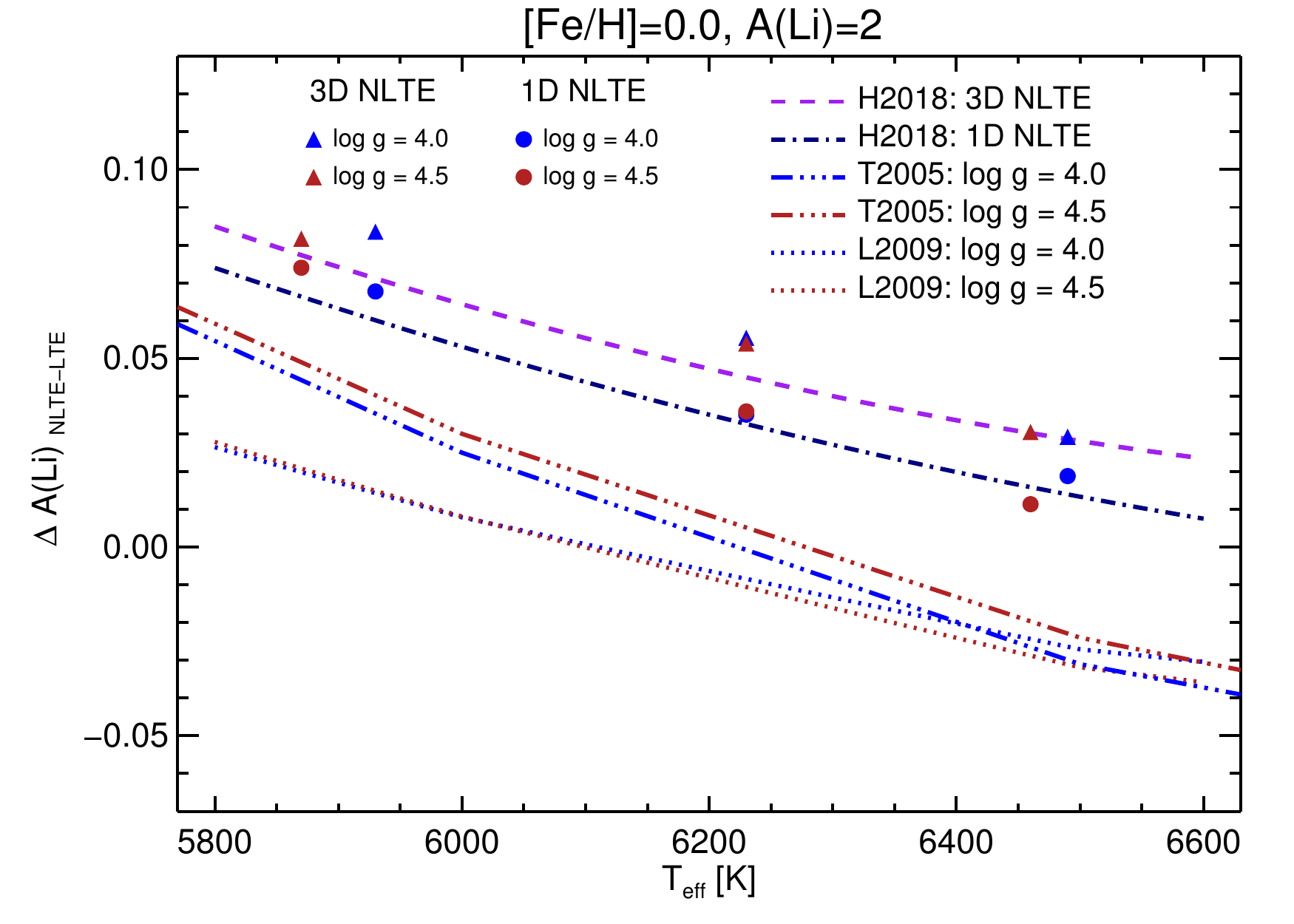}
\includegraphics[clip=true,width=250pt]{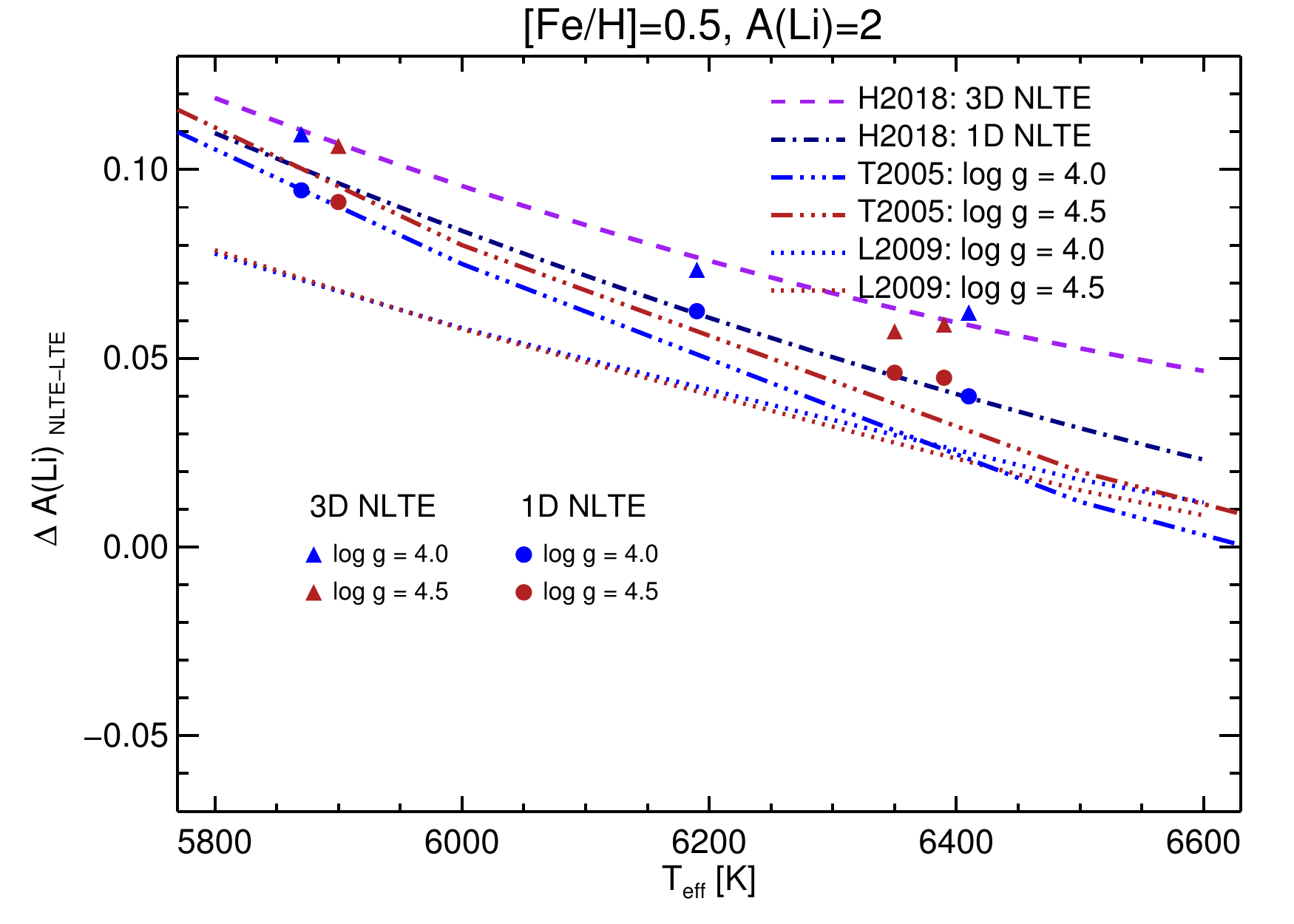}
\caption{Comparison of the 1D and 3D NLTE abundance corrections 
derived in this work with the 1D NLTE corrections from \cite{lind2009} (L2009, 
dotted lines) and  \cite{takeda2005} (T2005, 
dash-dotted lines)  for metallicities $-1.0$, $-0.5$, 
$0.0$, and $+0.5$ (from top left to bottom right). Red and blue triangles 
denote the individual 3D NLTE corrections, while red and blue circles denote 
the individual 1D NLTE corrections for \ali\ computed in this work (H2018) 
for \logg\ of 4.0 and  4.5, respectively. 
}
\label{fig_comp}
\end{figure*}

In order to distinguish between the contributions of 3D and NLTE effects in
the combined 3D NLTE correction, we repeated the same procedure for deriving
the 3D NLTE corrections but replacing the grid of 3D NLTE spectra for
\aliTd\,=\,2.0 and \qliTd\,=\,5\% with an identical grid of 1D NLTE spectra. A
rotational broadening of \vsini\,=\,2.0\,\kms\ was applied to both 1D NLTE and
1D LTE spectra. The 1D NLTE corrections for the \iso\ ratio are
very close to zero, having a mean value of $-0.10\%$ with a standard deviation 
of 0.08\%. This is expected as both LTE and NLTE line profiles are intrinsically
symmetric. On the other hand, the 1D NLTE corrections for \ali\ show a similar
trend as the 3D NLTE corrections, although with an offset to slightly smaller
values (see Fig.\,\ref{fig_comp}). We have derived an equation similar to the
Eq.\,(\ref{eq_li}) for the 1D NLTE case to facilitate the comparison with our
3D corrections and with the results from the literature (see below).

Figure\,\ref{fig_comp} shows the functions for estimating the 3D NLTE and 1D
NLTE \ali\ corrections derived in this work for metallicities \met\ = $-1.0$, $-0.5$, $0.0$, and $+0.5$, together with the results obtained by \cite{takeda2005} and \cite{lind2009}. Our 3D and 1D NLTE
corrections follow similar trends but with small offset towards
lower values (ranging between 0.01 and 0.03\,dex, depending on \met\ and 
\teff). This only small difference indicates that the 3D
effects slightly enhance the non-LTE effects, while the latter generally
dominate over the 3D effects in the combined 3D NLTE corrections.


The overall trends of our 1D NLTE corrections are similar to the trends
obtained by \cite{lind2009}, with an offset ranging between $0.02$
  and 0.06\,dex, depending mainly on the metallicity, while the curves of
\cite{takeda2005} have a slightly different slope. Such differences in the 1D
NLTE lithium abundance corrections, being $\sim$ 0.06\,dex at most, are not
unexpected due to the differences in the methods (model atmospheres, model
atom, numerical procedure for computing departure coefficients, etc.)  used by
different authors. Nevertheless, we performed a test to check if the
difference in the 1D NLTE \ali\ corrections could be explained by the
different model atmospheres used in the spectral synthesis in this work (1D
\mlhd\ models) and in \cite{lind2009} (MARCS models), respectively.  The test
was performed using a 1D NLTE spectrum computed from a MARCS model with \teff
=6500 K, \met =$-1$, \logg =4.0, assuming \alionednlte =2.0 and \isoonednlte
=5\,$\%$, and a corresponding grid of 1D LTE spectra computed for combinations
of nine \ali\ and nine \iso\ values, using the same MARCS model atmospheres
\citep{gustaffson2008} as in \cite{lind2009}. The MARCS NLTE spectrum was
fitted by the grid of MARCS LTE line profiles following the same fitting
procedure as described in Sect.~\ref{sec_fitting3d}. The resulting 1D LTE
\ali\ correction is $-0.055$\,dex, in very good agreement with the correction
of $-0.053$\,dex obtained for those stellar parameters by
\cite{lind2009}. This result confirms that the shift of $\sim$0.05\,dex
between our 1D NLTE lithium abundance corrections and the ones by
\cite{lind2009} (see Fig.\,\ref{fig_comp}, top left panel)
are likely due to the different model atmospheres adopted in the
spectral synthesis.

The analytical expressions given in \cite{sbordone2010} for \ali\ were derived
using 3D model atmospheres with metallicities $-2$ and $-3$ that do not
overlap with our grid. Since we do not observe a linear dependence of the
\ali\ 3D NLTE corrections on the metallicity, the extrapolated corrections
from \cite{sbordone2010} would be too uncertain for a useful comparison with our results. Therefore, we do not present such a comparison.

We compared the 3D NLTE corrections for \iso\ derived in this work with the
corrections presented by \cite{steffen2012} and found very good
agreement. This was expected, since the model atmospheres and methods used in
these two works are similar. The present study extends the work of 
\cite{steffen2012}, using a 3D NLTE grid with more metallicity points around the solar value, and also investigating the dependence of the 3D NLTE corrections on \ali, \iso, \vsini, and the list of blend lines.

\section{Li abundance and \iso\ isotopic ratio in HD\,207129 and HD\,95456}
\label{sec_stars}

In this section, we analyze two solar-type stars, HD\,207129 and HD\,95456, in
order to determine \ali\ and the \iso\ isotopic ratios, at first with a 1D LTE
approach, and afterwards applying our already pre-computed 3D NLTE
corrections. These stars have been selected for this study because
the difference in their effective temperatures places them at two extreme
positions in our correction grid. They are part of a larger sample of stars
with high lithium content and low activity levels, for which high-resolution
and very-high-S/N spectra are available that allow a
sensitive analysis of the \iso\ ratio. We point out that
\cite{Gomesdasilva2014} performed long-term activity studies of HD~207129 and
HD~95456 using \ion{Ca}{II} H $\&$ K and H$\alpha$ lines in their HARPS
spectra \citep{mayor2003}, and found both stars to show very low activity
levels with no significant long-term variability.

\subsection{Observations and stellar parameters}

The observed spectra used in this work have been obtained with the HARPS
spectrograph at La Silla observatory (ESO, Chile) within the framework of the
HARPS GTO planet search program \citep[subsample HARPS-I,][]{mayor2003}. The
spectral resolution of R$\sim$115 000 and the high S/N of $\sim$2000 of the combined spectrum of each star provide the quality that is needed for analysis of the \ion{Li}{i}  $\lambda$670.8\,nm region in the context of the \iso\ measurements.    

The stellar atmospheric parameters (\teff, \logg, \met\ and \vmic) of both
stars were derived by \cite{sousa2008} and are summarized in 
Table~\ref{table_stellarpar}. The abundances of several  elements with blending lines present in the line lists (Si, Ca, Ti, V) were adopted from \cite{adib2012}, whereas the carbon and oxygen abundances were taken from \cite{suarez2017} and \cite{bertran2015}, respectively, and the nitrogen abundance was chosen to be equal to the carbon abundance. 
The elemental abundances adopted for the Sun, HD\,207129, 
and HD\,95456 are given in Table~\ref{t-abund}.


\begin{table}[htb!]
\caption{Stellar parameters of HD\,207129 and HD\,95456. }
\centering
\begin{tabular}{lll}
\hline\hline
Parameter       & HD\,207129    & HD\,95456  \\
\hline
\teff\ [K]      & 5937  $\pm$   13      &  6276   $\pm$ 22              \\
\logg\          & 4.49  $\pm$   0.02    &  4.35  $\pm$ 0.04                     \\
\met\           & 0.00          $\pm$ 0.01              &  0.16 $\pm$   0.02    \\
\vmic\  [\kms]  & 1.40  $\pm$           0.01&  1.40     $\pm$   0.02    \\
\vsini\ [\kms]& 2.21$\pm$ 0.24          &  3.28                  \\
$\Delta$\met    & $-0.025$ $\pm$ 0.063      &  $-0.068$ $\pm$ 0.067             \\
\hline
\end{tabular}
\label{table_stellarpar}
\tablefoot{\teff, \logg, \met, \vmic\ and their errors are adopted from
  \cite{sousa2008}, and the \vsini\ for HD\,95456 is taken from
  \cite{dm2015}. $\Delta$\met\ is the small iron abundance  correction
   derived in Sect.\,\ref{sec_rotbr} relative to the adopted metallicity 
   (third row).
  }
\end{table}

\subsection{Rotational broadening and iron abundance}
\label{sec_rotbr}
A list of atomic line data has been constructed for a number of isolated,
unblended  \ion{Fe}{i} lines taken from \cite{doyle2014}, \cite{tsantaki2013}
and from the VALD-v3 database \citep{kupka2011} to estimate the projected
rotational velocity (\vsini) and the possible small iron abundance correction
for the stars studied in this work. For each line of each star, a grid of nine
1D LTE Kurucz Atlas spectra was computed assuming an iron abundance range of $\pm$ 0.2\,dex around the adopted iron abundance (metallicity) value, with a step of 0.05\,dex.The spectra were computed using the 2014 version of the spectral synthesis code MOOG \citep{sneden1973}.

The fitting procedure was performed for the solar flux atlas from
\cite{kurucz2005}, and for the HARPS spectra of HD\,207129 and HD\,95456. The HARPS spectra have been locally normalized for each Fe line. We investigated the quality of the fits by eye and by means of a \chisq\ analysis. All the lines that seemed to be blended with other spectral features (at least in one of the two stars or the Sun), or which provided a non-satisfactory fit to the observations, were excluded from the sample. Eventually, we were left with 10 "clean"  \ion{Fe}{i}  lines which are well isolated and can be fitted well in the solar spectrum and the HARPS spectra of the two stars. Table~\ref{t-feI} lists the atomic data of the selected Fe I lines: wavelength, excitation potential (EP), oscillator strength (\loggf), and the reference to the source they were taken from for each individual  \ion{Fe}{i} line.


\begin{table}[htb!]
\caption{Chemical abundances $A(\mathrm{X})$ used in this work.}
\centering

\begin{tabular}{l  | c  | c  | c   }
\hline
\hline
\noalign{\smallskip}

Element  & Sun & HD 207129 & HD 95456 \\

\noalign{\smallskip}
\hline 
\noalign{\smallskip}

C  & 8.50 & 8.44 & 8.67 \\
O  & 8.76 & 8.69 & 8.93\\
Si & 7.51 & 7.55 & 7.70 \\ 
Ca & 6.34 & 6.39 & 6.54\\ 
Ti & 4.95 & 5.04 & 5.16 \\ 
V  & 3.93 & 4.03 & 4.17 \\
Fe & 7.52 & 7.495 & 7.612 \\

\hline
\end{tabular}
\tablefoot{The abundances of C, O and Fe for the Sun are adopted from
  \cite{caffau2011} and of Si, Ca, Ti and V from \cite{asplund2009}. For HD
  207129 and HD 95456, the abundances of C and O are taken from
  \cite{suarez2017} and \cite{bertran2015}, respectively, and the abundances of
  Si, Ca, Ti and V are from \cite{adib2012}. The Fe abundance 
  is computed in this work (Sect.~\ref{sec_rotbr}). %
  }
\label{t-abund}
\end{table}

Four parameters were varied to achieve the best fit: [Fe/H], \vsini, the
global wavelength shift ($\Delta v$), and the continuum level. Since it was
difficult to obtain both \vmac\ and \vsini\ individually from the fitting
procedure due to a `degeneracy' of the solution, a fixed Gaussian
macroturbulence velocity (\vmac) was assumed for the fitting. \vmac\ was
adopted from \cite{doyle2014} and used after the application of a factor to
convert the radial-tangential (RT) macroturbulence parameter ($V_{RT}$)
provided by these authors to the Gaussian macroturbulence parameter
\vmac\ that we use. This conversion factor was determined for the solar case
using the same solar \vsini\ and atmospheric parameters 
(including \vmic\ = 1.0\,\kms) as
assumed in \cite{doyle2014}. We derive a solar \met\ = $-0.02$ $\pm$ 0.06 and
\vmac\ = 2.12 $\pm$ 0.10 \kms\, resulting in the conversion formula
\vmac\ $\sim$0.66$V_{RT}$. The relation between the different macroturbulence
velocity models has been discussed in a recent work by \cite{takeda2017}, and
the conversion factor of $0.66$ that we derive is in agreement with their
results. For a given \vmac, each line profile has been broadened by nine
different \vsini\ values with the flux convolution approximation, assuming a
limb darkening coefficient of $\epsilon=0.6$ \citep{gray2005}. The
least-squares fitting method that we apply (MPFIT; \citealt{markwardt}) relies on
interpolation in the grid of the precomputed synthetic spectra with different
\met\ and \vsini. For each star, the best fit values of \met\ and \vsini\ were
averaged over the ten different \ion{Fe}{i} lines and the uncertainties were
derived from the standard deviations (cf.\, Table~\ref{t-feI}).

For HD\,207129, we obtain \vsini\,=\,2.21\,$\pm$ 0.24 \kms\, and an iron
abundance correction of $\Delta$\met\,=\,$-$0.025\,$\pm$ 0.063\,dex. For
HD\,95456, the radial-tangential macroturbulence is
$V_{RT}$\,=\,5.05\,\kms\ according to \cite{doyle2014}, which is large enough
to broaden the synthetic line profiles to such an extent that the fits were
not very different for \vsini\ between 0 and 3\,\kms. For instance, when we
use the \vmac\ value from \cite{doyle2014}, we obtain a formal \vsini\ of
0.88\,\kms. As we discuss in Section~\ref{fitting_harps}, the measured \ali\ and
\iso\ ratio for HD\,95456 essentially do not change with different
\vsini\ assumptions. We therefore prefer the \vsini\ value of 3.28\,\kms\ determined by \cite{dm2015}, based on a combination of Fourier transform and goodness-of-fit methods. We derive a small Fe abundance correction of $\Delta$\met\,=\,$-0.068$ $\pm$ 0.067\,dex (relative to the literature value of \met) for HD\,95456.


\begin{table*}[htb!]
\caption{\ion{Fe}{i} lines used for deriving \vsini\ and the iron abundance correction $\Delta$\met.}

\begin{tabular}{c c c c c | c r | c  r | c r  }
\hline
\hline
\multicolumn{5}{c}{Spectral line parameters} & \multicolumn{2}{c}{Sun} & \multicolumn{2}{c}{HD 207129} & \multicolumn{2}{c}{HD 95456} \\
\noalign{\smallskip}
\hline 
\noalign{\smallskip} Chemical   &Wavelength                     &EP             &log $gf$    & Source& \vmac\ & $\Delta$\met\ & \vsini\ & $\Delta$\met\ & \vsini\ & $\Delta$\met\ \\
Species         &$\lambda$ (\AA)                &(eV)   &(dex)          & &
(\kms\ ) & (dex) &  (\kms\ ) & (dex) &  (\kms\ ) & (dex) \\ \hline
\ion{Fe}{i}&6094.3800&4.650&$-1.566$& T13  & 2.14&$-0.088$& 2.53&$-0.088$&1.58&$-0.118$\\ 
\ion{Fe}{i}&6200.3190&2.609&$-2.437$& Do14 & 2.06&$-0.019$& 1.97&$-0.022$&0.00&$-0.091$\\ 
\ion{Fe}{i}&6481.8800&2.280&$-2.929$& T13  & 1.94&$-0.078$& 1.95&$-0.074$&0.00&$-0.130$\\ 
\ion{Fe}{i}&6593.8800&2.430&$-2.384$& T13  & 2.05&$-0.049$& 1.94&$-0.055$&0.00&$-0.134$\\ 
\ion{Fe}{i}&6699.1408&4.593&$-2.101$& Vald3& 2.23&$-0.078$& 2.14&$-0.119$&1.59&$-0.112$\\ 
\ion{Fe}{i}&6703.5655&2.759&$-3.160$& Vald3& 2.00&$ 0.068$& 2.04&$ 0.069$&0.61&$0.0419$\\ 
\ion{Fe}{i}&6725.3553&4.103&$-2.300$& Vald3& 2.23&$ 0.051$& 2.41&$ 0.052$&1.12&$0.0288$\\ 
\ion{Fe}{i}&6726.6658&4.607&$-1.094$& Vald3& 2.19&$ 0.021$& 2.39&$ 0.016$&1.40&$-0.037$\\ 
\ion{Fe}{i}&6810.2570&4.607&$-0.986$& Do14 & 2.15&$-0.044$& 2.24&$-0.049$&0.98&$-0.111$\\ 
\ion{Fe}{i}&6857.2490&4.076&$-2.150$& Do14 & 2.17&$ 0.010$& 2.53&$ 0.019$&1.51&$-0.015$\\ 
\hline  
Average  & & & & & $2.12$     &$   -0.021$ &    $2.21$ &     $-0.025$ &     $0.88$  &   $-0.068$\\ 
Std. dev.& & & & & $\pm 0.10$ &$\pm 0.056$ &$\pm 0.24$ &  $\pm 0.063$ & $\pm 0.68$  &$\pm 0.067$\\ 
\hline
\end{tabular}
\tablefoot{The lines and their parameters are adopted from the VALD-v3
  database \citep{kupka2011} (Vald3), \cite{tsantaki2013} (T13) or
  \cite{doyle2014} (Do14), as indicated in the fifth
  column. For the Sun, we used a fixed \vsini\ of 1.9~\kms in order to derive its \vmac, whereas for HD~207129 and HD~95456 we proceeded the other way round, fixing the \vmac\ values at 2.33 and 3.33~\kms, respectively, in finding their \vsini\ (see Sect.~\ref{sec_rotbr}). 
}
\label{t-feI}
\end{table*}

\subsection{Blend lines in the lithium $\lambda$670.8\,nm region }
\label{sec_linelist}
One of the challenges in \iso\ ratio measurements is the presence of several
other atomic and molecular lines overlapping with the resonance doublet. While
the contribution of these blends may be very small in metal poor stars, they
become more significant at higher metallicities and must be treated carefully
for stars having metallicities close to solar or higher.
There are several lists of atomic and molecular lines currently
available in the literature which have been carefully constructed
to reproduce the Li $\lambda$670.8\,nm region \mbox{(e.g., \citealt{mandell2004},
\citealt{ghezzi2009}}, \citealt{melendez2012}, \citealt{israelian2014}) and are
fitted for very sensitive measurements of the \iso\ isotopic ratio in
solar-type stars. It has already been demonstrated that the use of different
lists of atomic and molecular lines interfering with the \ion{Li}{i}
$\lambda$670.8\,nm doublet can lead to noticeable differences in the measured values of the \iso\ ratio \citep[e.g.,][]{israelian2003, mott2017}. We therefore perform our analysis using different line lists provided in the literature and compare the results. 

Specifically, we used three different line lists for the analysis of \ali\ and
\iso\ in the atmospheres of HD\,207129 and HD\,95456. In addition to G09 and
M12, we used the line lists constructed by \citet{israelian2014}, as
given in \citealt{mott2017}), hereafter, I14. The line lists were adjusted by
computing 1D LTE synthetic spectra and fitting the 670.8\,nm region in the
solar spectrum (G09) or in spectra of other stars (I14).  The above authors
assumed different elemental abundances from what we assume for the Sun
(Table \ref{t-abund}, col.\ 2). 
Specifically, we adopted the solar abundance values for 12
elements (Li, C, N, O, P, S, K, Fe, Eu, Hf, Os, Th) from \cite{caffau2011}
(their Table 5), and for other elements we used the internal solar elemental
abundances of the 2014 version of MOOG, which is based on the solar abundances
recommended  by \cite{asplund2009}.

A molecule that is present in all line lists representing the \ion{Li}{i}
670.8\,nm region is CN. The presence of several lines of this molecule is
important in the \lisix\ determination, and assuming the correct C, N, and O
abundances (because  O can influence the equilibrium of the molecular reaction
network) can be crucial. Particularly, the assumptions for solar abundances of
O and N were significantly different between the authors of the line lists
and our work. Therefore, we made some calibrations to eliminate the differences
arising from different assumptions of solar abundances while constructing the
line lists for performing our \iso\ analysis. In case of O and N, our
abundances were lower by up to 0.17 and 0.19\,dex, respectively, and by
0.06\,dex for C. In this case, we have scaled the \loggf\ values of all the
CN lines in this region by a constant factor. 
This factor was derived by fitting the solar flux atlas with a pre-computed
grid of 1D LTE synthetic spectra for the Sun. For each line list, a grid of 1D
LTE spectra was computed applying different factors (within the range of
$\pm$0.3\,dex with a step of 0.01\,dex) to the \loggf\ values of all the CN
lines. Then, for a given line list, we adopted the \loggf\ factor resulting
in the best fit ($\chi^{2}$ minimization) to the solar flux atlas. 
This factor was +0.20\,dex in case of the G09 and M12 line lists, 
and +0.19\,dex in case of the I14 line list. For elements other 
than C, N, and O present in the line lists, the small solar abundance 
differences were taken into account by applying corresponding corrections to 
the \loggf\ values of the individual  lines of each element. After this 
adjustment, all line lists are `normalized' to our adopted solar abundances.

In addition, the wavelength and the \loggf\ value of the \ion{V}{i} line at
670.81096\,nm have been updated according to the recently published values by
\cite{lawler2014}. This blend is positioned close to one of the
\lisix\ components and may affect the result of the \iso\ determination if
incorrect line parameters are assumed (see Sect.\,\ref{sec_si-v}).

\subsection{Spectral synthesis for \ali\ and \iso\ studies of HD\,207129 and HD\,95456}

In order to derive \ali\ and the \iso\ ratio in the two solar-type stars, we perform a standard 1D LTE analysis using the 2014 version of the spectral synthesis code MOOG \citep{sneden1973} together with Kurucz ATLAS9 1D model atmospheres \citep{kurucz1993}. 

A grid of 1D LTE synthetic spectra has been computed for each star. The
precomputed line profiles cover a range in lithium abundance defined by the
expected \ali\ (literature value, \citealt{dm2015}) $\pm$0.5\,dex with a 
step of 0.05\,dex, and a range in \qli\ between 0 and 20\% with a step of 1\%.   

For each star, the third dimension of the grid is defined by a variation of the C, N, O abundances. These abundances were scaled together by the same factor within a range of $\pm0.2$\,dex around the fiducial C, N, O abundances and with a step of 0.05~dex, such that the grid of spectra includes nine different CNO values. 
The \ion{Fe}{i} line at $\sim$670.74\,nm, being one of the strongest blends in
this region, can also play an important role in the \iso\ analysis. Therefore,
we extend our grid by computing the spectra with different \met\ values, which
vary around the fiducial Fe abundance by $\pm$ 0.1\,dex, with a step of 0.1\,dex for HD\,207129 and 0.05\,dex for HD\,95456. 

Thus, a four-dimensional (4D) grid of 1D LTE spectra is obtained for combinations of
21 \ali, 21 \iso, nine CNO abundances, and three (five) different
\met\ values, both for HD\,207129 and HD\,95456. In addition, these grids of
spectra for both stars were replicated for three different line lists (G09,
M12, I14). For the estimation of the systematic errors of \ali\ and \iso\ due to
uncertainties in effective temperature and surface gravity, the 4D
grids for both stars were also computed for model atmospheres with
\teff\,$\pm50$\,[K] and \logg\,$\pm0.1$, but only for line list G09, and for the Fe and CNO abundance correction factors from the best fit for that line list. In total, 37485 and 61299 1D LTE synthetic spectra were computed for HD\,207129 and HD\,95456, respectively.


\begin{figure*}[htb!]
\includegraphics[clip=true,width=250pt]{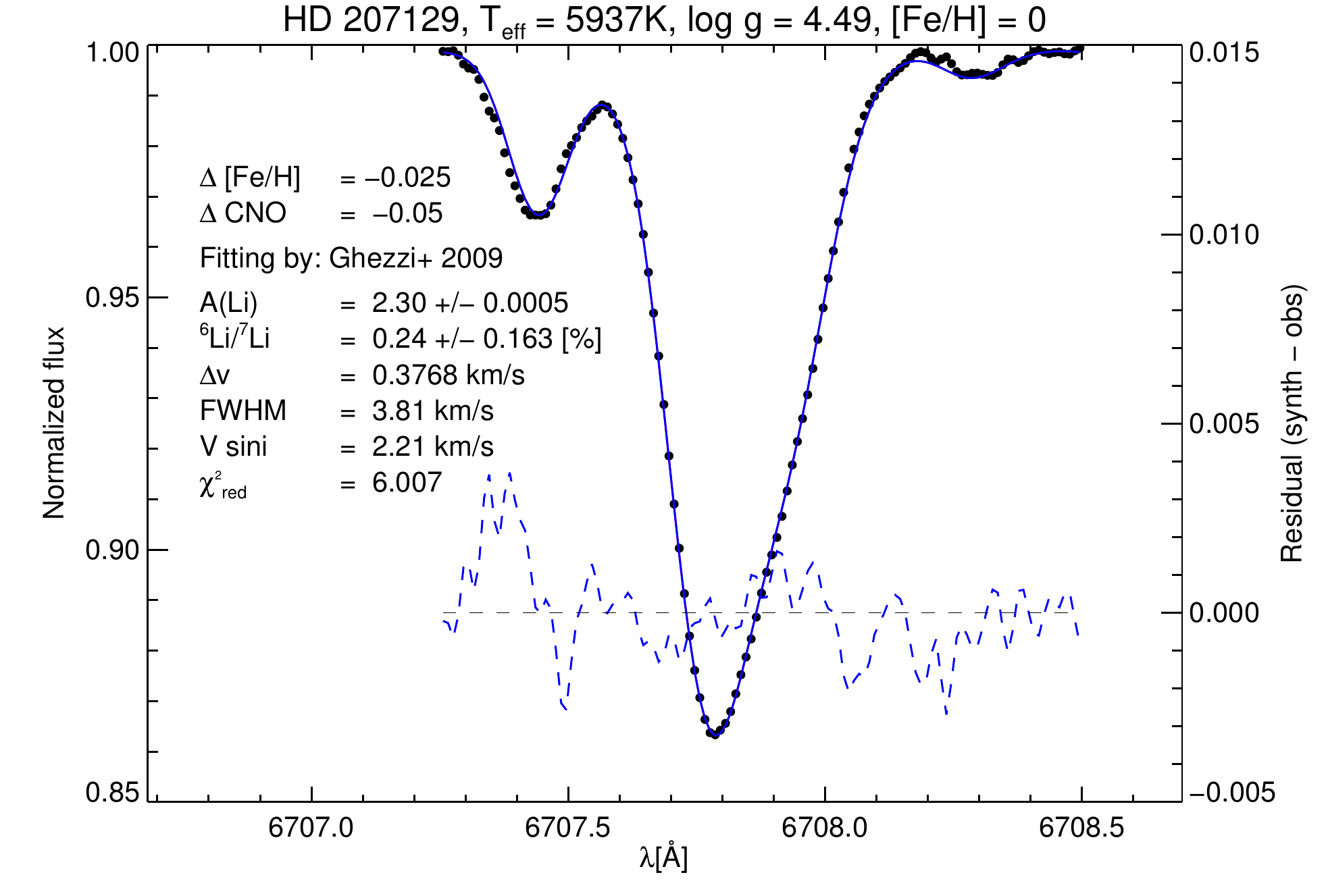}
\includegraphics[clip=true,width=250pt]{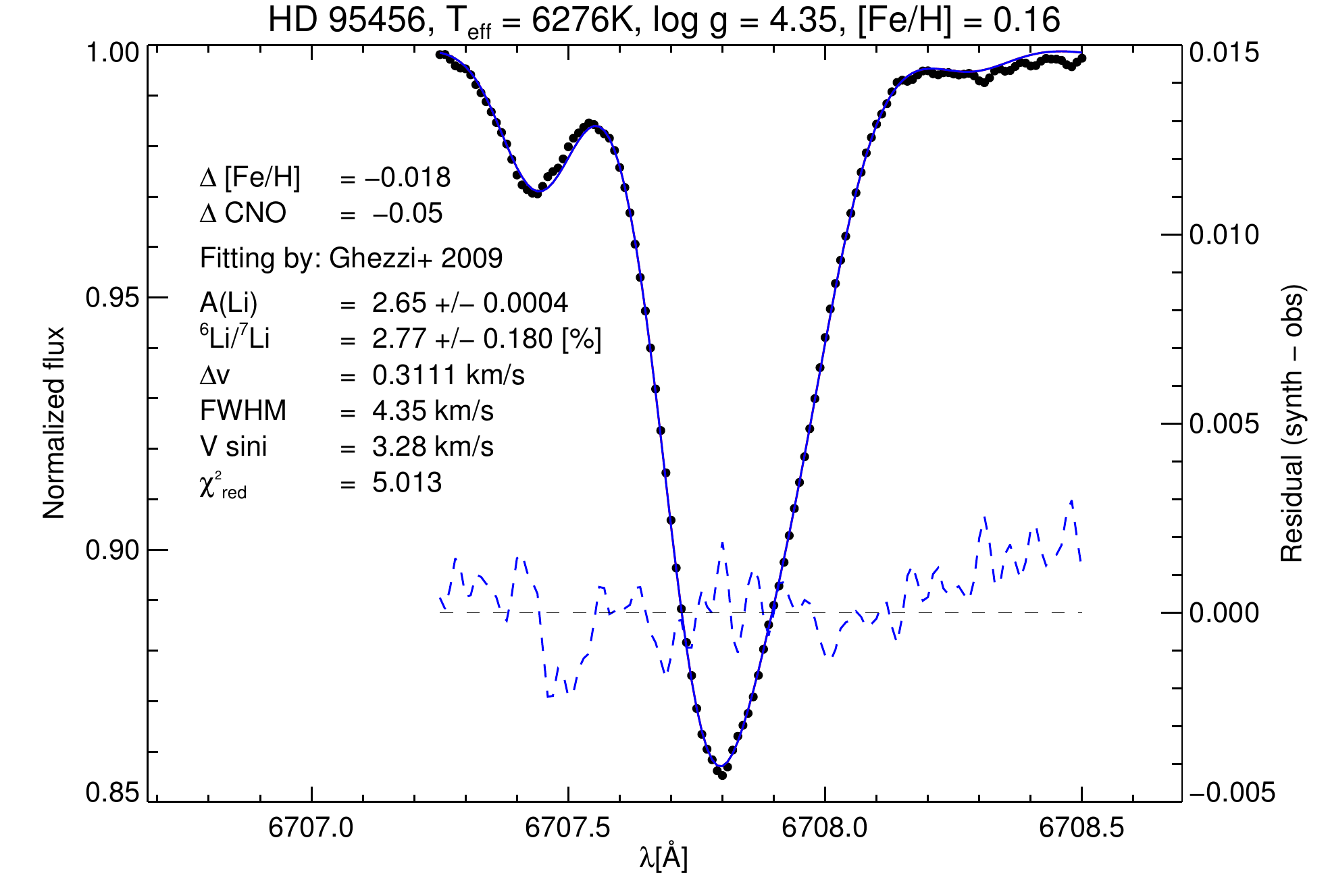}

\includegraphics[clip=true,width=250pt]{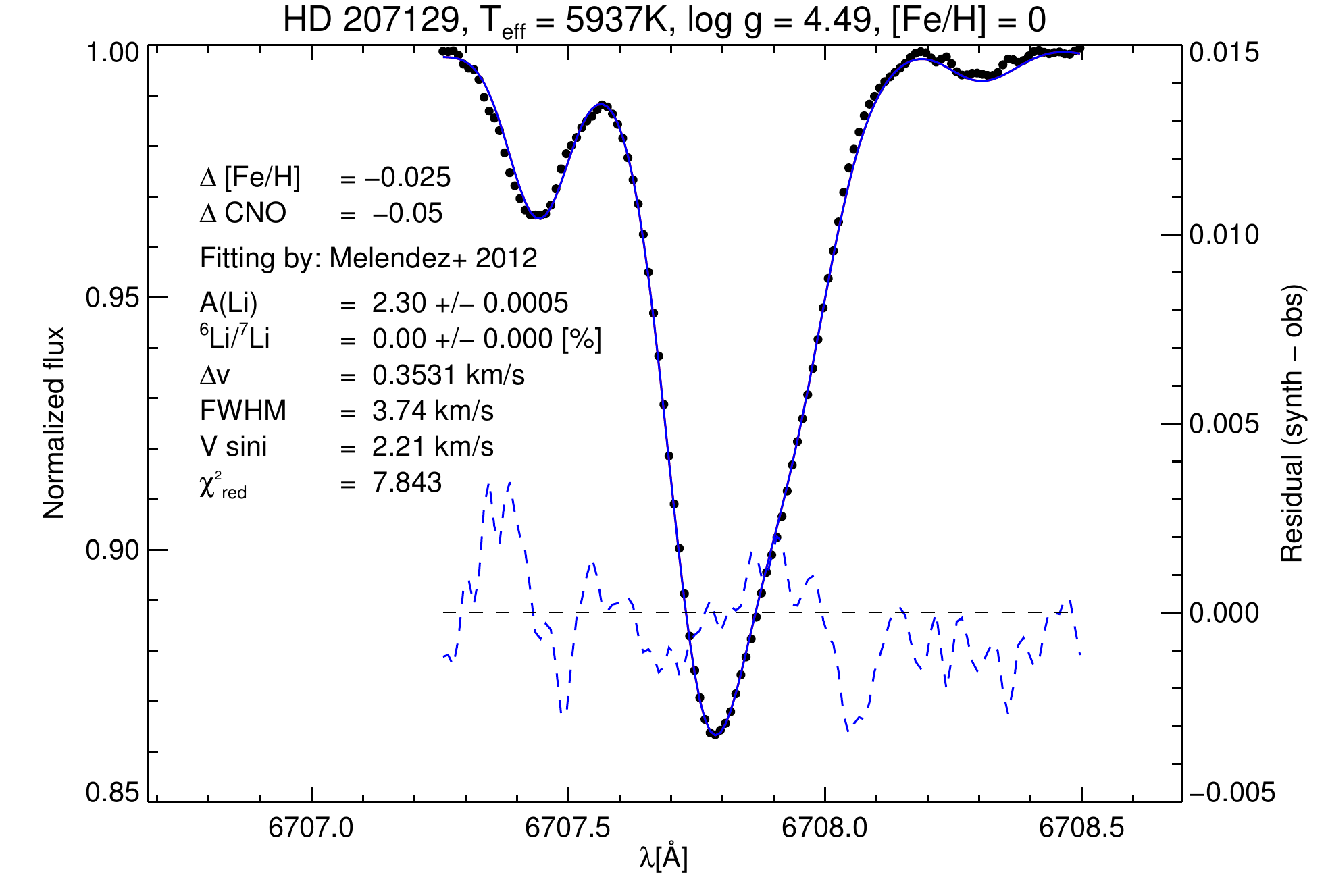}
\includegraphics[clip=true,width=250pt]{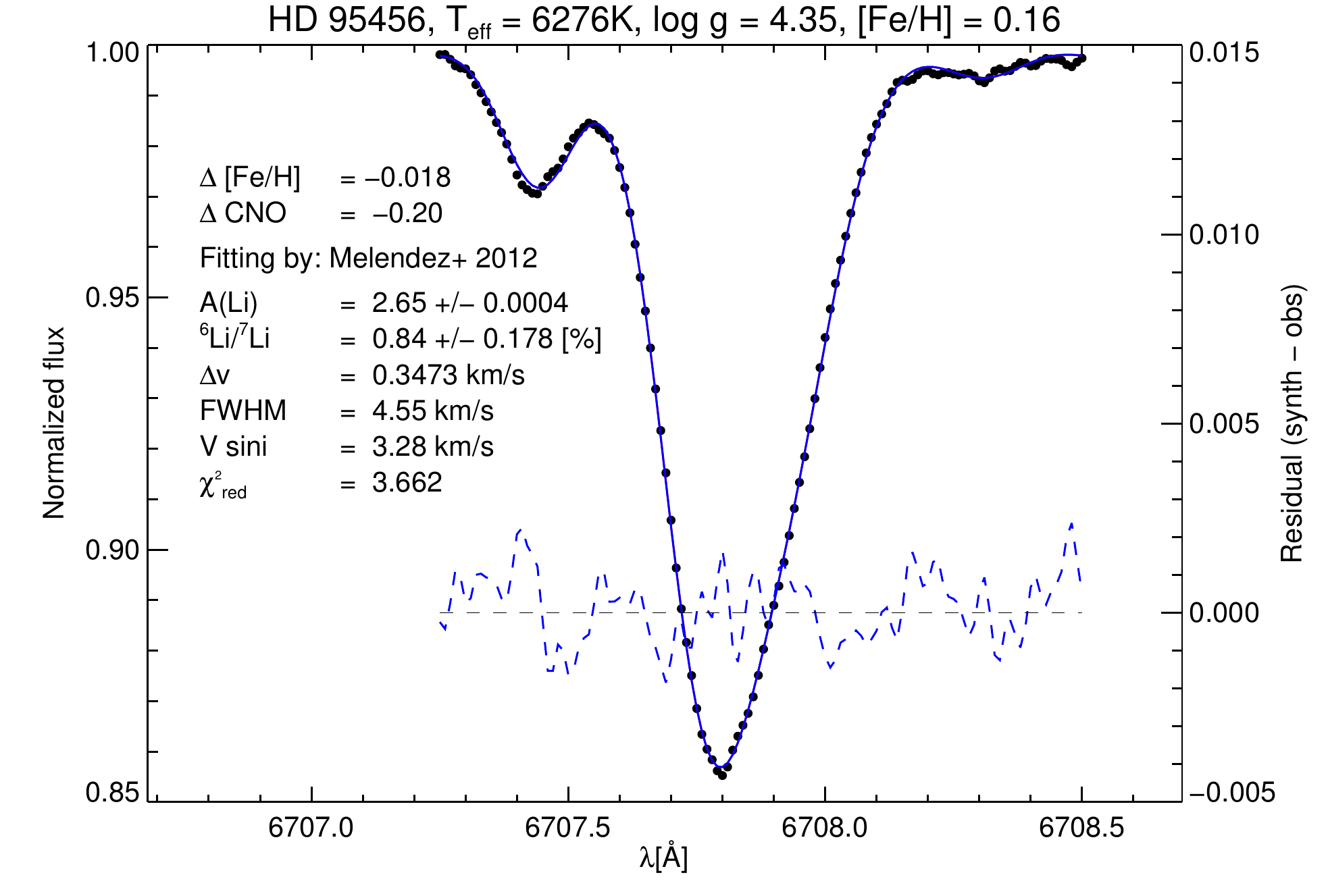}

\includegraphics[clip=true,width=250pt]{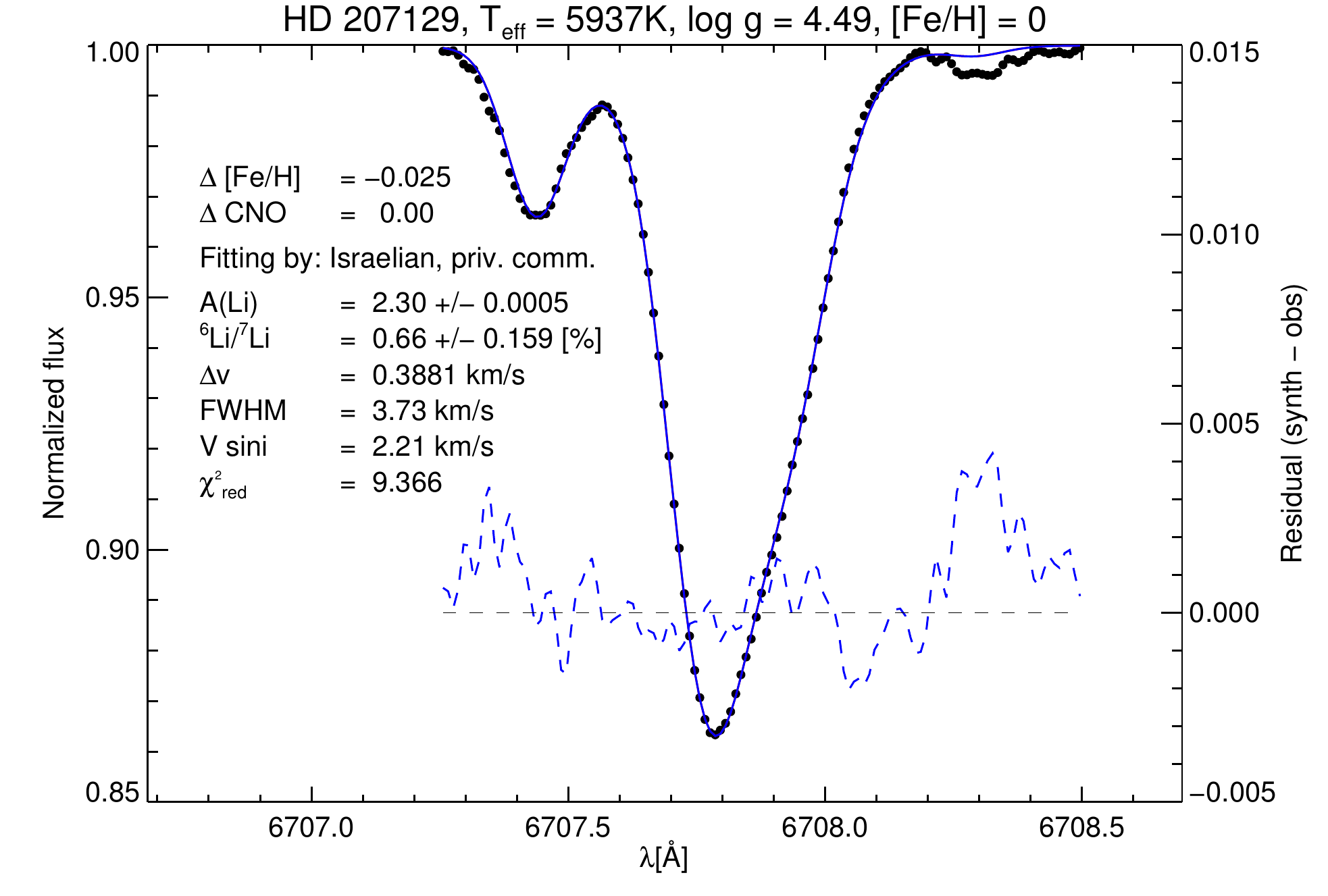}
\includegraphics[clip=true,width=250pt]{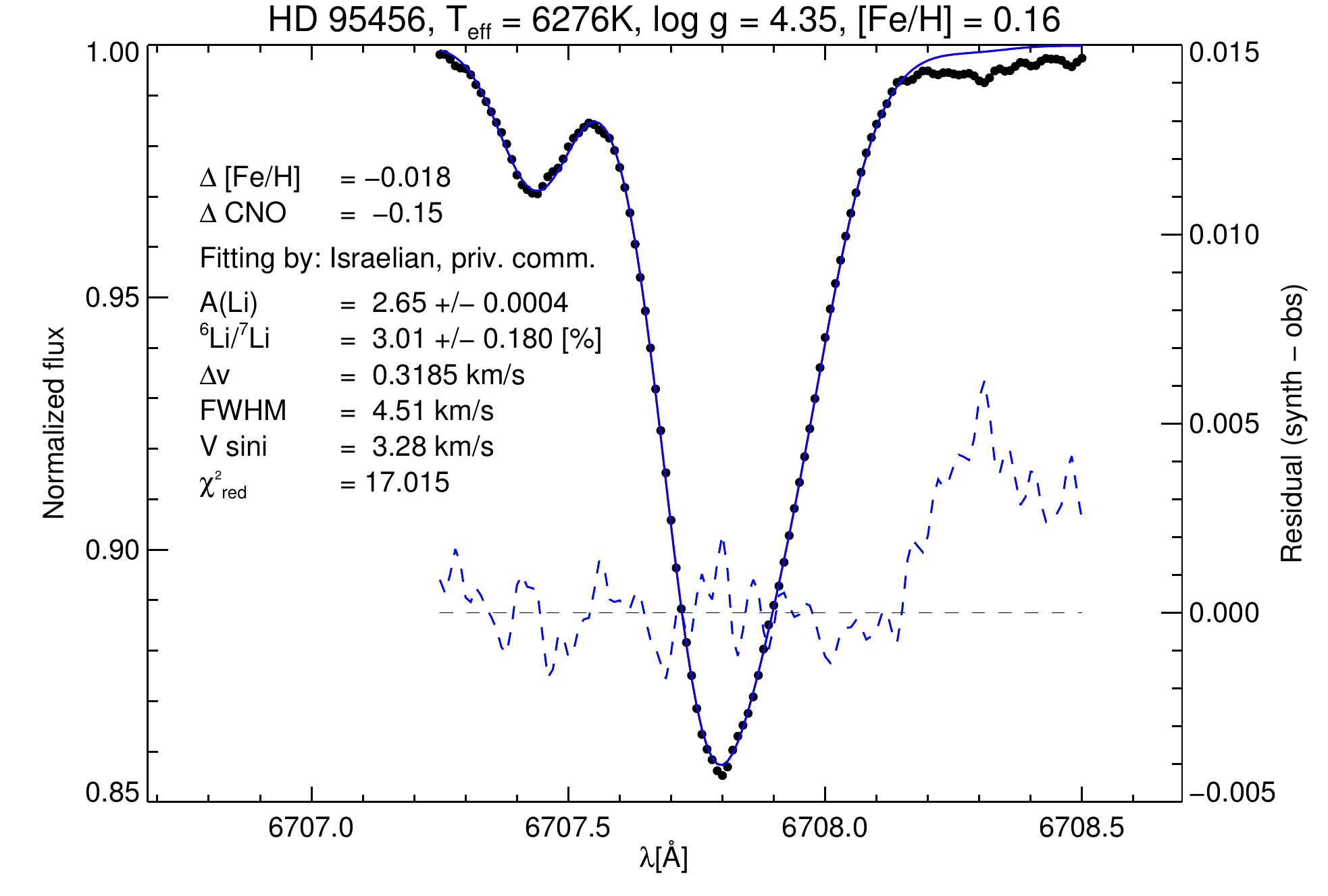}

\caption{
The best fitting 1D LTE spectrum (solid line) superimposed on the
HARPS spectrum (dots) for HD\,207129 (left panels)
and HD\,95456 (right panels). The synthetic line profiles are computed
adopting the line lists G09, M12, and I14 (from top to bottom), with 
modifications described in Sect.\,\ref{sec_linelist}. The right y-axis 
defines the scale of the residuals (dashed line).}
\label{bestfit_G09}
\end{figure*}


\begin{figure}[htb!]
\centering
\includegraphics[clip=true,width=250pt]{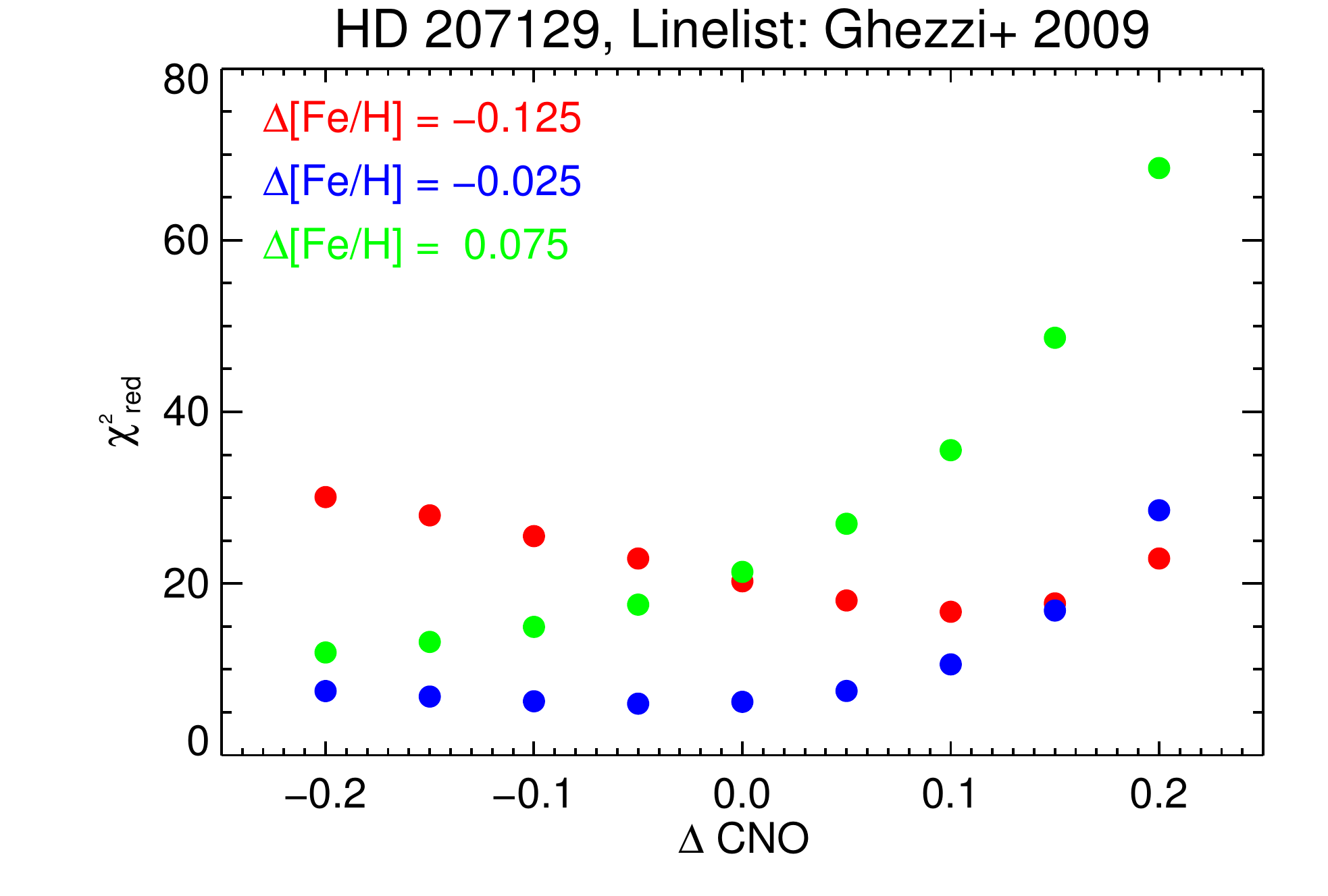}\\
\includegraphics[clip=true,width=250pt]{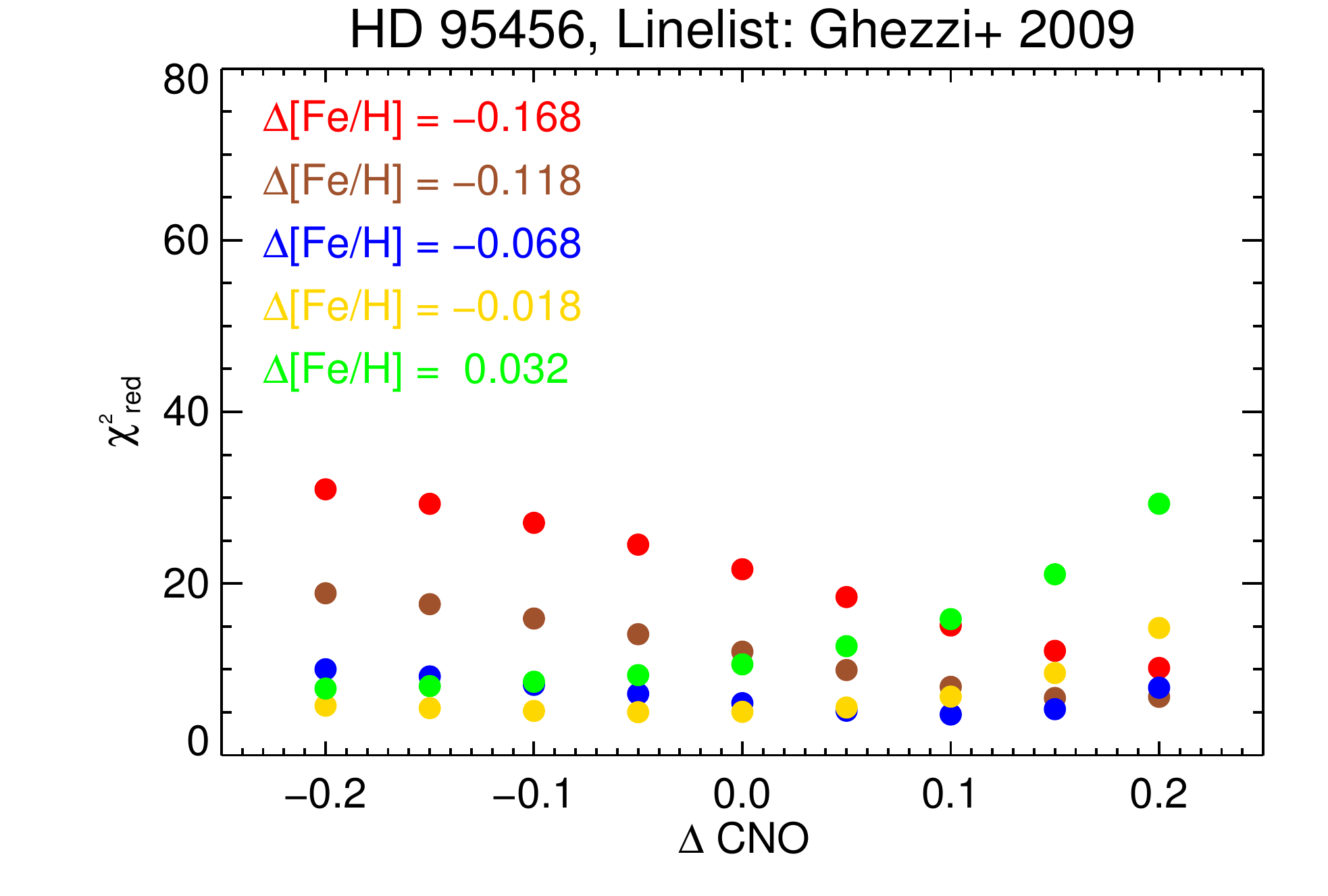}
\caption{
The best \chisq\ obtained by fitting the HARPS spectra  of HD\,207129 (upper panel) and HD\,95456 (lower panel) with grids of 1D LTE line profiles computed for combinations of different Fe and CNO abundances. The line list G09 is adopted after some modifications described in Section~\ref{sec_linelist}. For each point, the minimum \chisq\ is obtained by optimizing \ali\ and \iso\ at given \met\ and CNO abundance.}
\label{fit_cno_f_G09}
\end{figure}


\begin{table*}
\caption{Best fitting results for \ali\ and \iso\ ratio derived for HD\,207129
  and HD\,95456 using the line lists G09, M12 and I14 before and after the
  application of the 3D NLTE corrections. For reference, we also show the
  fitting result for the Sun obtained with line list G09.}
\label{table_Lires}
\begin{flushleft}
\begin{center}
\begin{tabular}{c c c c c r r r r | l r }
\hline 
\noalign{\smallskip}
\multicolumn{2}{l}{} & \multicolumn{7}{c}{1D LTE} & \multicolumn{2}{c}{3D NLTE}  \\

\noalign{\smallskip}
\hline 
\noalign{\smallskip} Star & Line list & \ali & \iso\ [\%] & \chisq &
$\Delta$\met & $\Delta$\,CNO & $\Delta$\,Si & $\Delta$\,V & \ali & \iso\ [\%] \\
\noalign{\smallskip}
\hline
\noalign{\smallskip}
\multicolumn{1}{ c }{\multirow{3}{*}{HD\,207129} } &
\multicolumn{1}{ c }{G09} & 2.30 & 0.2 & 6.0  & $-0.025$& $-0.05$ & --& --& 2.37$\pm$ 0.04 & $-0.9\pm$0.68 \hspace{\fill} \\ 
\multicolumn{1}{ c  }{}                        &
\multicolumn{1}{ c }{M12} & 2.30 & 0.0 & 7.8   &  $-0.025$ & $-0.05$ & -- & -- & 2.37 & $-1.1$ \hspace{\fill} \\
\multicolumn{1}{ c  }{}                        &
\multicolumn{1}{ c }{I14} & 2.30 & 0.7 & 9.4  &  $-0.025$ &  $0.00$  & -- & -- & 2.37 & $-0.4$ \hspace{\fill} \\ 
\noalign{\smallskip}
\hline
\noalign{\smallskip}
\multicolumn{1}{ c  }{\multirow{3}{*}{HD\,95456} } &
\multicolumn{1}{ c }{G09} & 2.65 & 2.8 & 5.0 & $-0.018$  & $-0.05$ & -- & -- & 2.70 $\pm$ 0.04 & $\;\;0.9\pm$1.36 \hspace{\fill} \\ 
\multicolumn{1}{ c  }{}                        &
\multicolumn{1}{ c }{M12} & 2.65 & 0.8 & 3.7 &  $-0.018$ & $-0.20$ & -- & -- & 2.70 & $-1.1$ \hspace{\fill} \\
\multicolumn{1}{ c  }{}                        &
\multicolumn{1}{ c }{I14} & 2.65 & 3.0 & 17.0 & $-0.018$  & $-0.15$ & -- & -- & 2.70 & $\;\;1.1$ \hspace{\fill} \\
\hline 
\noalign{\smallskip}
\multicolumn{1}{ c }{\multirow{1}{*}{Sun (Case A)} } &
\multicolumn{1}{ c }{G09} & 0.99 & 0.12 & 1.05  & $ 0.000 $& $0.000$ & $0.000$ & $0.000$ & $1.09\pm 0.03$ & $-0.7\pm2.0$  \hspace{\fill} \\ 
\multicolumn{1}{ c }{\multirow{1}{*}{Sun (Case B)} } &
\multicolumn{1}{ c }{G09} & 0.98 & 1.71 & 1.00  & $ -0.012 $& $0.073$ & $0.000$ & $0.000$ & $1.08\pm 0.03$ & $\;\;0.9\pm2.9$  \hspace{\fill} \\ 
\multicolumn{1}{ c }{\multirow{1}{*}{Sun (Case C)} } &
\multicolumn{1}{ c }{G09} & 0.98 & 0.71 & 0.94  & $-0.011 $& $ 0.073$ & $0.017$ & $-0.045$ & $1.08\pm 0.03$ & $-0.1\pm4.1$  \hspace{\fill} \\ 
\noalign{\smallskip}
\hline
\end{tabular}

\tablefoot{The 3D NLTE values are obtained by applying the 3D NLTE
  corrections for \ali\ and \iso\ to the 1D LTE results. $\Delta$\,CNO and
  $\Delta$\met\ represent the additional corrections applied to CNO and Fe
  abundances, respectively, to obtain the best fit. For the Sun only, 
  $\Delta$\,Si and $\Delta$\,V are further corrections of the silicon 
  and vanadium abundance in the final fit. The errors of \ali\ and \iso\ 
  for the two stars are computed for line list G09 and include the 
  uncertainties related to the different lists of blend lines (see
  Sect.~\ref{sec_fit_st}). For the Sun, the error in \ali\ corresponds 
  to a continuum placement uncertainty of $\pm 0.05\%$, while the error
  in \iso\ is the formal $1\,\sigma$ fitting error.}
\end{center}
\end{flushleft}
\end{table*}

\subsection{Fitting procedure for the HARPS spectra}
\label{fitting_harps}
To fit the observed spectra of the two target stars, we followed a procedure
similar to the one already described in Section \,\ref{sec_fitting3d}. The
HARPS spectra of HD\,207129 and HD\,95456 were fitted through interpolation
across a grid of 1D LTE line profiles computed with MOOG for line 
lists G09, M12 and I14. To achieve the best fit evaluated by means of a
\chisq\ minimization method, four free parameters were varied: \ali, \qli, the
global wavelength shift ($\Delta v$) and the FWHM of the Gaussian line
broadening, which accounts also for \vmac, while \vsini\ and \vmic\ are fixed
at the values given in Table\,\ref{table_stellarpar}. Additionally, for each
combination of stellar parameters and line list, we used the 1D LTE
spectra assuming different Fe, and CNO abundances. This allowed us to achieve
an even better fit, allowing for possible small abundance deviations from the
adopted values.
We applied a similar fitting procedure to derive the lithium 
abundance and \iso\ isotopic ratio in the Sun (Sect.\,\ref{sec_fit_sun}).

\subsection{Fitting results for HD\,207129 and HD\,95456}
\label{sec_fit_st}
Figure\,\ref{bestfit_G09} shows the best 1D LTE fit to the HD\,207129
(left panels) and HD\,95456 (right panels) HARPS spectra,
obtained with line lists G09, M12, and I14 (from top to bottom). The best fit with line list G09 is achieved for
$\Delta$\,\met\ = $-0.025$ and $\Delta$\,CNO = $-0.05$ for HD\,207129,
resulting in \ali\ of 2.30 and \iso\ of 0.2\% in this star. For HD\,95456, the
best fit with line list G09 is obtained for $\Delta$\,\met\ = $-0.018$ and for $\Delta$\,CNO =
$-0.05$, resulting in \ali\ = 2.65 and \iso\ = 2.8\%. The Gaussian \vmac\ for
which the best fit is achieved is 2.29 and 2.61\,\kms\ for
HD\,207129 and HD\,95456, respectively. For HD~207129, this \vmac\ is in good agreement with
the \vmac\ used for fitting the \ion{Fe}{i} lines given in
Table~\ref{t-feI} (Section\,\ref{sec_rotbr}). For HD~95456, the \vmac\ derived
with this procedure is lower, since we adopted a larger \vsini\ (Table~\ref{table_stellarpar}) than its resulting value from the fitting of the \ion{Fe}{i} lines (Sect.~\ref{sec_rotbr}). The errors given in Fig.\,\ref{bestfit_G09}
are the $1\,\sigma$ formal fitting errors. The 1D LTE best-fit values for \ali\ and \iso\ for all three line lists are
collected in Table\,\ref{table_Lires}. The different line lists result
in somewhat different \iso\ ratios for the same star. We adopt the results
obtained using line list G09 as the representative result of our analysis
since the \iso\ ratio obtained with this set of blends falls between the
values from the analysis with line lists M12 and I14. The uncertainty due to
the different line lists is included as one contribution to the total error 
(see below).

Figure\,\ref{fit_cno_f_G09} shows the best fitting \chisq\ values obtained by fitting the HARPS spectra of HD\,207129 (upper panel) and HD\,95456 (lower panel) with synthetic 1D LTE spectra computed for all the combinations of Fe and CNO abundances based on the preferred line list G09. From this figure, it is clear that the lowest \chisq\ values are obtained around the adopted abundances of Fe and CNO. We note that the best fits shown in Fig.\,\ref{bestfit_G09} are based on the optimal choice of \met\ and CNO derived from Fig.\,\ref{fit_cno_f_G09}.

The 3D NLTE corrections for the lithium abundance and \iso\ ratio for
HD\,207129  and HD\,95456 were obtained using the analytical approximations
developed in this work. The 3D NLTE corrections of \ali\ were computed as a
function of \teff, \met, and \ali, the 3D NLTE corrections of \iso\ as a function of \teff, \logg, \met, \vsini, \ali\ and \iso. We derive 3D NLTE \ali\ corrections of $+0.07$\,dex and
$+0.05$\,dex, and \iso\ corrections of $-1.1$ and $-1.9$ percentage points for HD\,207129 and HD\,95456, respectively. Afterwards, we applied these corrections to the measured 1D LTE best-fitting values of \ali\ and \iso. The derived values after such corrections are also presented in Table\,\ref{table_Lires}. 

We compute the errors of the 1D LTE lithium abundance and
\iso\ ratio by considering seven sources of uncertainty: uncertainties related to the choice of the effective temperature
($\bigtriangleup$\teff=$\pm$50 [K]) and surface gravity
($\bigtriangleup$\logg=$\pm$0.1) as well as the continuum placement (1.0 $\pm$ 0.05\%), the internal fitting error, the fitting
errors related to different Fe ($\Delta$\,\met\ is $\pm$0.1 for HD\,207129 and $\pm$0.05 for HD\,95456) and CNO abundance factors ($\Delta$\,CNO =$\pm$
0.05). These six errors are computed for the line list G09. The seventh error is related to the different lists of atomic and
molecular blends used in this work, and for both lithium abundance and
isotopic ratio is computed as the standard deviation of measured \ali\ and
\iso\, respectively, adopting line lists G09, M12, and I14. These seven
different error contributions are combined in quadrature and are given as the
total error in Table\,\ref{table_Lires}.

Finally, to test the dependence of our results on the used \vsini\ value, we applied a \vsini\ of 0.85\,\kms to the synthetic spectra of HD\,95456, as
measured in Section~\ref{sec_rotbr} using \ion{Fe}{i} lines, and repeated the 
fitting procedure for the best fitting values of Fe and CNO abundance for the 
line list G09. The resulting LTE values of \ali\,=\,2.65 and \qli\,=\,2.7\% 
are in very good agreement with the values derived assuming 
\vsini\,=\,3.28\,\kms, demonstrating that an accurate determination of the 
rotational broadening is not critical.

\subsection{Tests with the \ion{Si}{i} and \ion{V}{i} lines}
\label{sec_si-v}
As mentioned in Section~\ref{sec_linelist}, the measurement of \iso\ isotopic
ratio is sensitive to the list of atomic and molecular lines adopted for the
spectral synthesis. In the case of HD\,95456, we find a difference as high as
$\sim$ 2.2\%, where the lowest \iso\ is 0.8\% when the M12 line list is
adopted, and the highest \iso\ is 3.0\% when I14 is used (see Table\,\ref{table_Lires}). The differences among the line lists arise not only from the adopted atomic data of specific lines (wavelength, \loggf, excitation potential), but also from the different chemical elements present in each particular list of blends.  

The \ion{Si}{i} 670.8025\,nm line is of particular interest since it lies
very close to the \lisix\ feature and has been shown to have an impact on the
\iso\ ratio measurements (e.g., \citealt{israelian2001, israelian2003}). This 
line was first introduced by \cite{muller1975} in their analysis of the solar 
spectrum  and later investigated  in detail by \cite{israelian2003} in a
number of solar-type metal-rich stars. The latter authors have shown that the 
\ion{Si}{I} line at 670.8025 nm is the best candidate for this unidentified 
feature severely blended with the \lisix\ line. We also note that this line  has
been assigned different parameters in the three line lists we use in this work. While the wavelength differences are rather small (670.8023\,nm in G09 and M12, and 670.8025\,nm in I14), the \loggf\ values are quite different ($-2.91$ in G09, $-2.80$ in M12 and $-2.97$ in I14).

In a recent work by \cite{bensby2018}, the authors noted
that new laboratory measurements showed no sign of a
\ion{Si}{i} line at 670.8025\,nm, referring to a private communication
source (Henrik Hartmann). However, as it has been demonstrated
that a ``fictitious'' \ion{Si}{i} line with the assigned atomic parameters
is able to reproduce the unknown blend rather well
\citep{muller1975,israelian2003, mandell2004},
it is still the best choice for a reliable \iso\ analysis
in solar-type stars. By no means should the new measurements be
taken as justification to simply remove the ``fictitious'' \ion{Si}{i}
line from the list of blends, as long as the unknown feature has
not been unambiguously identified.

We tested to what extent the differences in the \ion{Si}{i} line parameters,
particularly in the \loggf\ value, can affect the \iso\ measurement. For this
purpose, we performed the same 1D LTE analysis as described in
Section\,\ref{fitting_harps}, but for different silicon abundance values at
fixed Fe and CNO abundance. For each silicon abundance, the grid of lithium
abundances covers 21 \ali\ times 21 \iso\ values. This test is performed for
HD\,95456 only, where the derived 1D LTE \iso\ ratio appears to be significantly different from zero.

\begin{figure}[htb!]
\centering
\includegraphics[clip=true, width=250pt]{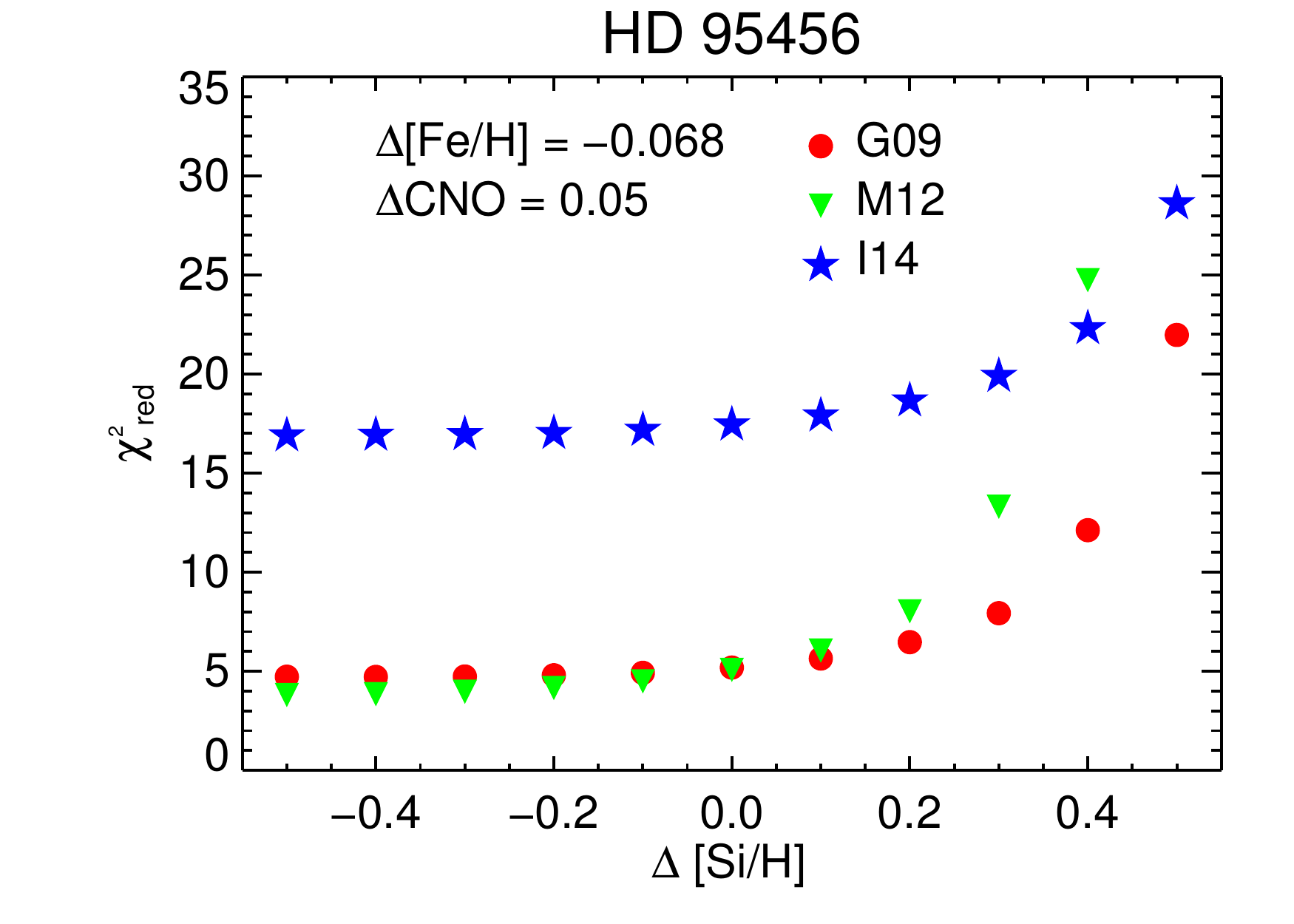}\\
\includegraphics[clip=true, width=250pt]{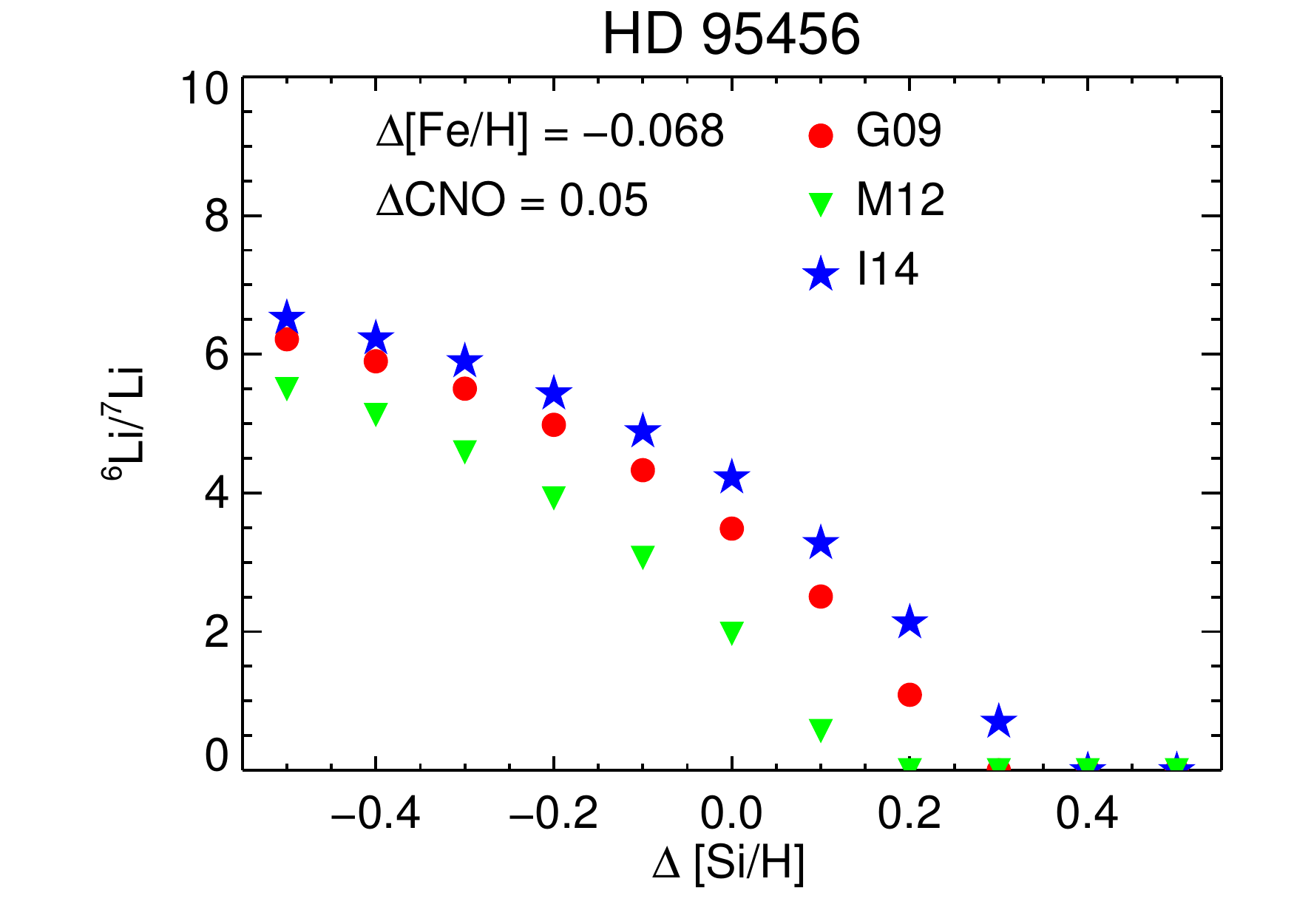}\\
\includegraphics[clip=true, width=250pt]{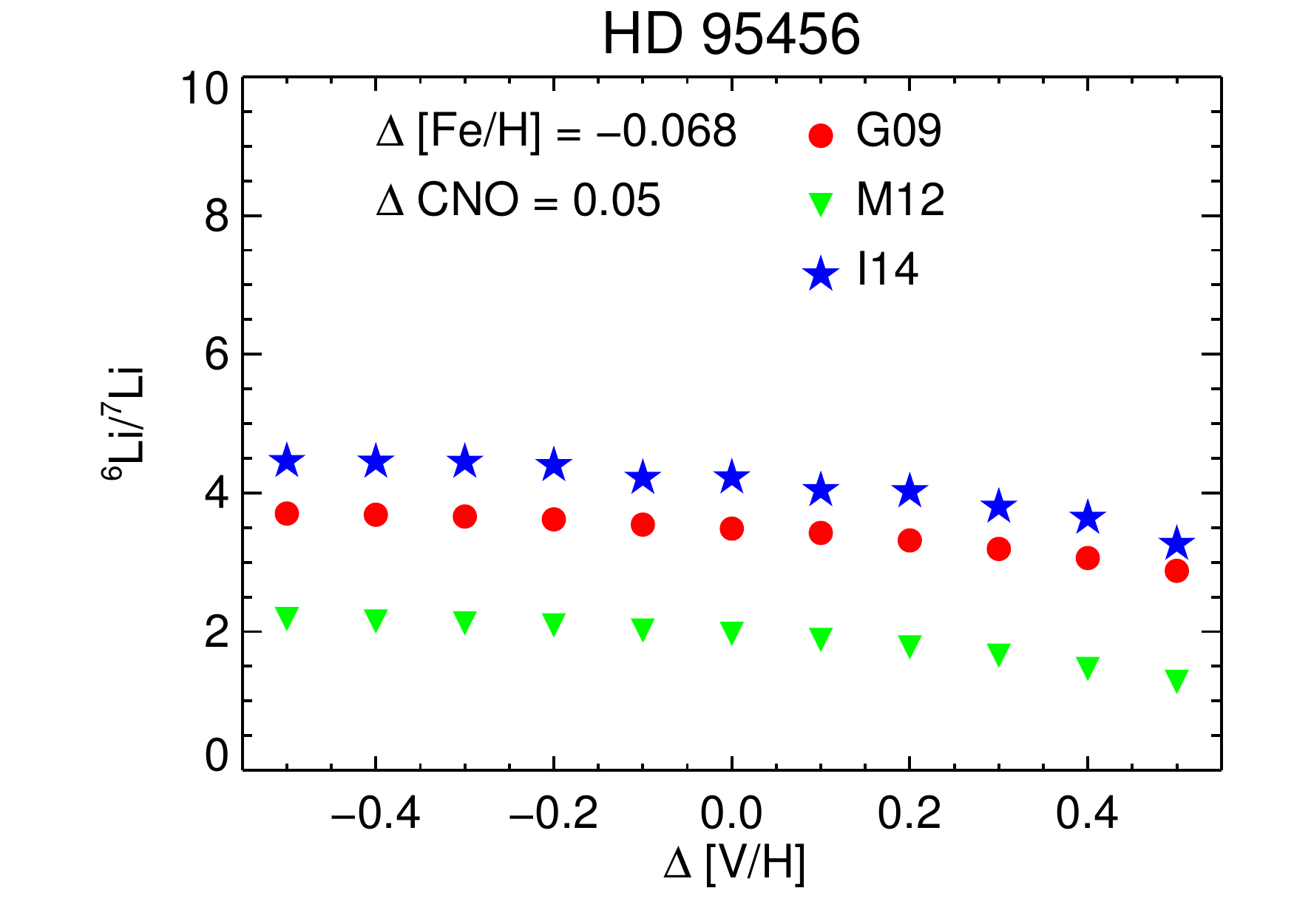}

\caption{The best-fitting \chisq\ (upper panel) and
  \iso\ (middle panel) for different Si abundances, and \iso\ 
for different V abundances (lower panel) for HD\,95456,
using line lists G09 (red circles), M12 (green triangles), and I14
(blue stars), with modifications as described in
Section~\ref{sec_linelist}. For each point, the minimum \chisq\ is obtained by
optimizing \ali\ and \iso\ at given abundances of CNO, Fe, and Si
  (upper and middle panels) or V (lower panel).
  }
\label{fig_Si_V}
\end{figure}

Figure~\ref{fig_Si_V} shows the \chisq\ (upper panel) of the best fit and the resulting
\iso\ ratio (middle panel) plotted versus the deviation (in\,dex) from the adopted silicon
abundance for line lists G09 (red circles), M12 (green triangles), and I14 (blue stars). It is worth noting that, when the silicon
abundance is reduced, the measured \iso\ ratio increases, showing that
changing the silicon abundance by just 0.1\,dex can alter the measured
\iso\ ratio by more than 1 percentage point for this star, while the reduced
\chisq\ value changes only marginally. Furthermore, after increasing the
Si abundance by a certain amount, depending on the line list ($+0.3$, $+0.2$,
and $+0.4$ dex for G09, M12 and I14, respectively), the measured \iso\ ratio
reaches 0\%, indicating that the potential \lisix\ feature is fully
(over-)compensated by the \ion{Si}{i} line. 

This test shows that the atomic data of this particular blend line are very
important for the measurements of the \iso\ ratio in solar-type stars, as
suggested by \cite{israelian2003}, and might be the main factor responsible for the
differences in the results achieved with the different line lists. It is worth
noting that the \loggf\ of this line was calibrated by \cite{israelian2003} in
 high-S/N spectra of several solar-type stars (as well as the Sun) with 
different effective temperatures and metallicities. 
In the alternative case of G09 and M12 line lists, the \loggf\ of this line and of
other blend features have been adjusted  in order to better reproduce the
solar spectrum. 

In fact, if we assume the same \loggf\ for the \ion{Si}{i} line in all the line lists, the results appear to be in very good agreement despite differences in the other adopted blends. This shows that differences in \loggf\ of the \ion{Si}{i} line are indeed the reason for the differences in the \iso\ isotopic ratio. We can assume that this is generally the case, at least for stars with atmospheric parameters similar to HD\,95456.

Because of the recently revised \loggf of the \ion{V}{i} blend, we repeat the
same test as above also for the \ion{V}{i} line. \cite{lawler2014} gave a
\loggf\ higher by 0.3 to 0.5\,dex, dependent on the line list considered. The resulting plot (Fig. ~\ref{fig_Si_V}, lower panel) indicates that this line does not considerably affect the \iso\ measurement,
giving differences of $\Delta q\sim$ 1 percentage point for a 1-dex change in the vanadium
abundance. Therefore, this line is not expected to change the \iso\ result
significantly, at least for stars with stellar parameters similar to HD\,95456.


\subsection{Fitting results for the Sun}
\label{sec_fit_sun}
We applied a similar fitting procedure to derive the lithium 
abundance and \iso\ isotopic ratio in the Sun by fitting the solar flux atlas 
of \cite{kurucz2005}. The 1D LTE synthetic line profiles were computed 
from a solar ATLAS9 model (\teff=5770\,K, \logg=4.44, \vmic=1.24\,\kms, 
$\alpha_{\rm MLT}=1.25$; see \citealt{kurucz1993}) with
MOOG for combinations of 21 \ali\ and 21 \iso\ values, assuming a rotational
broadening of \vsini\,=1.9\,\kms.

\begin{figure}[tb]
\centering
\mbox{\includegraphics[bb=10 0 550 330,clip=true,width=250pt]{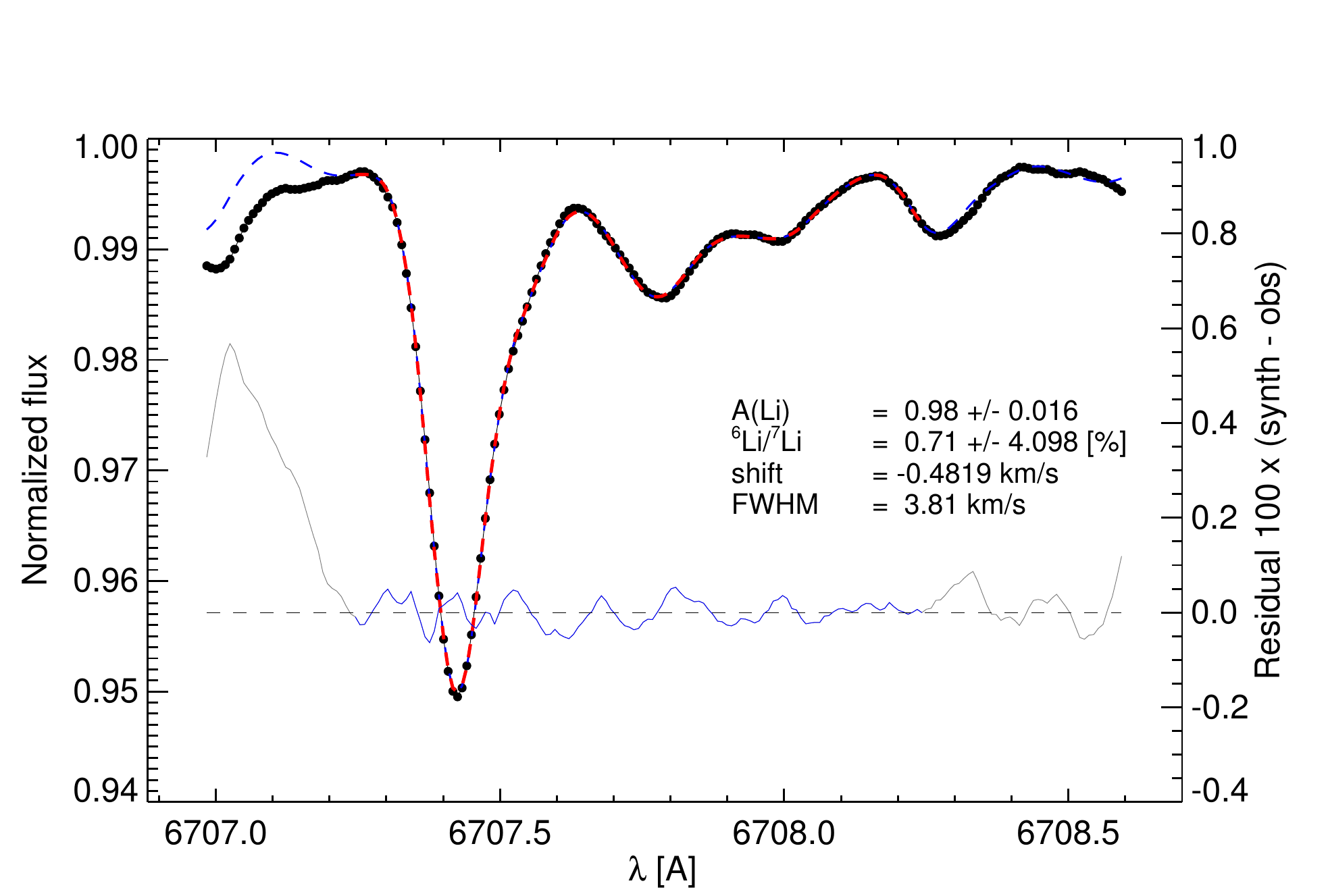}}\\
\caption{The best fitting 1D LTE ATLAS/MOOG spectrum (dashed line) 
superimposed on the solar flux atlas spectrum of \cite{kurucz2005} 
(black dots), fixing the continuum level at $0.9977$ 
(relative to the continuum placement of the flux atlas), and using 
line list G09 (with modifications described in Section~\ref{sec_linelist}) 
and slight adjustments in the strengths of the CN, Fe, Si, and V lines
(see Case C, Table\,\ref{table_Lires}) The right y-axis defines the scale 
of the residuals (thin blue line).}
\label{bestfit_sun_moog}
\end{figure}

Using line list G09 (with modifications described in 
Section~\ref{sec_linelist}) and fixing the continuum level at $c_1=0.9975$ 
(relative to the continuum placement of the solar flux atlas) as in
\cite{ghezzi2009}, the best fit was obtained for \ali\ = $0.99 \pm 0.01$\,dex
and \iso\ = $0.12 \pm 2.0$\% (Case A, Table.~\ref{table_Lires}), where the 
errors are the $1\,\sigma$ formal fitting errors. The derived 
isotopic ratio close to zero reflects the fact that \cite{ghezzi2009}
constructed their line list under the assumption that \iso=0.
An even better fit of the solar Li region can be achieved by
allowing small adjustments in the continuum level and in the abundances 
of Fe and CNO, as described above for the stellar fits. In this case, the best 
fit was obtained for $c_1=0.9977$, \ali\ = $0.98 \pm 0.015$\,dex 
and \iso\ = $1.71 \pm 2.9$\% (Case B, Table.~\ref{table_Lires}). Allowing 
in addition slight adjustments of the Si and V abundances leads to a marginal 
improvement of the fit, but with even larger formal fitting errors:
$c_1=0.9977$, \ali\ = $0.98 \pm 0.02$\,dex, \iso\ = $0.71 \pm 4.1$\% 
(Case C, Fig.~\ref{bestfit_sun_moog}).

For the Sun, our grid of 3D NLTE corrections suggests (after slight 
extrapolation) \delA\,$\sim$\,$+0.1$\,dex and \delq\,$\sim$\,$-0.73\%$,
while the direct calculation of the corrections from 3D NLTE 
and 1D LHD synthetic solar Li line profiles by the method described in 
Sect.\,\ref{sec_fitting3d} yields
\delA\,$\sim$\,$+0.1$\,dex and \delq\,$\sim$\,$-0.83\%$.

After applying the 3D NLTE correction for \ali\ to our 1D LTE best fit
result (Case C), we obtain \ali\,=\,1.08 $\pm$ 0.03, where the error is 
estimated from measurements with different continuum locations (best fit 
location $\pm$ 0.05\%). The \iso\ isotopic ratio obtained from the 
best 1D LTE fit (Fig.~\ref{bestfit_sun_moog}) is
$-0.1$\% after application of the 3D NLTE correction of $-0.83$\%, 
with a large formal fitting error of $\pm 4$\%.

The lithium abundance obtained in this way is in very good agreement with a 
recent 3D NLTE analysis of a very-high-resolution PEPSI spectrum of the 
\ion{Li}{i} $\lambda$\,670.8\,nm region of the Sun by \cite{pepsi2018}
 who measured  \ali\ = 1.09 $\pm$ 0.04. Their estimate of 
\iso\ = 1.4 $\pm$ 1.6\% agrees with our above \iso\ value within the large
error bars.

We point out that \cite{pepsi2018} also employed line list G09, which is
custom-made for 1D modeling and therefore leads to an inferior fit when used
unaltered for 3D spectrum synthesis. This is very likely the reason why the
derived \iso\ ratio is not fully consistent with our corrected 1D result, 
\mbox{\iso\,$\approx 0$}. 
A more consistent fit in 3D can only be achieved if the line list is
adapted to account for 3D effects, in particular adjusting the strength 
and wavelength of the silicon line (see Appendix\,\ref{ap_test_Li_Si}).
We present a detailed comparison of fine-tuned 3D and 1D fits to the solar 
\ion{Li}{i} $\lambda$\,670.8\,nm spectral region in
Appendix\,\ref{ap_bestfit_sun}.

\section{Summary and conclusions}
\label{Sec:conclusion}
We presented 3D NLTE corrections for lithium abundance, \ali, and the
\iso\ ratio, \qli, based on a six-dimensional grid of 3D NLTE spectra.
The stellar parameters defining the grid are effective temperatures, \teff\
(3 different values), gravity, $\log g$ (2), microturbulence velocity, 
$V_{\rm micro}$ (3), metallicity, \met\ (4), lithium abundance, \aliTd\ (3),
and isotopic ratio, \qliTd\ (3) (see Sect.\,\ref{sec_methods} for details). In
addition, four different $v\sin i$ values of rotational broadening were
applied to the 3D NLTE and 1D LTE spectra. For each list of atomic and
molecular blend lines (K95, G09, and M12), this results in a total of 
$2592$ 3D NLTE corrections for both \ali\ and \qli.
The results obtained using the line lists G09 and M12
depend strongly on the wavelength range selected for fitting the \ion{Li}{i}
670.8\,nm spectral region, as well as on whether the continuum placement is a
free or a fixed parameter. As a consequence, the 3D NLTE \iso\ corrections
depend on the atomic and molecular blends used for the spectral line
synthesis. Eventually, we rely only on the 3D NLTE corrections 
based on line list K95 (Li only). This definition is supported
by detailed fits of the solar Li spectral region with both 1D- and 3D-based
synthetic spectra computed with fine-tuned line lists 
(Appendix\,\ref{ap_bestfit_sun}).

The \iso\ ratio derived by fitting a given 3D NLTE spectrum
with the grid of 1D LTE line profiles is found to be always larger than the
input 3D NLTE values by $0.4$ to  $4.9\%$ for the
range of stellar parameters studied in this work. This implies that a
1D LTE spectral analysis leads to an overestimation of the $^6$Li/$^7$Li
isotopic ratio by up to 4.9 percentage points in solar-type stars covered by our grid. 
The 3D NLTE corrections, \delq, are therefore always negative.

The corrections show a systematic dependence on the
stellar parameters. For a given metallicity, generally \delq\ becomes 
larger with
higher \teff\ and lower \logg. For given \teff\ and \logg, they become larger
with higher metallicity, \ali,\ and isotopic ratio \qli. Moreover, we observe a
dependence of \delq\ on the applied rotational broadening: the higher the
\vsini, the smaller the corrections. Furthermore, the quality of the fit
improves assuming higher \vsini\ values applied simultaneously to the 
3D NLTE and 1D LTE spectra, leading to lower values of \chisq\ 
(but also implying larger fitting errors). 
On the other hand, we observe only a very weak variation  of \delq\ with the
small changes ($\pm$0.5 \kms) of the microturbulence parameter, and therefore
adopt the results obtained assuming the central \vmic\ values as
representative corrections. 

Similarly, the abundance corrections, \delA, show a systematic dependence on
the stellar parameters. For a given \met, they generally become larger towards
lower \teff. We observe small variations of \delA\ for different \logg\ values
(4.0 or 4.5) and \ali, whereas the dependence on \iso, \vmic, and \vsini\ is
negligible. The 3D NLTE correction for \ali\ turned out to be very similar for
different lists of the atomic and molecular data used for the spectral line
synthesis.
They are found to range between $-0.01$ and $+0.11$\,dex.

We provide analytical expressions (Eqs.\,\ref{eq_li} and \ref{eq_iso}) which
allow to estimate the 3D NLTE correction for \ali\ and \iso\ as a function of
$T_{\rm eff}$, $\log g$, and \met\ (see Section~\ref{sec_analytic}). These
mathematical functions are valid for the representative values
\ali\,=\,2.0, \qli\,=\,5\%,
and \vsini\,=\,2 \kms\ for a quick evaluation of the 3D NLTE corrections in
the investigated range of stellar parameters. For full details, 
we provide a
(electronic) table with the complete grid of the 3D NLTE corrections, and a
link to a web page that allows the user to compute the 3D NLTE corrections of
\ali\ as a function of \teff\, \met, and \ali, and the 3D NLTE 
corrections of \iso\ as a function of \teff, \logg, \met, \vsini, \ali\ and 
\iso.

Our analytical expressions, valid for solar-type stars, allow to account for
3D plus NLTE effects without the need of direct access to complex 3D NLTE
computations. This is particularly important when a large sample of stars
needs to be analyzed in terms of lithium abundance and isotopic ratio, since
3D NLTE analysis for even a single target is computationally demanding and
time consuming. The analysis of the observed spectra can be carried out at
first by using standard 1D LTE spectrum synthesis techniques. Afterwards, the
1D results can be corrected for the 3D NLTE effects by applying the
precomputed 3D NLTE corrections interpolated to the desired set of stellar
parameters.

In the second part of this work, we use high-quality HARPS spectra of two
solar-type stars in order to derive the \ali\ and \iso\ ratio in their
atmospheres. 
The stars were selected because they are located at opposite
sides of the investigated  \teff\ range where the corrections show the 
largest differences.
The lithium doublet in these two stars was first analyzed with
a standard 1D LTE approach, and subsequently corrected for 3D NLTE effects using
the pre-computed 3D NLTE corrections. After applying the 3D NLTE corrections, 
we obtain \ali\,=\,$2.37\pm 0.04$ and \iso\,=$-0.9\%\pm 0.7\%$ for
HD\,207129, whereas we derive \ali\,=\,$2.70\pm 0.04$ and 
\iso\,=\,$0.9\%\pm 1.4\%$ for HD\,95456. 
The correction for the \iso\ ratio is estimated to be \delq\,=\,$-1.1\%$ for 
HD\,207129 and \delq\,=\,$-1.9\%$ for HD\,95456.

In the case of HD\,95456, the $2\,\sigma$ detection of \lisix\ in the 
1D LTE analysis is turned into a clear non-detection after applying the 
3D NLTE correction. In conclusion, we do not find significant amounts of 
\lisix\ in either of the two stars. 

Additionally, we studied the impact of two weak absorption lines (\ion{Si}{i}, \ion{V}{i}) on our \iso\ ratio measurements in HD 95456, and
concluded that the \ion{Si}{i} 670.8025~nm feature is the most critical blend for this analysis, confirming the result of \cite{israelian2003}.

Finally, we have demonstrated that a list of blend lines that are adjusted 
(like G09) by fine-tuning \loggf-values and wavelength positions to yield 
a perfect fit to the solar Li spectral region with a particular 1D model 
atmosphere cannot be expected to produce an equally good fit when 
employed unaltered for 3D modeling. For this reason, fitting 3D model spectra 
directly to observations may not always be the most desirable approach.

\begin{acknowledgements} This work has made use of the VALD database, operated at Uppsala University, the Institute of Astronomy RAS in Moscow, and the University of Vienna. We thank the Leibniz-Association for supporting G.H. and A.M. through a SAW graduate school grant. JIGH acknowledges financial support from the Spanish ministry project MINECO AYA2014- 56359-P and from the Spanish Ministry of Economy and Competitiveness (MINECO) under the 2013 Ram\'on y Cajal program MINECO RYC-2013-14875. We thank J{\'a}nos Bartus for creating the web page calculator. Finally, we thank the anonymous referee for 
his/her critical questions which helped us to improve the presentation of our results.

\end{acknowledgements}

\bibliographystyle{aa}             
\bibliography{bibliography}

\begin{appendix}

\section{3D NLTE corrections for \ali\ and \iso}
\label{appendix:A}

\begin{figure}[htb!]
\centering
\includegraphics[clip=true,width=250pt]{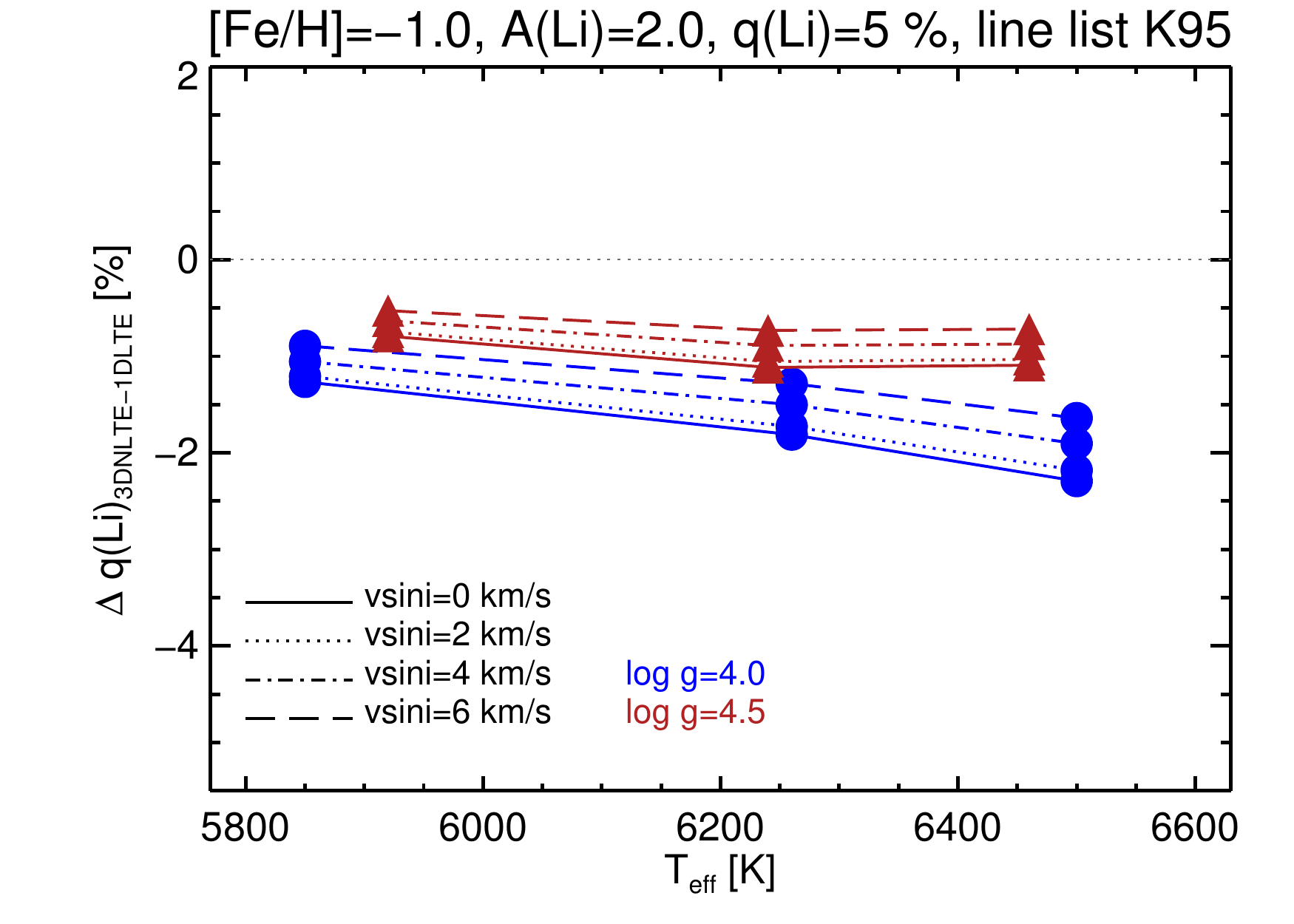}\\
\includegraphics[clip=true,width=250pt]{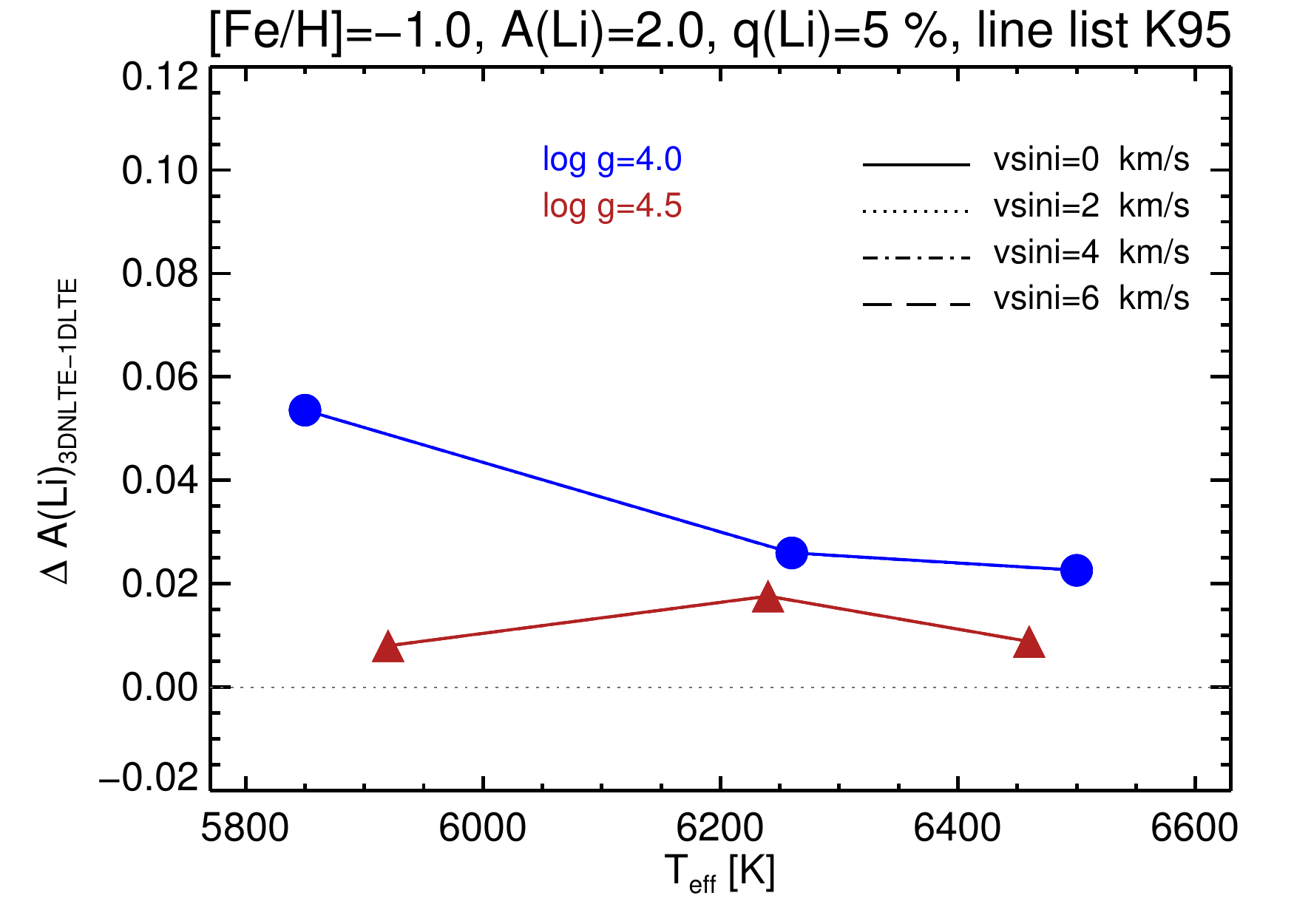}
\caption{3D NLTE corrections for \iso\ (upper panel) and \ali\ (lower panel) 
  versus $T_{\rm eff}$ for \met\,=\,-1.0, \ali=2.0, \qli=5\%, derived
  with line list K95. The blue circles and the red triangles correspond to $\log g$=4.0 and $\log g$=4.5, respectively. The corrections for different $v\sin i$ values are connected with lines of different styles.
}
\label{a1}
\end{figure}

\begin{figure}[htb!]
\centering
\includegraphics[clip=true,width=250pt]{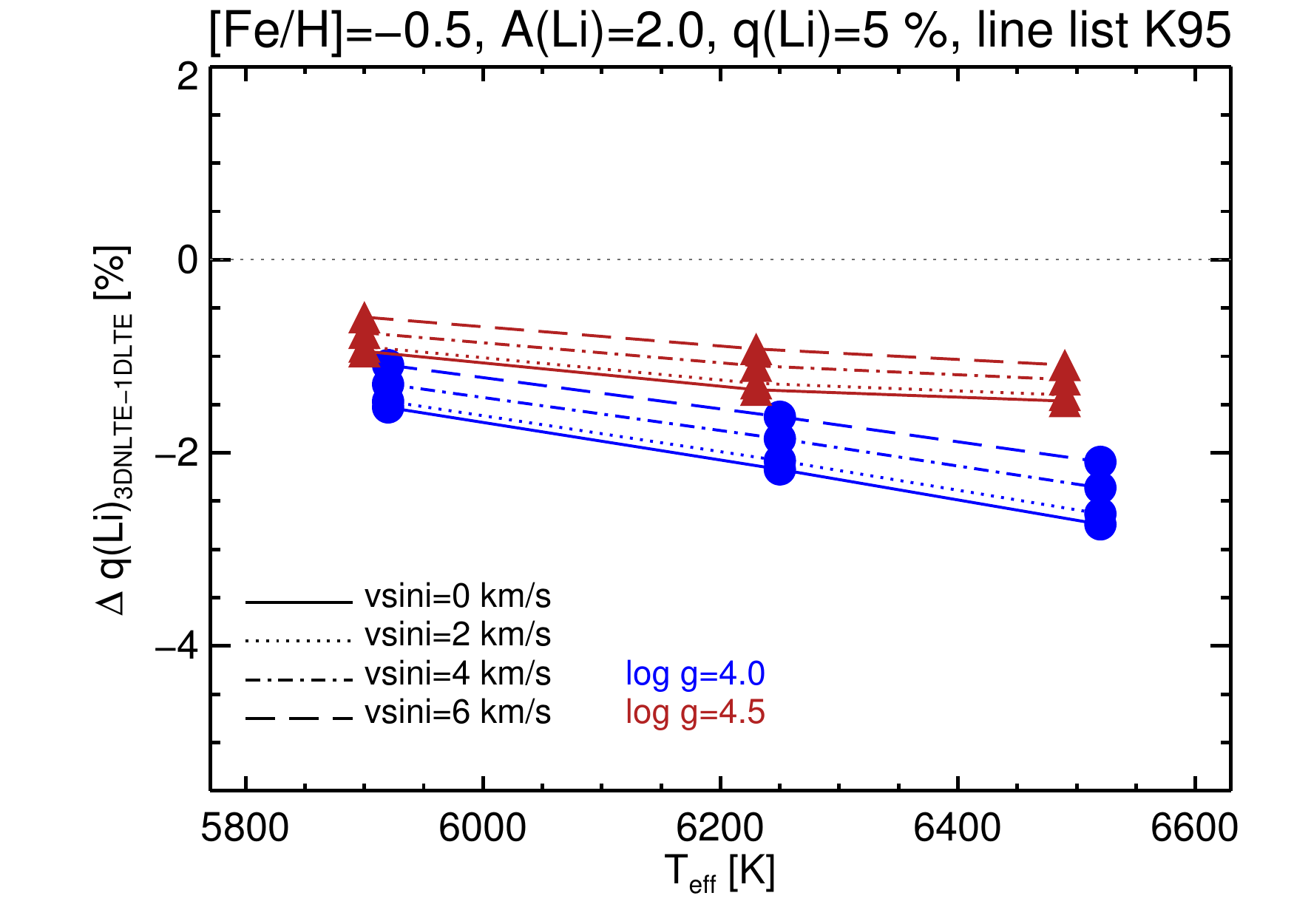}\\
\includegraphics[clip=true,width=250pt]{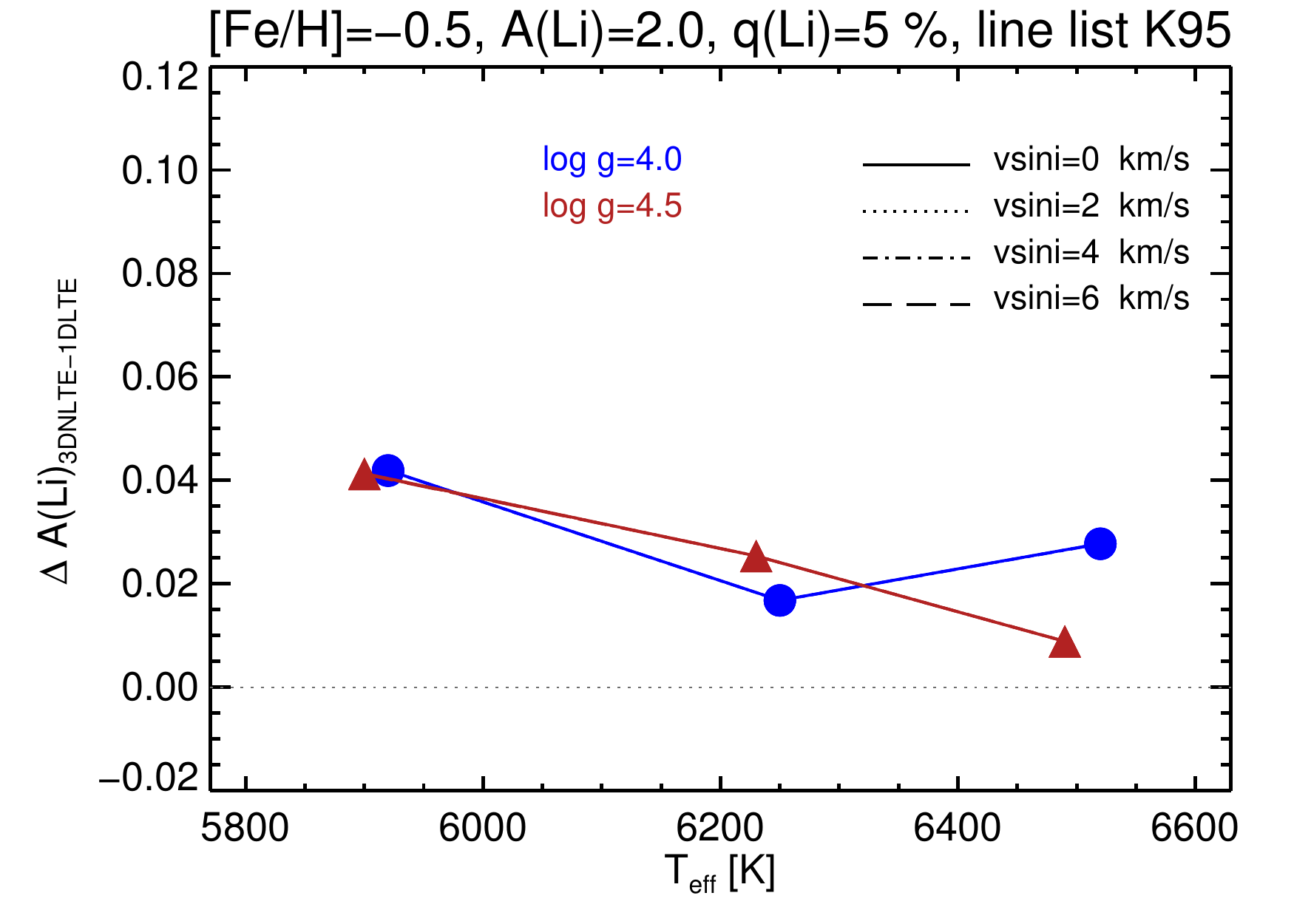}
\caption{As Fig.\,\ref{a1} but for \met\,=\,$-0.5$.
}
\label{a2}
\end{figure}

\begin{figure}[htb!]
\centering
\includegraphics[clip=true,width=250pt]{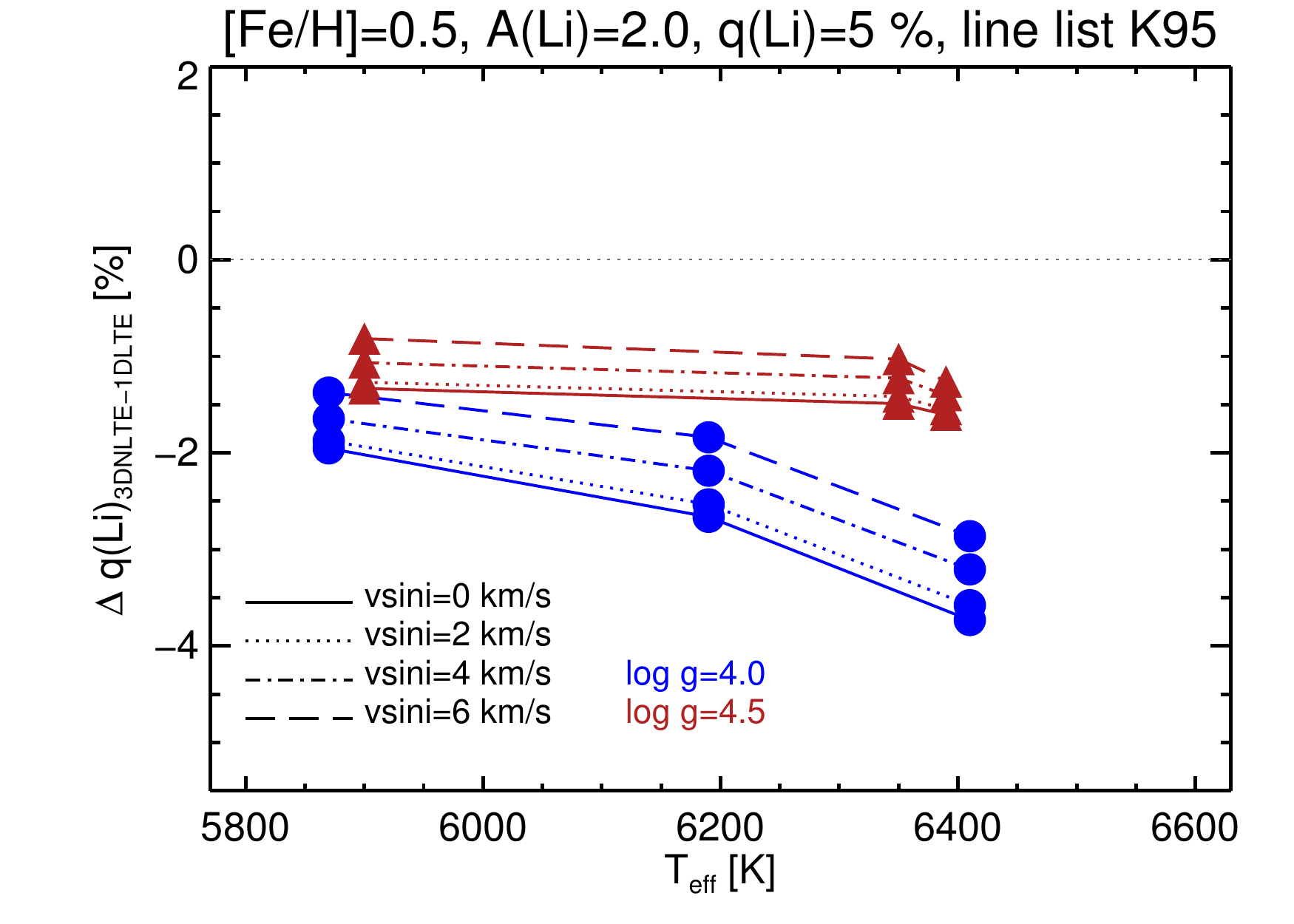}\\
\includegraphics[clip=true,width=250pt]{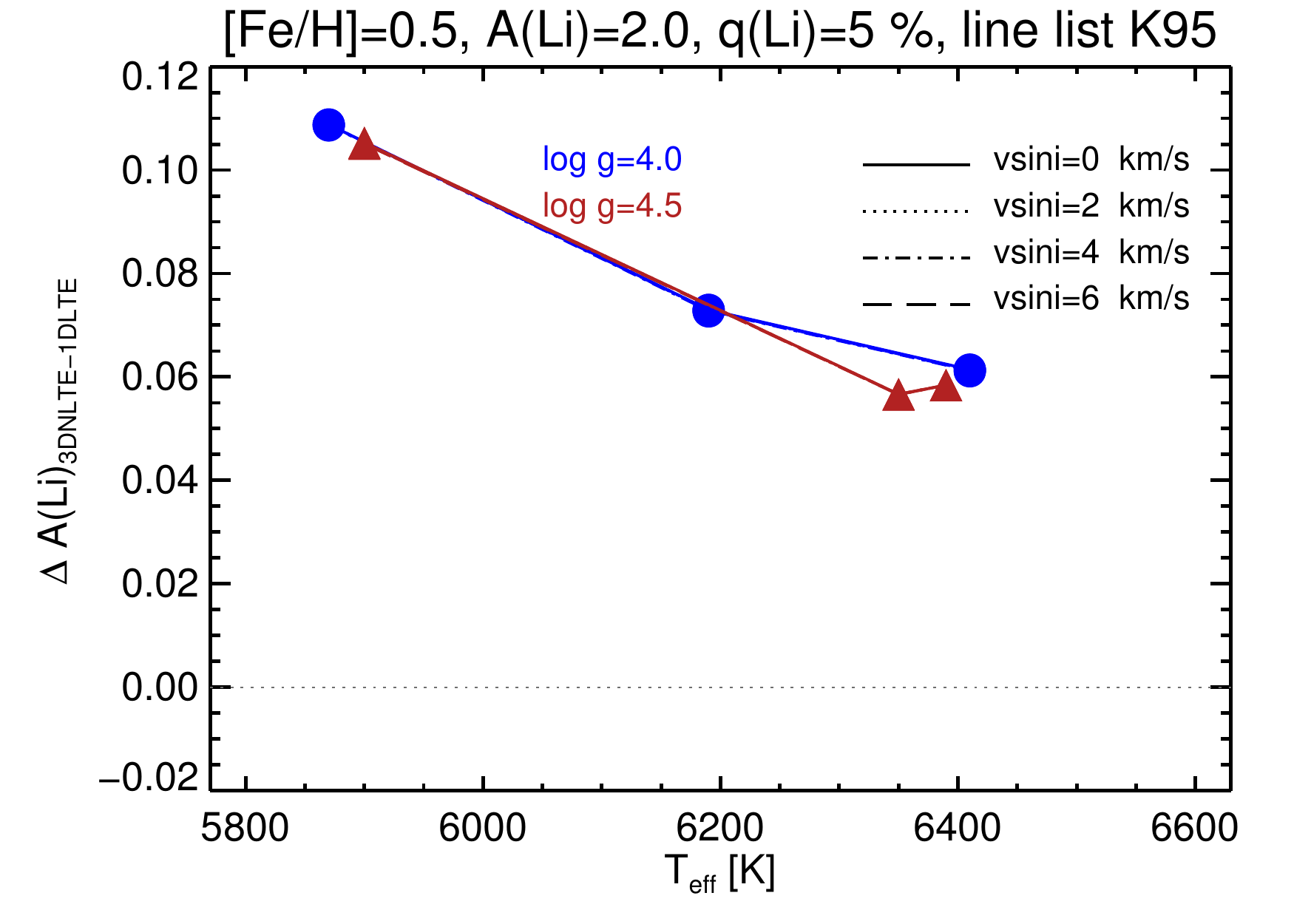}
\caption{As Fig.\,\ref{a1} but for \met\,=\,$+0.5$.
}
\label{a3}
\end{figure}

\begin{figure}[htb!]
\centering
\includegraphics[clip=true,width=250pt]{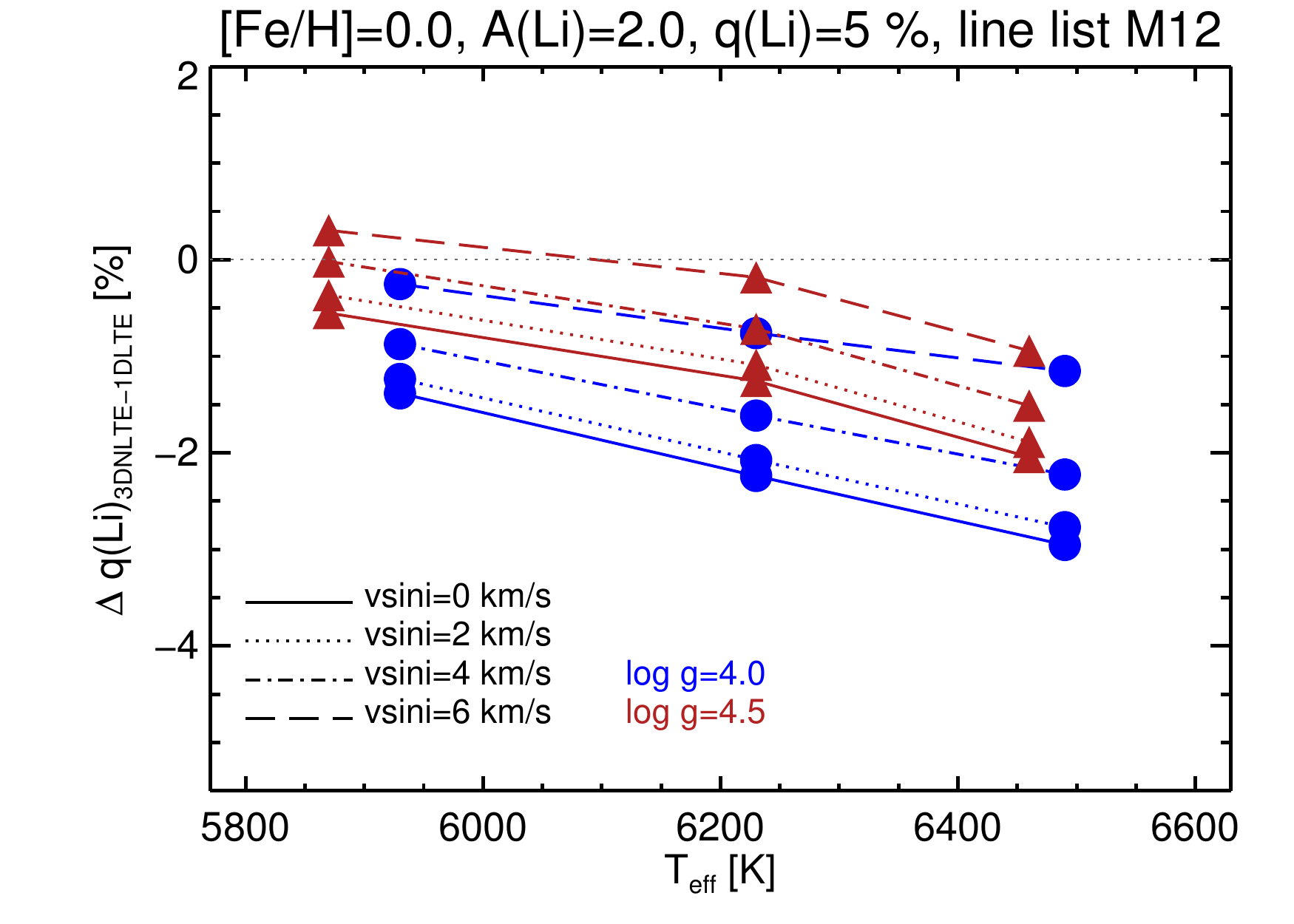}\\
\includegraphics[clip=true,width=250pt]{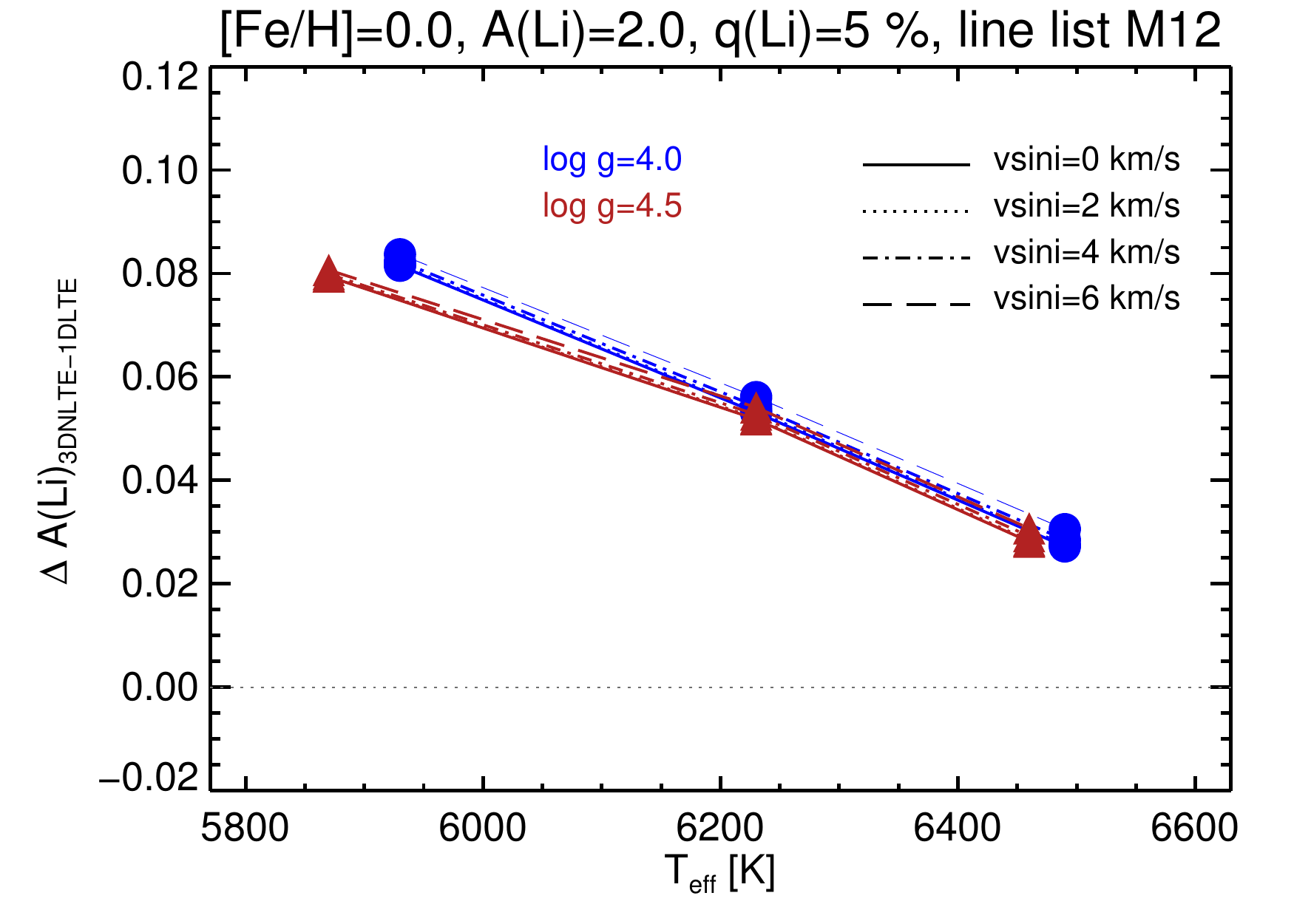}
\caption{Same corrections as shown in Figs.\,\ref{a1}--\ref{a3}, but 
derived with line list M12 for \met\,=\,$0.0$. 
}
\label{a4}
\end{figure}

Figures \ref{a1}--\ref{a3} present the plots of 3D NLTE corrections 
\delq\ (upper panels) and \delA\ (lower panels) derived with line list K95
versus the effective temperature for three metallicities of our 3D NLTE
spectral grid (\met\,=\,$-1.0$, $-0.5$ and $0.5$, respectively). 
Figure \ref{a4} shows the same 3D NLTE corrections derived with line 
list M12 for  metallicity \met\,=\,$0.0$.

\clearpage

\section{Test case solar \ion{Li}{i} + \ion{Si}{i} blend}
\label{ap_test_Li_Si}

This numerical experiment is to demonstrate that the corrections
\delA\ and \delq\ derived from fitting a 3D NLTE Li line profile
with overlapping blend lines by a grid of corresponding 1D LTE spectra are
very similar to the corrections obtained from fitting the pure Li
3D NLTE line profile with a grid of 1D LTE pure Li spectra, provided
that the blend lines in the 1D spectrum synthesis are adjusted to correct
for 3D effects.

We model the Li doublet in its full complexity, including all hyperfine
structure levels of both \lisix\ and \liseven. The blends are represented 
by a single silicon line (\ion{Si}{i}\,$\lambda\, 670.8025$\, nm) which 
is known to have a critical influence on the derived \iso\ ratio (see
Sect.\,\ref{sec_si-v}). The Sun is a good test case since the solar Li
abundance (\ali\,=\,1.1) is at least ten times lower than in our target 
stars (\ali\,$\ga$\,2.3) and hence the \ion{Si}{i} line has an even larger 
impact on the Li profile.

As a first step, we fit the pure \liseven\ 3D NLTE line profile computed
with \nlte\ + \linfor\ from a 3D solar \cobold\ model with a grid of
1D LTE pure \lisix+\liseven\ spectra based on a solar \mlhd\ model. The fitting 
parameters are (i) the Li abundance \ali, (ii) the \iso\ isotopic ratio \qli, 
(iii) the width of a global Gaussian line broadening kernel \textit{FWHM}, 
and (iv) a global Doppler shift $\Delta v$. The same procedure 
is repeated with the single \ion{Si}{i} line, which is modeled in LTE both in 3D
and in 1D, assuming the same atomic line parameters as in line list G09.  
Here the fitting parameters are $A$(Si), \textit{FWHM}, 
and $\Delta v$. The best fit results are summarized in Table\,\ref{t-3D-1D}
(Fit 0 and Fit 1).
We notice that the 3D LTE abundance correction for the Si line is
$\Delta A$\,(Si)\,=\,$+0.0419$\,dex, the absolute convective blue shift is
$\Delta v$\,=\,$-0.399$\,\kms, and the shift relative to Li is 
$\delta v$\,=\,$-0.399 + 0.188 = -0.211$\,\kms.

\begin{figure}[htb!]
\centering
\includegraphics[bb=2 0 555 330,clip=true,width=250pt]{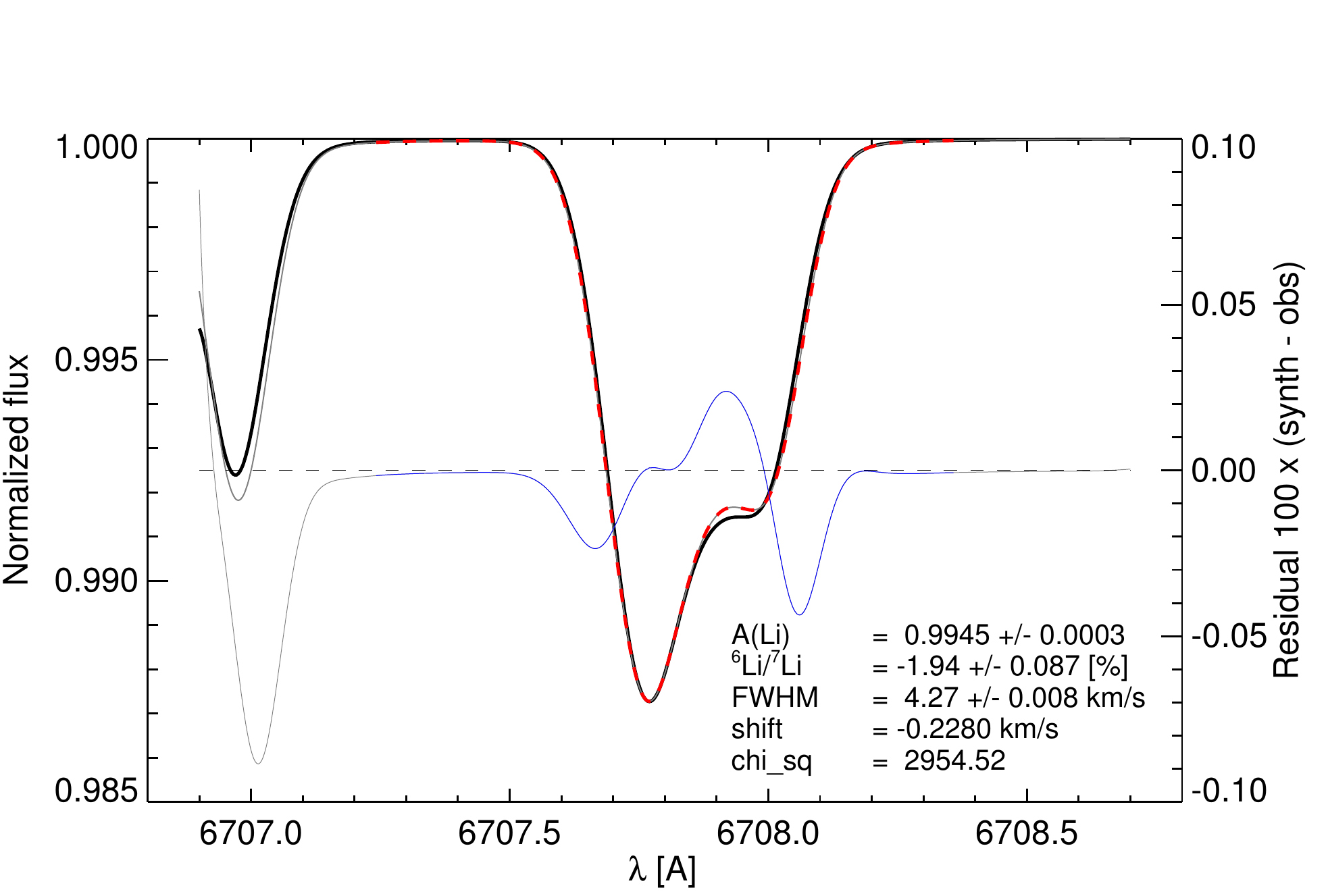}\\
\includegraphics[bb=2 0 555 330,clip=true,width=250pt]{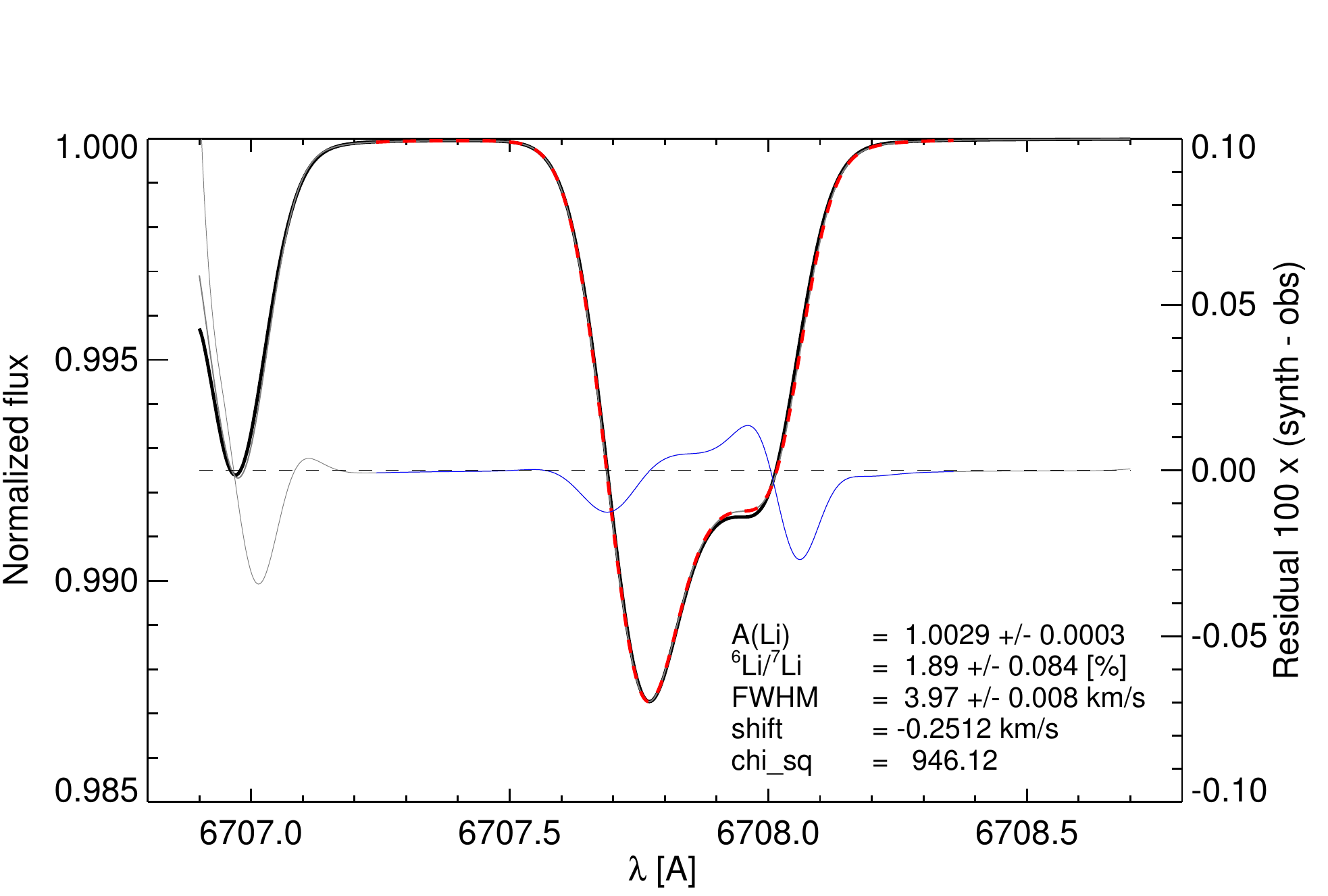}\\
\includegraphics[bb=2 0 555 330,clip=true,width=250pt]{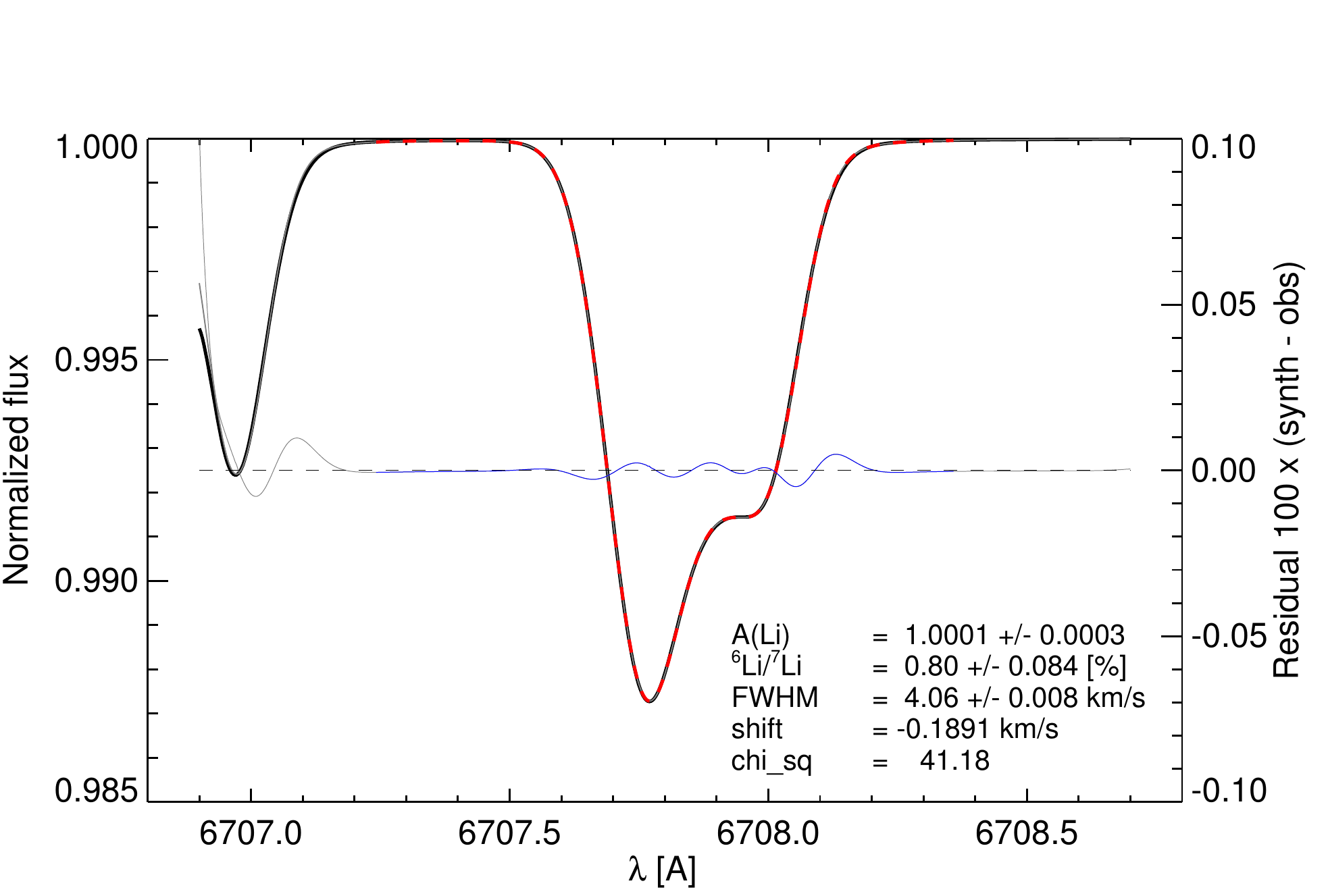}\\

\caption{Best 1D fits (dashed line) to the 3D composite \liseven\ (NLTE) 
+ Si (LTE) synthetic solar line profile (black solid line) achieved with 
different grids of 1D LTE \lisix+\liseven+Si profiles. From top to bottom, 
the panels show cases 2a, 2b, and 2c (see text and Table\,\ref{t-3D-1D} for 
details). The right y-axis defines the scale of the residuals (thin blue line).}
\label{bestfits_sun_LiSi}
\end{figure}

In a second step, we fit the 3D composite spectrum of Li (NLTE) plus
Si (LTE) with a grid of 1D LTE composite spectra covering a range in
\ali\ and \iso, but assuming the same fixed silicon abundance as in
the 3D spectrum synthesis ($A$(Si)=$7.55$). Under these constraints, 
the best fit is found for an isotopic ratio of \iso\,=\,$-1.9$\%.
However, this fit (Fit 2a) is rather poor (see top panel of 
Fig.\,\ref{bestfits_sun_LiSi} and Table\,\ref{t-3D-1D}). 

The fit can be improved if we reduce the silicon abundance by 
$\Delta A$(Si)\,=\,$-0.0419$\,dex, thus accounting for the 3D abundance 
corrections indicated by Fit 1. The resulting fit (Fit 2b) is significantly
better and indicates \iso\,=\,$+1.9$\%. Still, the fit is clearly 
unsatisfactory (see middle  panel of Fig.\,\ref{bestfits_sun_LiSi}).

A much better fit can be obtained if we allow in addition for a wavelength
shift of the Si line of $\delta v$\,=\,$-0.211$\,\kms\ in the 1D spectrum
synthesis to compensate for the convective blue shift of Si relative to Li. 
The solution for this setup (Fit 2c, shown in the lower panel of
Fig.\,\ref{bestfits_sun_LiSi}) is very acceptable. The derived \iso\
ratio is now essentially the same as the one indicated by the analysis of
the pure Li case (cf.\ Fit 0 and Fit 2c in Table\,\ref{t-3D-1D}).

The conclusion of this numerical experiment is that a list of blend lines,
adjusted (like G09) by fine-tuning \loggf-values and wavelength positions to
yield a perfect fit to the solar Li spectral region with a particular 1D model
atmosphere, cannot be expected to produce an equally good fit when employed
unaltered for 3D modeling. It is therefore not surprising that the 3D analysis
of the solar Li region with unmodified line list G09 produces a fit of
somewhat inferior quality compared to the original 1D analysis by
\cite{ghezzi2009}, and leads to somewhat different results for \iso.
Consistent results in 3D can only be achieved if line list G09 is adjusted for
3D effects. In this case, we expect that
\iso\,(3D)\,$\approx$\,\iso\,(1D)$-0.8$\%
(cf.\ Appendix\,\ref{ap_bestfit_sun}).

\begin{table*}[htb!]
\caption{Summary of 3D - 1D fitting results.}

\begin{tabular}{l l c  c  |  c  c  c  c  c  c  c}
\hline
\hline
\noalign{\smallskip}

Fit ID & Model & Line       & Spectrum & \ali  & \qli & $A$(Si) & $\Delta v$\,(Si) & \textit{FWHM} & $\Delta v$ & $\chi^2$ \\
       &       & formation  &          & (dex) & (\%) & (dex)   & (\kms)           & (\kms)         & (\kms)       \\

\noalign{\smallskip}
\hline 
\noalign{\smallskip}
  & 3D  & non-LTE & \lisix+\liseven   & $1.1000$ & $0.00$ & $0.0$ & -- & $0.500$ & $\;\;0.0000$ & -- \\
0 & 1D \mlhd  & LTE & \lisix+\liseven & $1.0003$ & $0.83$ & $0.0$ & -- & $4.031$ & $-0.1877$ & $17.2$ \\ 
\hline \noalign{\smallskip}
  & 3D        & LTE & Si & -- & --  & $7.5500$ & -- & $0.500$ &  $\;\;0.0000$ & -- \\
1 & 1D \mlhd  & LTE & Si & -- & --  & $7.5081$ & -- & $4.085$ &  $-0.3988$ & $11.9$ \\ 
\hline\hline \noalign{\smallskip}
   & 3D  & mixed & \lisix+\liseven+Si     & $1.1000$ & $0.00$ &  $7.5500$ & $0.00$ & $0.500$ & $\;\;0.0000$ & --\\
2a & 1D \mlhd  & LTE & \lisix+\liseven+Si & $0.9945$ & $-1.94$ & $\;7.5500^\ast$ & $\;0.00^\ast$ & $4.268$ & $-0.2280$ & $2954.5$ \\ 
\hline \noalign{\smallskip}
   & 3D  & mixed & \lisix+\liseven+Si     & $1.1000$ & $0.00$ & $7.5500$ & $0.00$ & $0.500$ & $\;\;0.0000$ & -- \\
2b & 1D \mlhd  & LTE & \lisix+\liseven+Si & $1.0029$ & $1.89$ & $\;7.5081^\ast$ & $\;0.00^\ast$ & $3.971$ & $-0.2512$ & $946.1$ \\ 
\hline \noalign{\smallskip}
   & 3D  & mixed & \lisix+\liseven+Si     & $1.1000$ & $0.00$ & $7.5500$ & $ 0.00$ & $0.500$ & $\;\;0.0000$ & -- \\
2c & 1D \mlhd  & LTE & \lisix+\liseven+Si & $1.0001$ & $0.80$ & $\;7.5081^\ast$ & $-0.211^\ast$ & $4.058$ & $-0.1891$ & $41.18$ \\ 
\hline \noalign{\smallskip}
\hline
\end{tabular}
\tablefoot{Rows for 3D model specify the input parameters for the 3D spectrum 
synthesis; rows for 1D model give the results of the best fit (right part of 
the Table), except numbers marked by an asterisk which are fixed input 
parameters. $\Delta v$\,(Si) (column 8) gives the artificial wavelength shift 
(converted to velocity) of the \ion{Si}{i} line introduced in the line list 
for the 1D LTE spectrum synthesis. The relative quality of the fits can be
judged by comparing their $\chi^2$ (given on an arbitrary absolute scale).}
\label{t-3D-1D}
\end{table*}

\section{Fitting the solar \ion{Li}{i} $\lambda\,670.8$\,nm region
with 1D\,LHD and 3D\,\cobold\ synthetic spectra}
\label{ap_bestfit_sun}

In this Appendix, we derive the solar lithium abundance and \iso\ ratio
by fitting the \ion{Li}{i} $\lambda\,670.8$\,nm spectral region of solar 
flux spectrum \citep{kurucz2005} with synthetic spectra 
computed from a solar 1D LHD model (\teff=5780\,K, \logg=4.44, 
\vmic=1.0\,\kms, $\alpha_{\rm MLT}$\,=\,$1.0$) and a 3D \cobold\ model
atmosphere, respectively. The purpose of this investigation is to test
the expected relation between the 1D and the 3D results.

In both cases, the spectrum synthesis is 
performed with \linfor, applying a fixed rotational broadening of 
\vsini\,=\,$1.8$\,\kms in the flux convolution approximation. 
As before, the Li doublet includes all hyperfine structure levels of both
\lisix\ and \liseven\ and is treated in NLTE (LTE) when computing 
the  grid of spectra from the 3D (1D LHD) models.

The grid of 1D LHD spectra is computed with line list G09.
The fitting of the solar flux spectrum with this grid follows the same 
methodology as the fitting with the ATLAS/MOOG spectra described in 
Sect.\,\ref{sec_fit_sun}. For a fixed continuum level, the basic fitting 
parameters are (i) the Li abundance \ali, (ii) the \iso\ isotopic ratio 
\qli, (iii) the width of a global Gaussian line broadening kernel 
\textit{FWHM}, and (iv) a global Doppler shift  $\Delta v$. In addition,
we allow for small adjustments of the strengths (abundances or \loggf\ values)
of the CN, \ion{Fe}{i}, \ion{Si}{i}, and  \ion{V}{i} lines to further 
optimize the fit.

\begin{figure}[tb]
\centering
\mbox{\includegraphics[bb=10 0 550 330,clip=true,width=250pt]{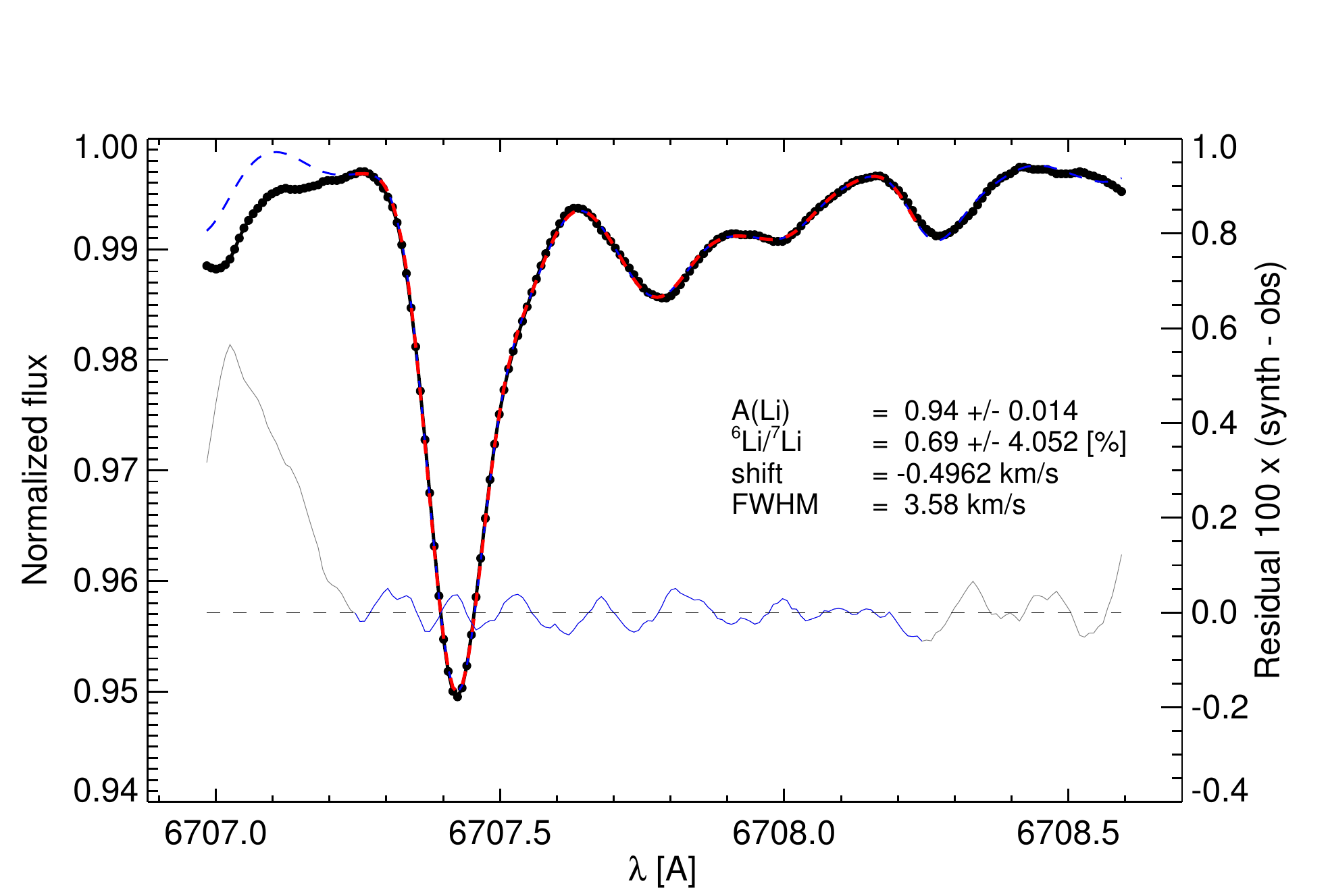}}
\caption{The best fitting 1D LTE LHD spectrum (red dashed line) 
superimposed on the solar flux spectrum of \cite{kurucz2005} 
(black dots), fixing the continuum level at $0.9977$ 
(relative to the continuum placement of the flux atlas), and using 
line list G09 with slight adjustments in the strengths of the CN, Fe, 
Si, and V  lines (\textit{Case~I}, Table\,\ref{t-solarfits}) 
The right y-axis defines the scale of the residuals (thin blue line).}
\label{bestfit_sun_lhd1}
\end{figure}

In \textit{Case~I}, we fix the continuum at the same level as for the best 
fit with the ATLAS/MOOG spectra (Case\,C, Fig.\,\ref{bestfit_sun_moog}),
$c_1=0.9977$ (relative to the continuum placement of the solar flux atlas). 
The best fit with our grid of 1D LHD synthetic spectra  if found for
\ali\,=\,$0.94$ and \qli\,=\,$0.69\%$ (see Fig.\,\ref{bestfit_sun_lhd1}). 
These results can be compared to those of Case~C, \ali\,=\,$0.98$ and 
\qli\,=\,$0.71\%$. While the derived \iso\ ratio is practically identical, the 
lithium abundance indicated by the LHD fit is slightly lower. This must be 
related to slight differences in the temperature stratification of 
the solar LHD model relative to the ATLAS9 model.

Next we fit the observed spectrum with a grid of synthetic spectra
derived from the \cobold\ 3D solar atmosphere, now treating the Li 
doublet in NLTE. Adopting the same continuum level as for the 1D 
fitting, we try out different assumptions about the chemical
abundances and the wavelengths of the blend lines. 

\begin{figure}[tb]
\centering
\mbox{\includegraphics[bb=10 0 550 330,clip=true,width=250pt]{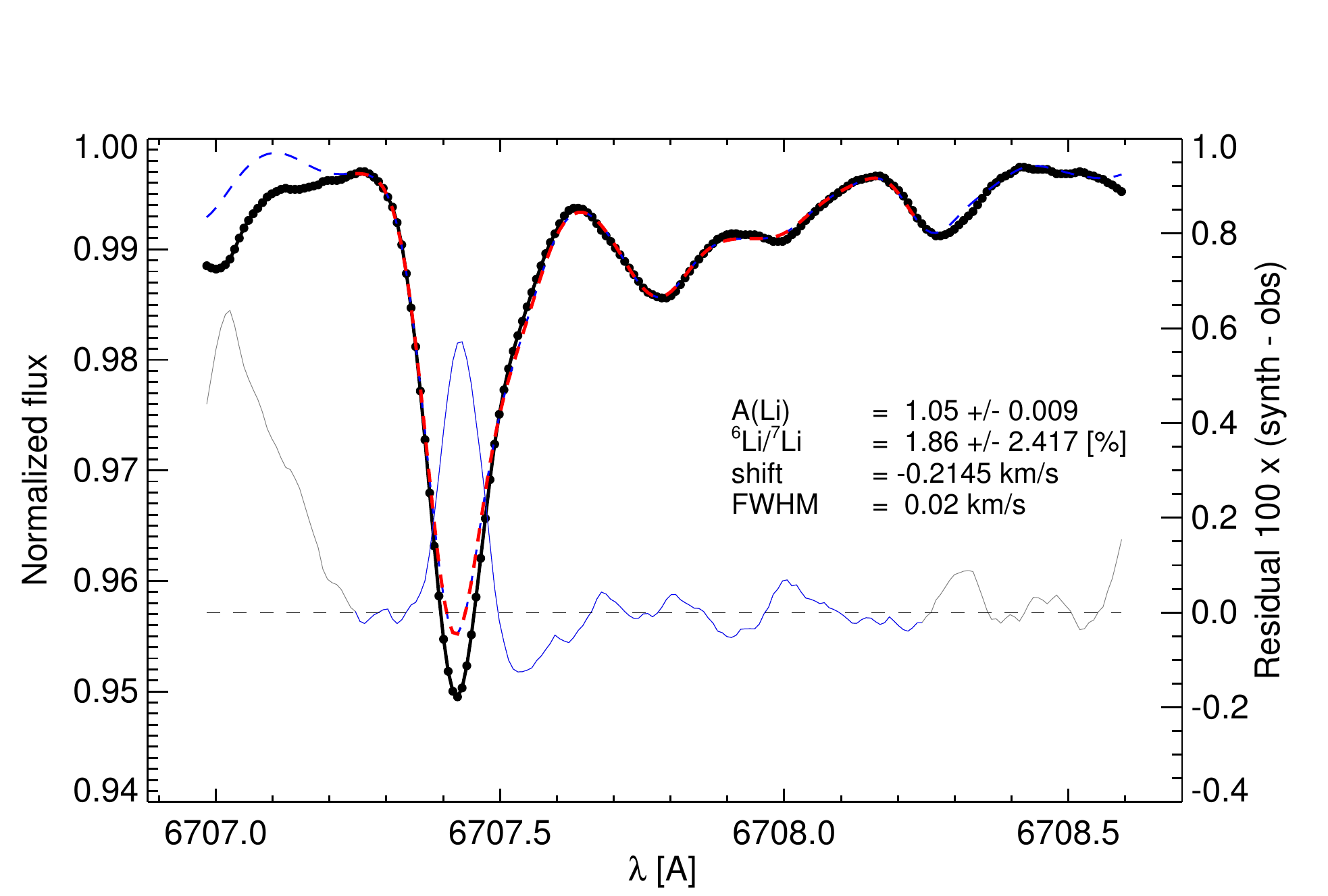}}
\mbox{\includegraphics[bb=10 0 550 330,clip=true,width=250pt]{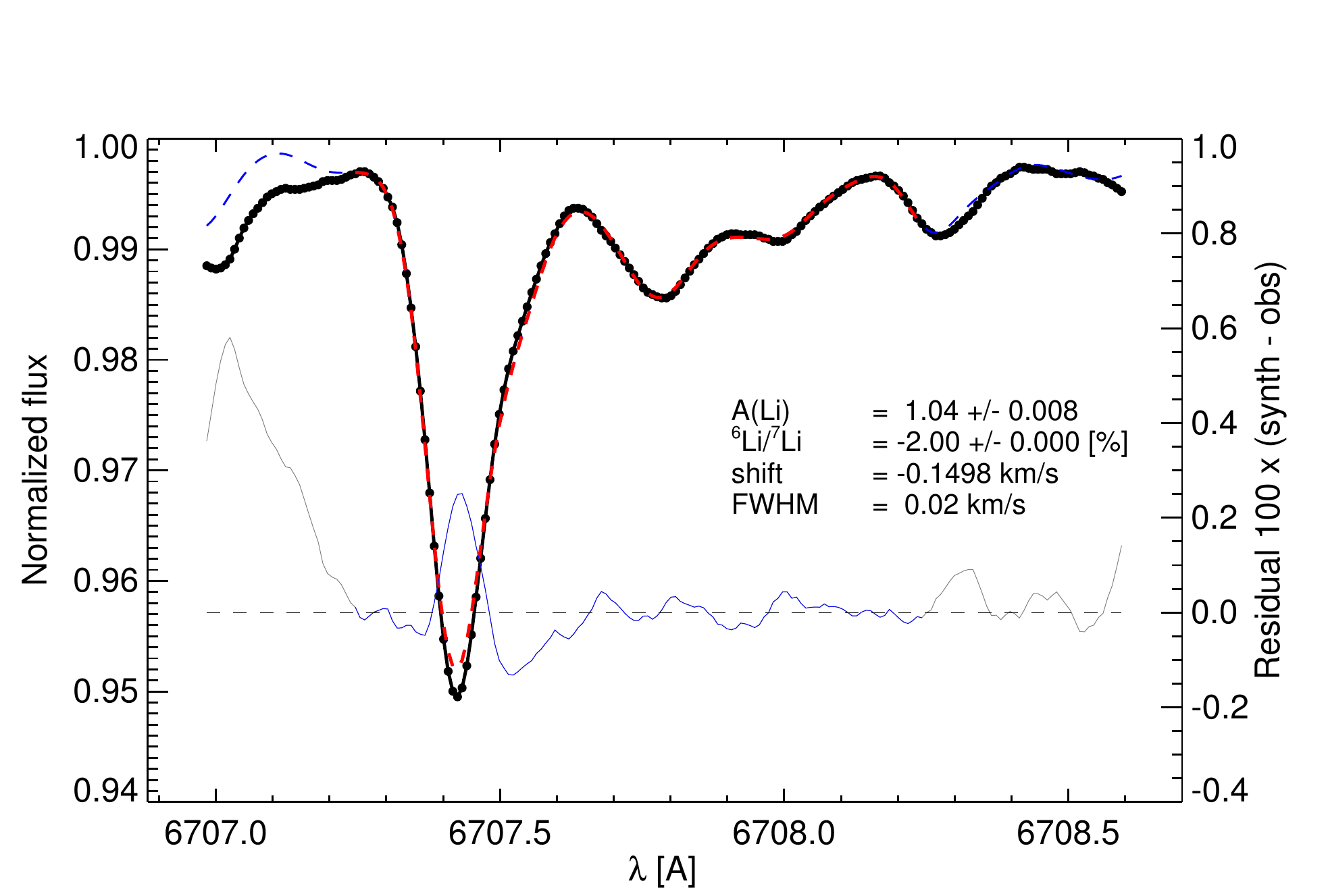}}
\mbox{\includegraphics[bb=10 0 550 330,clip=true,width=250pt]{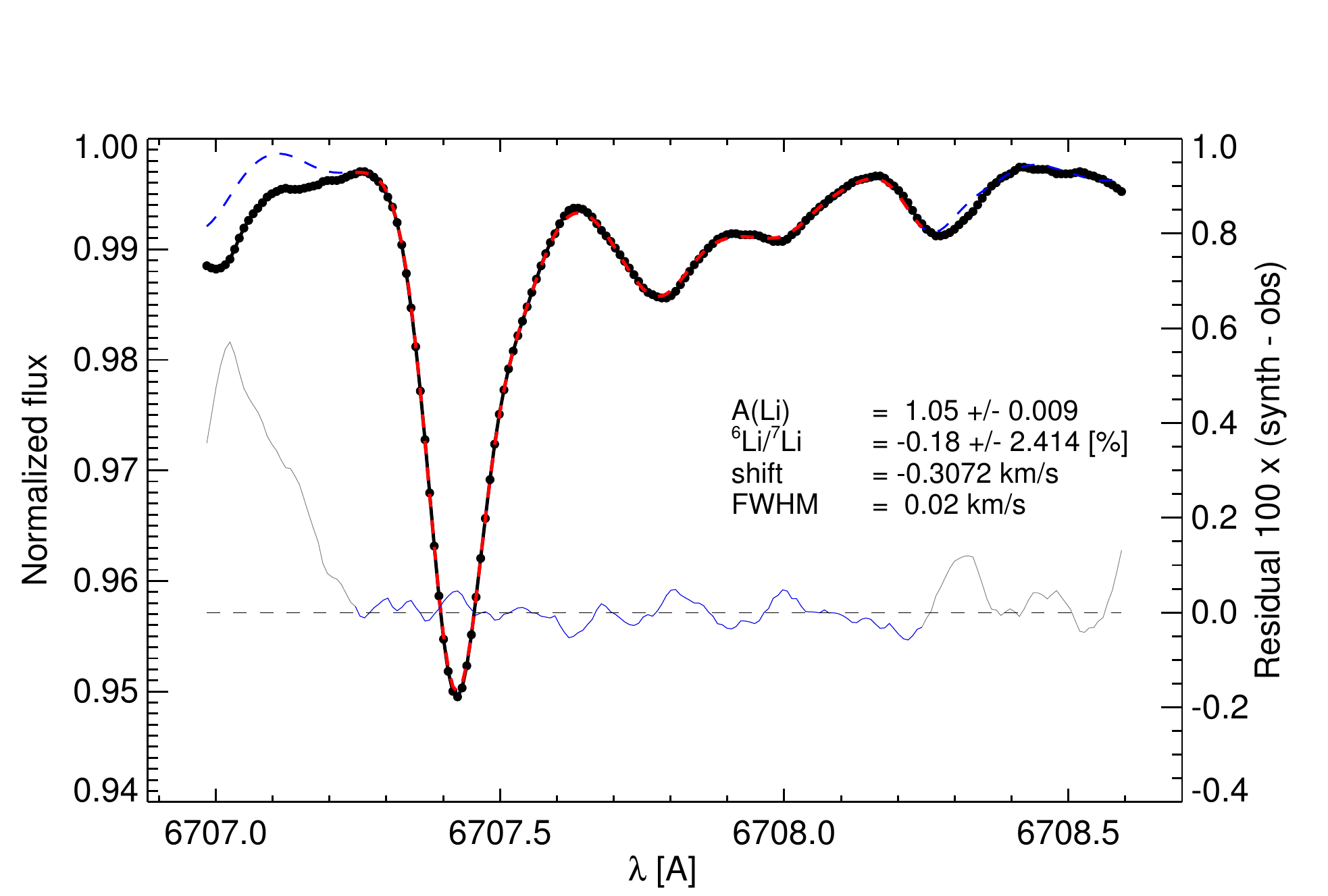}}
\caption{Best fits of the solar flux spectrum of \cite{kurucz2005} 
(black dots) with different sets of 3D NLTE \cobold\ spectra (red dashed line),
adopting the same continuum level as in Fig.\,\ref{bestfit_sun_lhd1}.
From top to bottom, \textit{Cases~IIa, IIb}, and \textit{IIc} are all based on 
line list G09, but assume different chemical abundances and rest wavelengths 
of the blend lines (see text and Table\,\ref{t-solarfits} for details).
The right y-axis defines the scale of the residuals (thin blue line).}
\label{bestfit_sun_3d}
\end{figure}

In \textit{Case~Ia}, we use the same line list (G09) as in \textit{Case~I}
together with the same 1D-optimized individual abundances obtained from 
the best fit with the 1D LHD spectra. Under these constraints, 
the best fit requires \ali\,=\,$1.05$ and \qli\,=\,$1.86\%$. As 
expected, the fit is rather poor (see Fig.\,\ref{bestfit_sun_3d}), 
top panel).

At the next stage (\textit{Case~Ib}), we adjust the strengths of the CN, Fe,
Si, and V blend lines according to the 3D-1D abundance corrections inferred
from fitting the 3D synthetic line profiles of the individual blend lines by a
grid of corresponding 1D spectra of different abundance, as described for the
Li and Si lines in Appendix\,\ref{ap_test_Li_Si}. The fit with these adjusted
abundances is shown in Fig.,\ref{bestfit_sun_3d} (middle panel), indicating
\ali\,=\,$1.04$ and a strongly negative \iso\ ratio of less than $-2\%$. While
the fit is generally improved with respect to \textit{Case~Ia}, the Fe+CN
blend at $\lambda\,670.74$\,nm is by far not reproduced to the level of
precision achieved with the 1D LHD fit (cf.\ Fig.\,\ref{bestfit_sun_lhd1}).

A 3D fit of comparable quality can be accomplished by readjusting the rest
wavelengths of the main blend components according to their differential
convective Doppler shifts. The latter are again deduced from fitting the
synthetic 3D line profiles by their 1D counterparts, as described for the Li
and Si lines in Appendix\,\ref{ap_test_Li_Si}.  In \textit{Case~Ic}, we
applied both these `theoretical' convective wavelength shifts and related
abundance corrections to the CN, Fe, Si, and V blend lines. Indeed, the
applied wavelength shifts\footnote{The asymmetric 3D profile of the 
Fe line at $\lambda\,670.74$\,nm is represented by two components
in 1D.} lead to a dramatic improvement of the fit (see
Fig.\,\ref{bestfit_sun_3d}), bottom panel). Now the best solution is found for
\ali\,=\,$1.05$ and \qli\,=\,$-0.18\%$.  We notice that the \iso\ ratio found
from this fine-tuned 3D fit is now in very good agreement with the result from
the LHD fit after application of the \delq\ correction ($0.69 - 0.83 =
-0.14\%$).
 
We performed the same exercise for a different continuum level,
$c_1=0.9980$ (\textit{Case~II}), which corresponds closely to the continuum
placement adopted for the analysis of the PEPSI spectrum in \cite{pepsi2018}.
We verified (\textit{Case~IIx}) that we obtain very similar results when 
adopting their set of abundances: \ali\,=\,$1.09$ and \qli\,=\,$1.77\pm 2.2\%$, 
compared to \ali\,=\,$1.09$ and \qli\,=\,$1.44\pm 1.6\%$ in \cite{pepsi2018}. 
Quantitatively, the two series of fits at slightly different continuum levels
show a systematic offset in the  absolute values of \ali\ and \qli, but otherwise
follow the same qualitative trends and lead to the same conclusions. Details of 
all fits discussed above can be found in Table\,\ref{t-solarfits} .

In summary, the fits of the solar spectrum by grids of 1D LHD and 
3D \cobold\ synthetic spectra computed with the full G09 line list
justify our definition of the 3D NLTE
corrections for \ali\ and \qli\ based on the comparison of 
3D NLTE and 1D LTE  Li-only line profiles. 
This conclusion is not limited to the solar case but should apply 
even more to the target stars of this paper, where a strong lithium
feature dominates over relatively weak blend lines.
 
We chose the Sun for demonstration because all involved
spectral lines are weak and therefore the combined spectrum may be
constructed by linear superposition of the individual components.
The presence of stronger lines would require a full spectrum 
synthesis of the complete spectrum for each individual configuration.
Under these conditions, a fine-tuned 3D (NLTE) fit to the observed spectra 
of our target stars would be next to impossible. However, the exercise 
presented in this Appendix also strongly suggests that using 3D model
spectra directly for the fitting is not really the most desirable 
approach since it requires non-trivial adjustments in the line lists 
that are fine-tuned  by 1D methods.

\begin{table*}[htb!]
\caption{Summary of the results of fitting the Li region of the solar flux
spectrum \citep{kurucz2005} with 1D LHD and 3D \cobold\ synthetic spectra.}

\begin{tabular}{l l c  c  l |  c  c  r }
\hline
\hline
\noalign{\smallskip}

Fit ID & Model & Continuum  & Abundances & Wavelengths & \ali  & \qli & $\chi^2$ \\
       &       & level      &            &             & (dex) & (\%) &          \\

\noalign{\smallskip}
\hline 
\noalign{\smallskip}
I & 1D \mlhd  & $0.9977$      & adjusted           & G09 & $0.94$ & $\;\;0.69$ &   $8.2$ \\ 
Ia  & 3D \cobold\ & $0.9977$  & as Fit I           & G09 & $1.05$ & $\;\;1.86$ & $251.3$ \\
Ib  & 3D \cobold\ & $0.9977$  & Fit I $+\Delta$\,A & G09 & $1.04$ & $<-2.0\;\;$ &  $53.6$ \\
Ic  & 3D \cobold\ & $0.9977$  & Fit I $+\Delta$\,A & G09 $+\Delta\lambda$ & $1.05$ & $-0.18$ &   $7.3$ \\
\hline \noalign{\smallskip}
II & 1D \mlhd & $0.9980$       & adjusted             & G09 & $0.96$ & $\;\;1.67$ &    $7.9$ \\ 
IIa  & 3D \cobold\ & $0.9980$  & as Fit II            & G09 & $1.06$ & $\;\;2.88$ & $247.3$ \\
IIb  & 3D \cobold\ & $0.9980$  & Fit II $+\Delta$\,A  & G09 & $1.06$ & $-0.89$ &  $54.0$ \\
IIc  & 3D \cobold\ & $0.9980$  & Fit II $+\Delta$\,A  & G09 +$\Delta\lambda$ & $1.06$ & $\;\;0.88$ &   $6.1$ \\
IIx  & 3D \cobold\ & $0.9980$  & SIS18              & G09 & $1.09$ & $\;\;1.77$ &   $29.5$ \\
\hline \noalign{\smallskip}
\hline
\end{tabular}
\tablefoot{Col.(4): $\Delta$\,A symbolizes the individual 3D abundance
  corrections; `SIS18' indicates the same abundances
as used in \cite{pepsi2018}. Column (5): G09 denotes the line list
of \cite{ghezzi2009}, $\Delta\lambda$ symbolizes the individual convective line shifts. The relative quality of the 
fits can be judged by comparing their $\chi^2$ (given on a common absolute scale).}
\label{t-solarfits}
\end{table*}

\end{appendix}

\end{document}